\documentclass[aps,prb,superscriptaddress,reprint,amsmath,amssymb]{revtex4-2}

\usepackage{graphicx}% Include figure files
\usepackage{dcolumn}% Align table columns on decimal point
\usepackage{bm}% bold math
\usepackage{hyperref}

\hypersetup{
    colorlinks=true,
    linkcolor=black,
    citecolor=black,
    filecolor=black,
    urlcolor=black,
}

\begin{document}

\title{Programming bistability in geometrically perturbed mechanical metamaterials}

\author{Yingchao Peng}
 \thanks{Equal contributors}
 \affiliation{Viterbi School of Engineering, University of Southern California, CA 90089}
\author{Imtiar Niloy}
 \thanks{Equal contributors}
 \affiliation{Department of Civil Engineering, Stony Brook University, NY 11794}
\author{Megan Kam}
 \affiliation{Department of Civil Engineering, Stony Brook University, NY 11794}
\author{Paolo Celli}
 \email{paolo.celli@stonybrook.edu}
 \affiliation{Department of Civil Engineering, Stony Brook University, NY 11794}
\author{Paul Plucinsky}
 \email{plucinsk@usc.edu}
 \affiliation{Viterbi School of Engineering, University of Southern California, CA 90089}

\date{\today}% It is always \today, today,

\begin{abstract}
Mechanical metamaterials capable of large deformations are an emerging platform for functional devices and structures across scales. Bistable designs are particularly attractive since they endow a single object with two configurations that display distinct shapes, properties and functionalities. We propose a strategy that takes a common (non-bistable) metamaterial design and transforms it into a bistable one, specifically, by allowing for irregular patterns through geometric perturbations of the unit cell and by leveraging the intercell constraints inherent to the large deformation response of metamaterials.  We exemplify this strategy by producing a design framework for bistable planar kirigami metamaterials starting from the canonical rotating-squares pattern. The framework comprises explicit design formulas for cell-based kirigami with unprecedented control over the shape of the two stable states, and an optimization methodology that allows for efficient tailoring of the geometric features of the designs to achieve target elastic properties as well as shape change. The versatility of this framework is illustrated through a wide variety of examples, including non-periodic designs that achieve two arbitrarily-shaped stable states. Quantitative and qualitative experiments, featuring prototypes with distinct engineering design details, complement the theory and shine light on the strengths and limitations of our design approach. These results show how to design bistable metamaterials from non-bistable templates, paving the way for further discovery of bistable systems and structures that are not simply arrangements of known bistable units.

\vspace{5px}
\normalsize{\textbf{This article may be downloaded for personal use only. Any other use requires prior permission of the author and the American Physical Society. This article appeared in}: \emph{Physical Review Applied} 22, 014073 (2024) \textbf{and may be found at}: \url{https://doi.org/10.1103/PhysRevApplied.22.014073}}
\end{abstract}

\maketitle

The past few decades have ushered in a paradigm shift in the way structural instabilities are perceived: once something to avoid, researchers now design mechanical systems and structures to reversibly buckle as part of their functionality~\cite{reis2015perspective, hu2015buckling, kochmann2017exploiting}. Bistable systems, possessing two morphologically-distinct stable states, are pervasive in nature and attractive in engineering. They are key to how a venus fly trap collects its prey~\cite{forterre2005venus} and a  beetle unfurls its wings~\cite{faber2018bioinspired}, and are leveraged in deployable space structures~\cite{Pellegrino2001}, soft robots~\cite{chen2018harnessing, tang2020leveraging, chi2022bistable}, MEMS devices~\cite{saif2000tunable, cao2021bistable}, and the like. They are even the cornerstone to popular everyday objects like PopSockets phone holders and toys like PopIt fidgets and jumping poppers~\cite{lapp2008exploring, pandey2014dynamics}.

Bistability comes in a variety of modalities.  It is achieved by purely geometric means through tailoring the design of thin-wall structures, like arches and shells~\cite{vangbo1998analytical}, or by designing clever ensembles of structural elements connected by pins or flexural joints~\cite{rafsanjani2016bistable}. It also emerges through a combination of geometry and material-induced rigidity, as in the case of 
bistable shells made of composite materials~\cite{iqbal2000bi}, or through prestress, as when multistable shells are obtained from pre-stretched 
strips~\cite{chen2012nonlinear}. Bistability can even be facilitated by nuanced features in origami and kirigami, including hinges with limited motion range~\cite{li2021metamorphosis, zhang2022kirigami} and creases with directional bias~\cite{silverberg2014using, hanna2014waterbomb}. 

\begin{figure}[t!]
\centering
\includegraphics[scale=0.99]{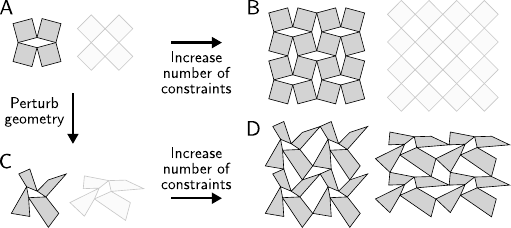}
\caption{Geometrical route to bistability. (A) Rotating-squares unit cell, with its mechanism motion shaded in light gray. (B) Periodic and overconstrained version of (A). (C) Perturbed version of (A). (D) Bistable structure obtained by tiling perturbed unit cells, in its first and second stable states.}
\label{fig:idea}
\end{figure}

Here we focus on purely geometrical routes to bistability, since design principles in this setting are broadly applicable to a variety of scales, materials, and manufacturing methods. The archetype of a purely geometrical bistable structure is the von Mises truss~\cite{vonMises1923}, featuring two inclined elastic bars connected at a hinge and constrained by pin supports. When this structure is loaded at its apex, the bars compress and eventually snap into an inverted tent shape, a second stable state by symmetry. This basic principle leads to a wealth of bistable systems. Cleverly arranged von Mises trusses form the basis for bistable planar lattices~\cite{Haghpanah2016} and kirigami~\cite{rafsanjani2016bistable}. Other complex multistable systems arise by exploiting snapping arches and domes ~\cite{seffen2006mechanical, schioler2007space, chen2011synthesis, Shan2015, Restrepo2015, ion2017digital, chen2017integrated, liu2019architected},  or more sophisticated bistable units like the square twist origami cell \cite{silverberg2015origami}. The underlying philosophy to all these works is the same: the key designer input to a complex bistable system is a bistable building block.

Our work seeks to break away from this prevailing philosophy. Towards this goal, it is notable that many motifs found in origami are bistable even though their basic building blocks are floppy. Examples include Kresling, waterbomb and helical tubes \cite{jianguo2015bistable,kuribayashi2006self,feng2020helical}, hypar \cite{filipov2018mechanical} and the typical origami flasher \cite{lang2016single, arya2021origami}. However, as is often the case with origami, most of these examples have artistic origins and their bistability is serendipitous. The question of what makes them, or any  ``pattern'',  bistable remains largely unexplored.

Our objective is to highlight the role of \textit{geometric perturbations} as a fundamental ingredient for bistability and to show that such perturbations, when suitably applied to fairly generic families of patterns, can be used as a versatile platform to design bistable metamaterials with a wide range of target properties.  Fig.~\ref{fig:idea} illustrates the key ideas applied to a prototypical 2D morphing metamaterial called the rotating squares pattern \cite{grima2000, tang2017design}. A single unit cell of this pattern (Fig.~\ref{fig:idea}A) is capable of changing its shape through a mechanism \cite{Pellegrino1986} or floppy mode \cite{lubensky2015phonons} given by counter-rotating its panels about the central slit. Symmetry, in turn, makes the intercell constraints redundant: periodic tiling of the cell (Fig.~\ref{fig:idea}B) yields a pattern that exhibits the same such mechanism. Breaking the design symmetry, however, reveals an interesting dichotomy: the cell continues to possess a floppy mode (Fig.~\ref{fig:idea}C), but the overall pattern does not (Fig.~\ref{fig:idea}D).  The intercell constraints are generically no longer redundant. Our main insight is that, while generic perturbations yield monostable patterns, careful geometric tuning of these perturbations can turn mechanism-based designs into bistable ones (Fig.~\ref{fig:idea}D).

There has been a thrust in recent years to go beyond metamaterials made of canonical unit cells and provide optimization tools to explore the geometry-property relationships in these systems.  Singh and van Hecke~\cite{singh2021design} and Deng et al.~\cite{deng2022inverse} show that optimizing the geometric features of perturbed rotating squares designs can yield a rich range of target elastic properties.
%, including designs that behave as pseudo-mechanisms or have multiple stable configurations. 
Mahadevan and colleagues \cite{choi2019programming,dudte2016} use global optimization frameworks to produce non-periodic generalizations of well known origami and kirigami metamaterials with target shapes. Hard-encoding design rules in a metamaterial (e.g., for the panels to rotate about flexible hinges or folds), as in Refs.~\cite{feng2020designs,dang2022inverse,dudte2021additive,walker2022algorithmic,dudte2023additive}, yield marching algorithms that improve the optimization schemes, enabling further demonstrations of programmability.

Our work builds on these ideas under the lense of bistability and with particular emphasis on  practical (reduced-order) design tools that can guide experiments at the conceptual/prototyping phase of design.  We start by showing how to hard-encode bistability in a large class of 2D periodic metamaterials composed of   repeating unit cells of panels and slits,  termed \textit{planar kirigami} herein and elsewhere~\cite{choi2019programming, choi2021compact, zheng2022continuum}, as opposed to kirigami that rely on out-of-plane buckling~\cite{rafsanjani2017buckling, yang2018multistable}.  We then introduce an optimization framework for bistable planar kirigami that incorporates a reduced-order model for the elastic energy, allowing us to tune the designs to achieve target morphing and elastic properties. A suite of representative examples and corresponding experiments follows.  We explore  examples ranging  from classical monolithic planar kirigami configurations to pin-jointed panel systems and truss-based  analogs to test the applicability of our design and optimization strategies in a variety of settings.  In each case,   the experiments validate the bistable behavior of the patterns and certain
qualitative features of their elastic properties, but also highlight how our theory can guide but not fully replace high-fidelity models and prototyping. Finally, we  showcase the versatility of our design approach by extending it to non-periodic  systems with complex shape-change.

\section*{Design formulas for bistable planar kirigami}

\begin{figure}%[tbhp]
\centering
\includegraphics[scale=0.99]{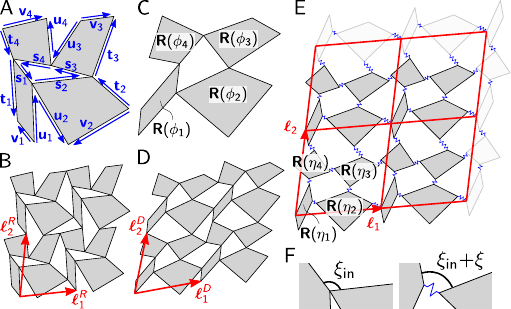}
\caption{Perturbed rotating-squares kirigami and design recipe. (A) Unit cell in its first stable state, indicating the design vectors, and (B) periodicity constraint. (C), (D) Second stable states of (A) and (B), respectively, obtained by rotating each panel by an angle $\phi_i$. (E) Elastic energy model. From (B), each panel is rotated and translated periodically to produce a homogeneous effective deformation with Bravais lattice vectors $\boldsymbol{\ell}_1$ and $\boldsymbol{\ell}_2$. As illustrated, the stored energy is calculated assuming linear springs between the separated panels. (F) Details of the two bottom-left panels of (B) and (E), indicating a change in opening angle.}
\label{fig:design}
\end{figure}

We begin by developing a general recipe for bistable planar kirigami comprised of a repeating unit cell of four quad panels and four quad slits, obtained through geometric perturbations of the rotating squares. The recipe amounts to  a compact design formula for bistability, which we explain using Fig.~\ref{fig:design} as a guide. Fig.~\ref{fig:design}A  shows a generic unit cell representing the first stable state of a quad kirigami design, with the panels and slits labeled by 2D vectors $\mathbf{s}_i$, $\mathbf{t}_i$, $\mathbf{u}_i$ and $\mathbf{v}_i$, $i =1,\ldots,4$.  Fig.~\ref{fig:design}B illustrates its periodicity using the 2D Bravais lattice vectors  $\boldsymbol{\ell}_1^R$ and $\boldsymbol{\ell}_2^R$. Fig.~\ref{fig:design}C shows the unit cell of the second stable state. Each panel in the cell is  rotated in the plane by a right-hand rotation $\mathbf{R}(\phi_i)$ of angle $\phi_i$, as shown, taking the initial cell vectors to deformed ones by the transformations $\mathbf{s}_i \mapsto \mathbf{R}(\phi_i) \mathbf{s}_i$, $\mathbf{t}_i \mapsto \mathbf{R}(\phi_i) \mathbf{t}_i$, $\mathbf{u}_i \mapsto \mathbf{R}(\phi_i) \mathbf{u}_i$ and $\mathbf{v}_i \mapsto\mathbf{R}(\phi_i)\mathbf{v}_i$,  $i = 1,\ldots,4$. Finally,  Fig.~\ref{fig:design}D shows 2D Bravais lattice vectors $\boldsymbol{\ell}_1^D$ and $\boldsymbol{\ell}_2^D$, quantifying the periodicity in the second stable state.  All of these vectors  and rotation angles  are subject to a variety of constraints for a compatible design, including  equality constraints that enforce periodicity and ensure  the vectors form closed loops about each panel and slit, as well as  inequality constraints that ensure the panels are convex quadrilaterals that do not  overlap. We enumerate all the compatibility conditions in \textit{Supplemental Material, Section S1.A} and manipulate them into forms  broadly useful for design in \textit{Supplemental Material, Sections S1.B and C}~\cite{suppl}. 

The  key result is a design formula  that parameterizes all the equality constraints. It takes the form
\begin{equation}
\begin{aligned}\label{eq:theorem1}
&\begin{bmatrix} \mathbf{s}_3 \\ \mathbf{s}_4 \\ \mathbf{t} \\ \mathbf{u} \\ 
\mathbf{v} \end{bmatrix} = \mathbf{D}(\boldsymbol{\phi}) \begin{bmatrix} \mathbf{s}_1 \\ \mathbf{s}_2 \\ \boldsymbol{\ell} \end{bmatrix}, 
  \end{aligned}
\end{equation}
where $\mathbf{t}, \mathbf{u}$ and $ \mathbf{v}$ stack the corresponding $\mathbf{t}_i, \mathbf{u}_i$ and $\mathbf{v}_i$ design vectors into 8 component arrays, and $\boldsymbol{\ell}$ does likewise for the four Bravais lattice vectors $\boldsymbol{\ell}_{1}^R, \ldots, \boldsymbol{\ell}_2^D$. The $28\times12$ matrix $\mathbf{D}(\boldsymbol{\phi})$, concretely linking these arrays, is a lengthy nonlinear expression of the rotation angles $\boldsymbol{\phi} = (\phi_1, \ldots, \phi_4)$. Its explicit formula is provided in  Eq.~[S21] of \textit{Supplemental Material, Section S1.B}~\cite{suppl}.

Eq.~[\ref{eq:theorem1}] organizes a wealth of information for designing and tuning bistable kirigami structures. The right-side contains the designer inputs. It includes the Bravais lattice vectors, which are the natural descriptors for target maximum stretch, Poisson's ratio, and shearing between the two stable states.  It also contains eight additional degrees of freedom (DOFs) through the four rotation angles $\boldsymbol{\phi}$ and design vectors $\mathbf{s}_1$ and $\mathbf{s}_2$. Each can be tuned to achieve, for instance, a desirable energy barrier between the two stable states.  The  remaining parameters  describing the designs in Fig.~\ref{fig:design} are all stacked on the left-side of Eq.~[\ref{eq:theorem1}] and thus are fully  determined from these designer inputs.

\section*{Optimization framework for elastic tuning}
An appealing aspect to this characterization is that it marries naturally with standard optimization tools to furnish a versatile design framework for tuning bistability. Assume a designer has in mind two stable states, obtained by prescribing the Bravais lattice vectors in the reference $\overline{\boldsymbol{\ell}_{1}^R}, \overline{\boldsymbol{\ell}_{2}^R}$ and deformed $\overline{\boldsymbol{\ell}_{1}^D}, \overline{\boldsymbol{\ell}_{2}^D}$ configurations. Since there are eight additional parameters on the right-side of  Eq.~[\ref{eq:theorem1}], the design can be optimized to achieve any general objective that can be written as a minimization problem:
\begin{equation}
\begin{aligned}
\label{eq:fobj}
\min ~& \big\{ f_{\text{obj}} ( \mathbf{s}_1, \mathbf{s}_2, \boldsymbol{\phi}) 
~ \big| ~ \mathbf{g}_{\text{ineq}} ( \mathbf{s}_1, \mathbf{s}_2, \boldsymbol{\phi}) \geq 0 \big\}. 
\end{aligned}
\end{equation}
In this formulation, $\mathbf{g}_{\text{ineq}} ( \mathbf{s}_1, \mathbf{s}_2, \boldsymbol{\phi})$ lists all the inequality constraints that are necessary and sufficient for the pattern to have convex  panels and slits in its reference and deformed stable states. These constraints, which are written out explicitly in \textit{SI  Appendix, Section S1.C}, are nonlinear in all their arguments. Thus, Eq.~[\ref{eq:fobj}] describes a constrained nonlinear optimization for which Matlab's \verb+fmincon+ toolbox provides several well-developed and efficient numerical tools to find local minimizers. In other words, this optimization framework is "ready-made" for engineering design.   See the  flowchart in Fig.S3 and \textit{Supplemental Material, Section S2.E}~\cite{suppl} for additional details on numerical aspects of this framework.

All that remains now is to prescribe an objective function for the optimization. We are particularly interested in objective functions that can assess and optimize  a variety of features of the stored elastic energy of the kirigami design. The challenge is that calculating an elastic energy based on high fidelity modeling, like FEM or even bar-hinge based modeling~\cite{Liu2017}, is not efficient and thus creates a bottleneck in the optimization process. We instead develop an elastic model that can be implemented directly into Matlab and evaluated using its fast solvers.

Our approach is formulated in detail in \textit{Supplemental Material, Section S2.A}~\cite{suppl} and illustrated in Fig.~\ref{fig:design}E. We model the corner points of the kirigami pattern as linear springs of zero rest length and unit stiffness and allow the panels to rotate and translate by a periodic motion that matches a bulk deformation  expressed by the Bravais lattice vectors $\boldsymbol{\ell}_1, \boldsymbol{\ell}_2$ in the figure. This deformation elongates the springs, generating an elastic energy expressed in terms of the panel rotations and translations, and the Bravais lattice vectors. 

After minimizing out the translations and lattice vectors in \textit{Supplemental Material, Section S2.B}~\cite{suppl}, we obtain the revealing form for the energy
\begin{equation}
    \begin{aligned}\label{eq:springEnergyMain}
        &E_{\text{spr}}(\eta_1, \eta_2, \eta_3, \eta_4) = \\
        &\qquad \begin{bmatrix}\sum_{i = 1,\ldots, 4} \mathbf{R}(\eta_i) \mathbf{s}_i \\  \sum_{i = 1,\ldots, 4} \mathbf{R}(\eta_i) \mathbf{t}_i \\  \sum_{i = 1,\ldots, 4} \mathbf{R}(\eta_i) \mathbf{u}_i \\ \sum_{i = 1,\ldots, 4} \mathbf{R}(\eta_i) \mathbf{v}_i \end{bmatrix} \cdot \mathbf{G} \begin{bmatrix}\sum_{i = 1,\ldots, 4} \mathbf{R}(\eta_i) \mathbf{s}_i \\  \sum_{i = 1,\ldots, 4} \mathbf{R}(\eta_i) \mathbf{t}_i \\  \sum_{i = 1,\ldots, 4} \mathbf{R}(\eta_i) \mathbf{u}_i \\ \sum_{i = 1,\ldots, 4} \mathbf{R}(\eta_i) \mathbf{v}_i \end{bmatrix}
    \end{aligned}
\end{equation}
for an $8\times8$ symmetric and positive definite  matrix $\mathbf{G}$ and the four panel rotations $\mathbf{R}(\eta_1), \ldots, \mathbf{R}(\eta_4)$, as shown in the figure. The heuristics behind this energy are as follows.  Prior to deformation, the pattern's four types of slits satisfy $\sum_{i=1,\ldots,4} \mathbf{s}_i = \mathbf{0}, \ldots, \sum_{i=1,\ldots,4}\mathbf{v}_i = \mathbf{0}$ because slits form closed loops. However, the deformed loops $\sum_{i = 1,\ldots,4} \mathbf{R}(\eta_i)\mathbf{s}_i, \ldots,\sum_{i = 1,\ldots,4} \mathbf{R}(\eta_i)\mathbf{v}_i$ are typically broken ($\neq \mathbf{0})$ under the panel motions. Eq.~[\ref{eq:springEnergyMain}] employs these broken loops as the fundamental measures of elastic strain in the pattern. The matrix $\mathbf{G}$ in this formula quantifies how the slits influence each other elastically. Its components range between values 0 and 1 independent of the krigami design, and are reported in Eq.~[S36] of \textit{Supplemental Material, Section S2.B}~\cite{suppl}.

A final minimization allows us to quantify the elasticity of a bistable kirigami design in terms of a single kinematic variable:
\begin{equation}
    \begin{aligned}
        &E_{\text{act}}(\xi, \mathbf{s}_1, \mathbf{s}_2, \boldsymbol{\phi})  =\\
        &\qquad \min_{\eta_3, \eta_4} \big\{  E_{\text{spr}}(0, \xi, \eta_3, \eta_4)  ~\big| ~\mathbf{s}_3, \mathbf{s}_4, \mathbf{t}, \mathbf{u}, \mathbf{v} \text{ solve  Eq.~[\ref{eq:theorem1}]} \big\} .
    \end{aligned}
\end{equation}
We call this energy the \textit{actuation energy}. It depends kinematically only on the angle $\xi$ shown in Fig.~\ref{fig:design}F, describing the relative rotation between the first and second panel of each unit cell as the overall pattern is actuated. It is non-negative and satisfies  $E_{\text{act}} = 0$ when $\xi = 0$ and $\xi = \phi_2 - \phi_1$, reflecting the bistability hard encoded by  Eq.~[\ref{eq:theorem1}].

This actuation energy integrates seamlessly with the optimization framework in Eq.~[\ref{eq:fobj}], allowing us to efficiently explore and tune elastic properties of the kirigami design. In this work, we demonstrate this capability by optimizing the designs based on two properties of  $E_{\text{act}}$ through objective functions of the form 
\begin{equation}
    \begin{aligned}\label{eq:fobjFormula}
        f_{\text{obj}}(\mathbf{s}_1, \mathbf{s}_2, \boldsymbol{\phi}) &= \underbrace{c_b |E_b(\mathbf{s}_1, \mathbf{s}_2, \boldsymbol{\phi}) - E_{b}^{\text{targ}}|^2}_{\text{target energy barrier}}  \\
        &\quad + \underbrace{ c_1|k_{1}(\mathbf{s}_1, \mathbf{s}_2, \boldsymbol{\phi}) - k_1^{\text{targ}}|^2}_{\text{target stiffness}}.
    \end{aligned}
\end{equation}
 The first term tunes the designs so that the energy barrier between the designed stables states,  $E_b(\mathbf{s}_1, \mathbf{s}_2, \boldsymbol{\phi}) = \max_{\xi\in(0, \phi_2-\phi_1)} E_{\text{act}}(\xi, \mathbf{s}_1, \mathbf{s}_2, \boldsymbol{\phi})$, is driven towards a specified target $E_{b}^{\text{targ}} \geq 0$,  reflecting the amount of work needed to actuate the pattern from one stable state to the other.   The second term optimizes for the stiffness $k_1(\mathbf{s}_1, \mathbf{s}_2, \boldsymbol{\phi}) = \partial_{\xi} \partial_{\xi} E_{\text{act}}(0,\mathbf{s}_1, \mathbf{s}_2, \boldsymbol{\phi})/(\lambda'(0))^2$  of the first  stable state with respect to a \textit{characteristic stretch} $\lambda(\xi) = \tfrac{| \mathbf{u}_1 -\mathbf{v}_1 + \mathbf{R}(\xi) (\mathbf{u}_2 - \mathbf{v}_2)|}{|\mathbf{u}_1 -\mathbf{v}_1 + \mathbf{u}_2 - \mathbf{v}_2|}$ that takes the value $\lambda(0) =  1$ in the first stable state  
 and $\lambda(\phi_2 - \phi_1) = |\boldsymbol{\ell}_1^D|/|\boldsymbol{\ell}_1^R|$ in the second one (see \textit{Supplemental Material, Section S2.D}~\cite{suppl} for more details). The designs are tuned by driving this  stiffness  towards a specified target $k_1^{\text{targ}} \geq 0$, allowing us to control whether we want this stable state to be "locked-in" or have some give at its typical performance loads.   Design tradeoffs are expected.  For example, it is not usually possible to achieve a design that has both a high stiffness  but a low overall energy barrier between the states. The numerical parameters $c_b, c_{1} \geq 0$ express the desired importance of each  term during an optimization. Though not done here, terms like the maximum force and/or the stiffness of the second stable state can also be included in the objective function.

\section*{Representative example}

\begin{figure*}[htb]
\centering
\includegraphics[scale=0.99]{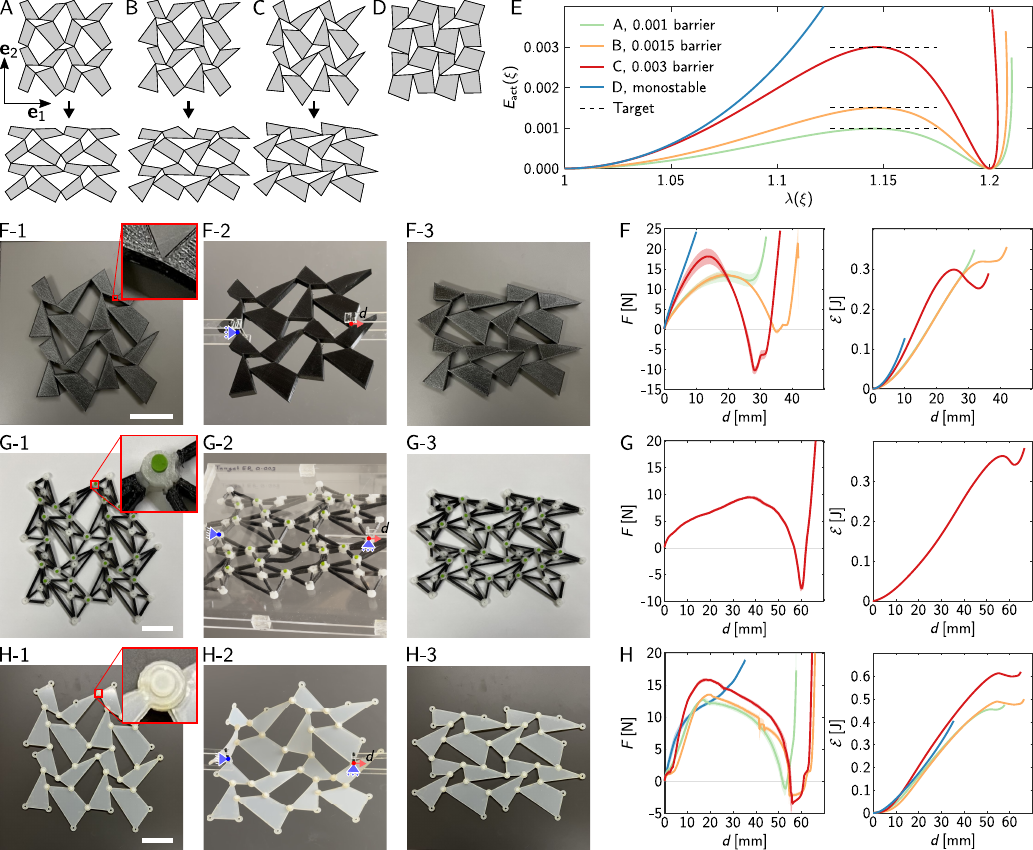}
\caption{Representative example and demonstration of bistability. (A-C) First and second stable states of the optimized patterns with energy barriers 0.001, 0.0015 and 0.003, respectively. Even though we represent them in 2$\times$2 cell versions, these patterns are periodic. (D) Monostable geometry obtained by randomly perturbing the rotating squares pattern. (E) Theoretical energy landscape of the patterns in (A-D), with horizontal lines indicating the target energy barriers; the legend shown here is valid throughout this figure. (F) Experimental results for a monolithic physical realization of the representative patterns, which involves compliant hinges; solid lines are averages of three tensile tests, and shaded areas represent the standard deviation; energy curves are obtained by numerically integrating the average force-displacement curves. (F-1, F-2, F-3) First stable state, snapshot of the deformation and second stable state, respectively, for a monolithic specimen with 0.003 barrier. (G) Same as (F), but for a physical realization of the pattern involving skeletal, bi-material panels connected via perfect pins and capable of in-plane deformation via buckling of the internal beams. (H) Same as (F), but for a physical realization involving thin, pin-jointed mono-material panels that bend out-of-plane during deformation. Scale bar: 5\,cm.}
\label{fig:rep}
\end{figure*}

We illustrate the optimization framework by tuning kirigami designs to achieve  a variety of  energy barriers under a prototypical square-to-rectangle transformation.   In the optimization, the Bravais lattice vectors  are set at    $\overline{\boldsymbol{\ell}_{1}^R}= \mathbf{e}_1, \overline{\boldsymbol{\ell}_{2}^R}= \mathbf{e}_2$ and $\overline{\boldsymbol{\ell}_{1}^D}=1.2 \mathbf{e}_1$, $\overline{\boldsymbol{\ell}_{2}^D}=0.8 \mathbf{e}_2$ to encode the effectively square and rectangular stable states, and the moduli  in  Eq.~[\ref{eq:fobjFormula}]  are taken as $c_b = 1$ and $c_{1} = 0$ to focus on optimizing for a target energy barrier. Fig.~\ref{fig:rep}A-C shows three optimized designs, obtained 
 by prescribing the target energy barrier from left to right as $E_b^{\text{targ}} = 0.001,   0.0015 , 0.003$  and performing the minimization in Eq.~[\ref{eq:fobj}] in each case. A plot of actuation energy $E_{\text{act}}(\xi)$ versus  stretch $\lambda(\xi)$ in Fig.~\ref{fig:rep}E shows that each design achieves its target energy barrier.
 A randomly generated monostable design in Fig.~\ref{fig:rep}D is included in Fig.~\ref{fig:rep}E as another point of comparison.

To test the validity of the optimization framework, we fabricate a series of prototypes for all the aforementioned  designs and examine their  elastic energy  and bistability experimentally. In comparing the theory to experiments, it is important to recognize that our mechanical model for the actuation energy is based on simplifying assumptions that  enable  us to efficiently optimize over  the purely geometric parameters of the designs to achieve "some notion" of  target elastic properties. The goal is to  provide design guidance in this large parameter space that gets trends right. Specifically, for a specified  fabrication strategy,  we expect that the energy barrier of the design in Fig.~\ref{fig:rep}A is smaller than that of Fig.~\ref{fig:rep}B, which in turn is smaller than that of Fig.~\ref{fig:rep}C. We cannot, however, say much more than that:  localized buckling, hinge elasticity, friction, viscoelasticity, and out-of-plane deformation can influence the elastic behavior of these systems, none of which are accounted for in our model. 

We propose three different fabrication strategies that differ in the way the energy barriers manifest as deformation, allowing us to explore the interplay between  the fine design details inherent to prototyping and  the theoretical predictions of bistability.
In all cases, the specimens are made of 2$\times$2 unit cells and are tested in tension via a universal testing system using custom fixtures.  Fig.~\ref{fig:rep}F-H  show the raw force-displacement curves for tension tests of the fabricated samples, as well as  their stored energy curves (obtained by integration of the force curves). The color scheme for the curves distinguishes the different designs, just as in Fig.~\ref{fig:rep}E. Additional details on the fabrication and experimental procedures, and on the dimensions of specimens, are given in \textit{Supplemental Material, Sections S4, S5}~\cite{suppl}. We only report experimental results in the main text; \textit{Supplemental Material, Section S7}~\cite{suppl} reports finite-element results that complement these  findings.

In the first incarnation of our designs, we   3D print  thick monolithic specimens made of "soft" TPU-95, which represents the most conventional way to fabricate these metamaterials~\cite{rafsanjani2016bistable, tang2017design, choi2019programming, wu2023situ}. Notably, each sample exhibits negligible out-of-plane deformations due to its large thickness, but  also has elastic hinges that offer some resistance to the relative rotations between panels. Fig.~\ref{fig:rep}F-1 shows the specimen with the largest (0.003) theoretical energy barrier in its first stable state; Fig.~\ref{fig:rep}F-2 shows a snapshot of its in-plane deformation process during a tension test, with the distortions mostly concentrated in the hinge regions; Fig.~\ref{fig:rep}F-3 shows the second stable state, which is qualitatively similar to the theoretical one, while featuring some localized bending near the hinges. This case is clearly bistable ---  the force-displacement curve dips below zero, resulting in two clear energy minima. However, bistability is far from  guaranteed because  hinge elasticity counteracts the geometric energy barriers that support a second stable state.   In fact, the other two cases with the smaller theoretical energy barriers of 0.001 and 0.0015 are not bistable, as indicated by their force and energy curves. 

Our second fabrication strategy eliminates the hinge elasticity that opposes bistability, while keeping the actuation essentially planar. To do this, we 3D print ``skeletal" and bi-material panels and assemble them  via actual pin joints. The printed panels are composed of soft TPU bars and stiff nylon hinge regions to ensure that the deformations concentrate in the bars  rather than in the neighborhood of the pin joints. We also place the specimens between clear acrylic plates during testing to  prevent out-of-plane deformation. Fig.~\ref{fig:rep}G-1 shows the 0.003  design fabricated in this fashion in its first stable state; Fig.~\ref{fig:rep}G-2 shows an intermediate  state during testing, illustrating how the deformation within the panels manifests as in-plane bending and buckling of the bars; Fig.~\ref{fig:rep}G-3 shows the second stable state, which now matches the theoretical one. The experimental force-displacement curves  for this specimen clearly indicate bistability. Note that the presence of pin-joints allows the specimen to be stress-free in its second stable state even though the curve shows that the energy is non-zero in this state. We attribute the tilted energy curve to friction and material viscoelasticity, which dissipate energy during the tests. While this second incarnation is a better candidate to demonstrate bistability for a broad range of designs, fabrication proved challenging and time consuming; for this reason, we only report results for the 0.003 specimen. 

The final incarnation also features pin joints, but the panels are now much thinner and laser cut out of PETG. As shown in Fig.~\ref{fig:rep}H-1 to H-3 for the 0.003 sample, these specimens transition between stable states via out-of-plane bending of the panels. The force and energy curves show that the patterns behave as expected --- all the theoretically bistable designs are indeed bistable and the magnitude of their energy barriers trends with that of the theory. In particular, the energy barrier for the 0.003 specimen is larger than the 0.0015 one, which is in turn larger than the 0.001 case.  Here,  significant frictional losses due to the panels pushing against each other and the rivets during out-of-plane deformation cause the energy curves to display non-zero values at the second equilibrium, even though these states are stress-free.

Overall, these case studies  validate our purely geometric design and optimization tools, as they showcase a variety of bistable metamaterials  whose  shape change matches the theory and whose energy barriers match the trends of the theory. We envision that the synergy between  optimization and prototyping can be improved by introducing ``non-universal" features into the objective function (Eq.~[\ref{eq:fobjFormula}]) that depend on the choice of fabrication strategy, although we do not pursue this further.

\begin{figure*}[!htb]
\centering
\includegraphics[scale=0.99]{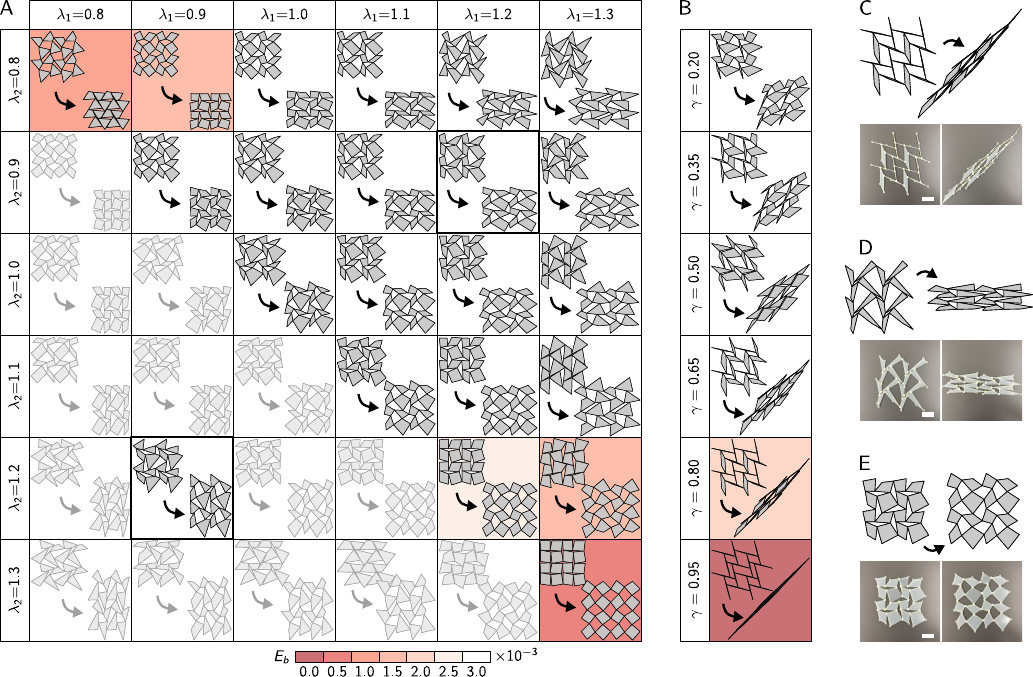}
\caption{Design explorations. (A) Matrix of designs optimized to attain axial shape transformation (from a square to another square or rectangle) of various degree, with a target energy barrier of 0.003. Designs are color-coded according to the actual value of the barrier achieved during optimization, as indicated in the legend below (B). Shaded designs in the lower diagonal are simply rotations of the corresponding designs in the upper diagonal. (B) Suite of designs that undergo a shear-type shape transformation. $\lambda_1$, $\lambda_2$ and $\gamma$, which control the magnitude of the shape change, are defined in Eq.~\ref{eq:axialShear}. (C-E) First and second stable states of various patterns, with their mono-material physical realizations. (C) Pattern designed to undergo extreme shear-type morphing, with $\gamma=0.8$; (D) Extreme non-auxetic morphing; (E) Extreme auxetic morphing. In all experimental images, scale bar: 5\,cm.}
\label{fig:explore}
\end{figure*}

\section*{Exploring the design and optimization space}

 We now highlight the richness of the design space by producing  bistable kirigami patterns that exhibit  a variety of axial and shearing shape changes. 
All examples correspond to reference lattice vectors $\overline{\boldsymbol{\ell}_1^R} = \mathbf{e}_1$ and $\overline{\boldsymbol{\ell}_2^R} = \mathbf{e}_2$ and achieve a second stable state given  by one of two parameterizations of the deformed lattice vectors 
 \begin{equation}\label{eq:axialShear}
     \begin{aligned}
         &\text{axial:} && \overline{\boldsymbol{\ell}_1^D} = \lambda_1 \mathbf{e}_1,  &&  \overline{\boldsymbol{\ell}_2^D} = \lambda_2 \mathbf{e}_2. \\&\text{shear:} &&  \overline{\boldsymbol{\ell}_1^D} = \mathbf{e}_1 + \gamma \mathbf{e}_2 ,&&  \overline{\boldsymbol{\ell}_2^D} = \mathbf{e}_2 + \gamma \mathbf{e}_1, 
     \end{aligned}
 \end{equation}
Fig.~\ref{fig:explore}A-E show designs obtained by optimizing the energy barrier using $c_b = 1, c_1 = 0$ and $E_b^{\text{targ}} = 0.003$ for a variety  of $\lambda_{1,2}$ and $\gamma$. 

Fig.~\ref{fig:explore}A, in particular, showcases a suite of designs corresponding to axial shape morphing with  $\lambda_{1}$ and $\lambda_2$ varied uniformly from $0.8$ (contraction) to $1.3$ (expansion) in a design matrix. The coloring scheme reflects whether the optimized design achieves the target $0.003$ energy barrier. As the coloring indicates, the richness of the design space depends significantly on the shape-morphing.  Non-auxetic designs where one side contracts and the other expands appear to be much more amenable to the large energy barriers than auxetic ones. In fact, the most extreme auxetic design -- the purely dilation one in the lower right corner of the figure -- is only a slight modification of the purely mechanistic rotating squares pattern, even though we optimize for a high energy barrier. This observation suggests that the rotating squares pattern is perhaps the singular template for extreme dilation in quad kirigami. Another interesting point concerns symmetry and non-uniqueness. As illustrated  in the design matrix,  every optimized design in the upper right  quadrant $(\lambda_1, \lambda_2) = (x,y)$ is related to one in the lower left $(\lambda_1, \lambda_2) = (y,x)$ by a $90^o$ rotation. These rotated designs are shaded in Fig.~\ref{fig:explore}A. In some of the less extreme cases, however, more than one design achieves the target energy and shape change. We illustrate this point by highlighting a $(\lambda_1, \lambda_2) = (0.9,1.2)$ optimized design that is distinct from the $(\lambda_1, \lambda_2) = (1.2,0.9)$ case shown.  

Fig.~\ref{fig:explore}B also showcases a suite of designs, this time for the shear case in Eq.~[\ref{eq:axialShear}], with $\gamma$ evolving uniformly from $ 0.2$ to $0.95$. Again the design space shrinks as the shear becomes more extreme, making it harder for the optimized design to achieve the target energy barrier $E_b^{\text{targ}} = 0.003$.  Note that the maximum shear in this setting is $\gamma_{\text{max}} = 1$, since this shape change takes  an effectively square reference domain to a line. Evolving the shear monotonically to this maximum leads to panels that degenerate to lines and slits to parallelograms.  A curious yet persistent observation in Fig.~\ref{fig:explore}A-B (and \textit{Supplemental Material, Section S2.F}\cite{suppl}) is that extremal shape change seems always to correspond to designs with parallelogram slits, i.e., designs known to always posses a single DOF mechanism \cite{zheng2022continuum}. Whether this  observation suggests  a universal relationship between  mechanism-based designs and bistable ones remains to be seen, but it is nonetheless compelling evidence of some sort of connection. 

Returning to prototyping, Fig.~\ref{fig:explore}C-E highlights the design and fabrication of  three examples of extreme shape change: a shear case, a non-auxetic case, and an auxetic, purely dilational case. In all cases, we employ the third fabrication strategy discussed above with pin-joints and laser cut PETG panels, since specimens made this way are easy to produce
and exhibit a "clean" performance.  As shown in the figure, each sample is  bistable and displays the predicted shape change. The supplementary videos listed and described in \textit{Supplemental Material, Section S6}\cite{suppl} provide further illustrations of the bistability of these samples.    

\begin{figure}[!htb]
\centering
\includegraphics[scale=0.99]{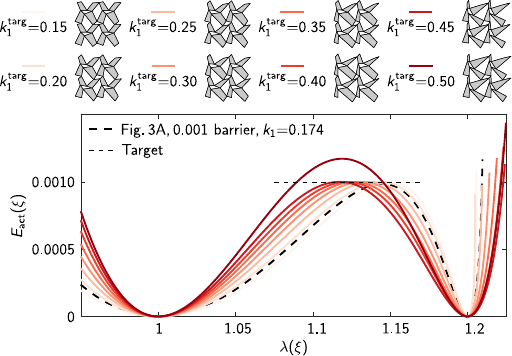}
\caption{Additional design explorations. Theoretical energy landscapes of patterns designed to display various target stiffness $k_1^{\text{targ}}$, while maintaining a barrier around 0.001. In the legend, we show the first stable states of all such patterns.}
\label{fig:explore2}
\end{figure}
  
 We end this section by showcasing designs tuned to achieve multiple  objectives at the same time. Going back to the prototypical square-to-rectangular transformation studied previously,  Fig.~\ref{fig:explore2} shows designs that have been optimized for both a target energy barrier and a target stiffness in its first stable state. Specifically, we fix $c_b = 1$, $c_1 = 0.002$ and $E_b^{\text{targ}} = 0.001$ and vary the target stiffness uniformly from $k_1^{\text{targ}} = 0.15$ to $0.50$ to produce eight optimized designs. As the plot indicates, we have control of both the stiffness and energy barrier over a wide range of the parameter space (from $k_1^{\text{targ}} =0.1$ to $0.45$). However, once $k_1^{\text{targ}}$ is sufficiently large, the energy barrier can no longer be held fixed. Instead it tilts up, reflecting a tradeoff between high stiffness and low energy.

\section*{Heterogeneous shape-change}

\begin{figure*}[htb]
\centering
\includegraphics[scale=0.99]{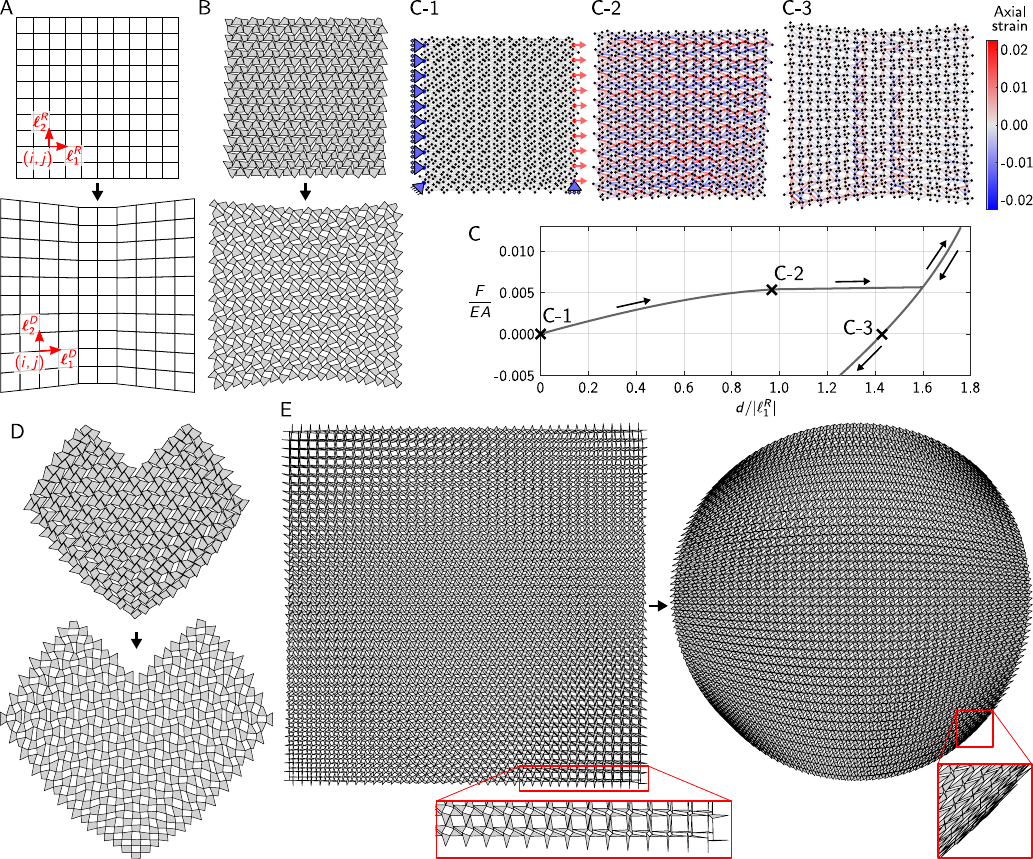}
\caption{Heterogeneous designs for complex shape change. (A-C) Example of a pattern designed to morph from a square to a bowtie shape. (A) Mesh grids of the two desired stable states, which give information on the lattice vectors to be prescribed to each unit cell. (B) Optimized pattern in its first and second stable states. (C) Numerical validation of bistability via a geometrically-nonlinear truss model, showing a force-displacement curve with arrows indicating the direction of loading/unloading and three snapshots of the deformation. Inset (C-1) indicates the boundary conditions for the simulation. (D) Beating heart pattern, in its first and second stable states. (E) Structure designed to morph from a square to a disc, in its first and second stable states.}
\label{fig:het}
\end{figure*}

We now go beyond periodic structures and show how our design and optimization methods can be integrated into a simple recipe to program bistable kirigami metamaterials with target heterogeneous shapes. We explain the approach through the square to bowtie transformation shown in Fig.~\ref{fig:het}A-C, before illustrating its versatility through the complex examples in Fig.~\ref{fig:het}D-E. 

In the transformation in Fig.~\ref{fig:het}A, the quad cells of the square mesh are mapped one-to-one  to the cells of the bowtie mesh through the lattice transformations $\boldsymbol{\ell}_1^R(i,j) \mapsto \boldsymbol{\ell}_1^D(i,j)$ and $\boldsymbol{\ell}_2^R(i,j) \mapsto \boldsymbol{\ell}_2^D(i,j)$, indexed by $(i,j)$ as shown. While  the bowtie mesh is not homogeneous, its lattice vectors vary slowly from cell to cell to produce the overall shape. We exploit these slow variations, and the key fact that the lattice vectors are designer inputs for bistability in Eq.~[\ref{eq:theorem1}], to obtain a  kirigami pattern capable of transforming from the square to the bowtie shape at very little stress.  

%These lattice vectors form the input to a marching algorithm

The general idea is as follows (see \textit{Supplemental Material, Section S3}\cite{suppl} for more details).  After an initialization step to seed the design of a  single kirigami cell, we produce a global kirigami pattern through a marching procedure that is fundamentally local.  At each $(i,j)$, indicating corresponding quads in the two meshes, we prescribe the lattice vectors for a bistable design in Eq.~[\ref{eq:theorem1}] as $\boldsymbol{\ell} = \boldsymbol{\ell}(i,j)$. Next, we choose the remaining DOFs in this equation as a set of minimizers $\mathbf{s}_1 =  \mathbf{s}_1(i,j), \mathbf{s}_2 = \mathbf{s}_2(i,j), \boldsymbol{\phi} =   \boldsymbol{\phi}(i,j)$ to Eq.~[\ref{eq:fobj}] with
\begin{equation}
\begin{aligned}
    f_{\text{obj}}(\mathbf{s}_1, \mathbf{s}_2, \boldsymbol{\phi}) = \big|(\mathbf{s}_1, \mathbf{s}_2, \boldsymbol{\phi}) - (\mathbf{s}^{\text{prev}}_1, \mathbf{s}^{\text{prev}}_2, \boldsymbol{\phi}^{\text{prev}})\big|^2,
\end{aligned}
\end{equation}
where $(\mathbf{s}^{\text{prev}}_1, \mathbf{s}^{\text{prev}}_2, \boldsymbol{\phi}^{\text{prev}})$ is from a previously computed neighboring cell.  Finally, we choose $\mathbf{s}_3 = \mathbf{s}_3(i,j),\ldots, \mathbf{v} = \mathbf{v}(i,j)$ to solve Eq.~[\ref{eq:theorem1}] for the above $(i,j)$ design variables. This recipe furnishes  a bistable kirigami cell that takes the shape of the two $(i,j)$-quads as its stable states. Iterating on it produces two global patterns, one with the desired overall  square shape and another with the desired bowtie. However, each has small gaps between neighboring unit cells due to the spatial variations of the lattice vectors (see Fig.~S5 in the \textit{Supplemental Material}\cite{suppl}). A final averaging step  glues each pattern together and furnishes the kirigami designs for the two shapes shown in Fig.~\ref{fig:het}B. 

The averaging part of the design procedure yields panels in the bowtie that are distorted slightly from their counterparts in the square pattern, meaning that the transformation is not stress-free. To verify bistability, we supplement the procedure with a truss based model of the pattern under loads in Fig.~\ref{fig:het}C, based on the bar-hinge model of Ref.~\cite{Liu2017}. The model assumes that  the square pattern is the stress-free reference configuration, that bars deform only axially, and that their material is linear elastic (see \textit{Supplemental Material, Section S7.D}\cite{suppl}). The pattern is then supported by rollers on its left boundary and loaded by a uniform set of horizontal nodal forces $F$ on its right (Fig.~\ref{fig:het}C-1). The overall horizontal displacement, denoted $d$, increases smoothly under force control until the configuration in Fig~\ref{fig:het}C-2, where it  jumps from $d/|\boldsymbol{\ell}_1^R| \approx 1$ to $d/|\boldsymbol{\ell}_1^R| \approx 1.6$ on a further increase of load.  Unloading after the jump, the curve crosses the zero force axis away from the origin, providing a demonstration of bistability. The bowtie shape of Fig.~\ref{fig:het}C-3, showing small residual strains, is the second equilibrium configuration.

The key point is not that we have identified a square-to-bowtie  bistable design, it is that the procedure is exceedingly simple and general. It only relies on the fact that we have a one-to-one "regular" quad mesh of the two stable states with lattice vectors that vary slowly from quad to quad (regular means that the meshes can be mapped  bijectively to a connected subset of the $\mathbb{Z}^2$ lattice). Such meshes are easy to obtain for a wide variety of shapes, so our procedure can be employed for a myriad of target bistable patterns.   Fig.~\ref{fig:het}D-E illustrate two such examples, a "beating heart" and a square-to-disc transformation.  

For the heart, we mesh the compact state (see Fig.~S7A in the \textit{Supplemental Material}\cite{suppl}) using  the  `Quasi-Structured Quad' setting from the freely available software Gmsh \cite{geuzaine2009gmsh}, and dilate this mesh to obtain the enlarged meshed state. The design procedure iterates through the cells in these meshes to produce the kirigami pattern.  We have studied a variety of dilation factors $\lambda \geq 1$. Increasing this factor monotonically leads to a compact state where the slits degenerate to lines (closed slits), and a dilated state where the slits tend towards being fully open. At sufficiently large  values for $\lambda$, the panels begin to overlap in the compact state,  violating the inequality constraints in Eq.~[\ref{eq:fobj}]. Fig.~\ref{fig:het}D shows a $\lambda = 1.25$ design obtained by our methods.  This level of dilation is actually quite impressive given the heterogeneity of the cells and  certain basic limitations of quad kirigami. The rotating squares pattern, for instance, transforms from its fully closed to fully open state by a uniform dilation of $\lambda_{rs} = \sqrt{2} \approx 1.41$, which likely sets the theoretical upper bound on dilation for these types of patterns.

Our last demonstration of heterogeneous morphing in Fig.~\ref{fig:het}E showcases a square-to-disc kirigami pattern. The quad mesh precursors are obtained by discretizing the square domain  uniformly on  $(-1,1)^2$ using a $40 \times 40$ set of square cells, and then deforming this mesh smoothly  to the unit disc by the Elliptical Grid mapping $(x,y) \mapsto (x\sqrt{1 - y^2/2},y\sqrt{1-x^2/2} )$. This mapping  spreads the mesh distortions smoothly from the interior to four singular points, corresponding to the deformed corners of the square (see Fig.~S7B in the \textit{Supplemental Material}\cite{suppl}). The design procedure accounts for this spreading by producing cells in the square whose slit area increases dramatically from its center to its corners; the dichotomy is reversed in the disc state. As shown in Fig.~\ref{fig:het}E, the corners of the square essentially collapse inward, closing their slits, to achieve the disc as the second stable state.  A key ingredient to this example's impressive display of shape change is the scalability of our methods. Since the design procedure is completely local, we can produce  kirigami designs with target shapes that involve thousands of  unit cells in a matter of minutes on a standard laptop.

\section*{Conclusions}

In summary, we have introduced a set of principles to rationally design and optimize bistable planar kirigami metamaterials, starting from a well known, non-bistable template. Importantly, the optmization tools incorporate a reduced-order elastic model, enabling on-demand design guidance for bistable kirigami with a rich variety of shape morphing capabilities and energy landscapes -- it provides experimentalists access to new designs in a matter of minutes. Optimized bistable designs are exemplified through a variety of fabrication strategies, with experimental results that largely match the shape change of the theory and the trends in the target elastic properties. Open questions remain. The fine details of  fabrication introduce features not present in the current design framework, e.g., hinge elasticity, beam buckling or panel bending. One challenge is to adapt the framework  to account for these details, while still remaining an efficient design tool. Another avenue concerns generality. The basic ingredients to our framework are geometric perturbations of a unit cell and intercell compatibility constraints.  As these ingredients are found across metamaterial templates, design formulas that hard encode bistability, like Eq.~[\ref{eq:theorem1}], are perhaps ripe for discovery in a wide range of metamaterials.

\vspace*{10pt}

\section*{Acknowledgments}

P.P. and Y.P. acknowledge support from the National Science Foundation (CMMI-CAREER-2237243) and from the Army Research Office (ARO-W911NF2310137). P.C. and I.N. acknowledge support from the National Science Foundation (CMMI-2045191). P.C. and I.N. thank Damiano Pasini and Christelle Combescure for constructive suggestions.

% \bibliography{PnasRef}% Produces the bibliography via BibTeX.


%%% SUPPLEMENT %%%

\clearpage
\widetext

% RESET FIGURE COUNTER TO 0
\setcounter{figure}{0}
\setcounter{equation}{0}
\setcounter{page}{1}
\setcounter{section}{0}
\renewcommand{\thefigure}{S\arabic{figure}}
\renewcommand{\thetable}{S\arabic{table}}
\renewcommand{\theequation}{S\arabic{equation}}
\renewcommand{\thesection}{S\arabic{section}}
\renewcommand{\thesubsection}{S\arabic{section}.\Alph{subsection}}
\renewcommand{\thepage}{S\arabic{page}}
\makeatletter

\section*{Supporting Information Text for ``Programming bistability in geometrically perturbed mechanical metamaterials''}

\section{Theory for designing bistable planar kirigami}

\begin{figure}[htb!]
\centering
\includegraphics[scale=1]{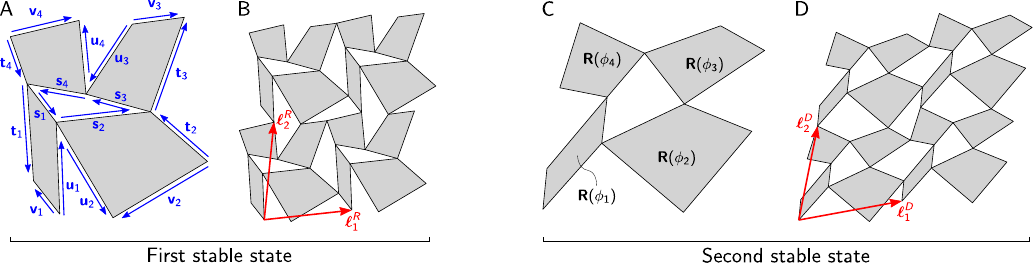}
\caption{Perturbed rotating-squares kirigami and design recipe (repetition of parts of Fig.~1 from the main manuscript). (A) Unit cell in its first stable state, indicating the design vectors, and (B) periodicity constraint. (C), (D) Second stable states of (A) and (B), respectively, obtained by rotating each panel by an angle $\phi_i$.}
\label{fig:designSup}
\end{figure}

\subsection{Design parameters and compatibility conditions} Fig.\;\ref{fig:designSup} illustrates the  2D vectors $\mathbf{s}_1, \ldots, \mathbf{v}_4, \boldsymbol{\ell}_{1}^R, \ldots, \boldsymbol{\ell}_{2}^D$ and  2D rotations 
\begin{equation}
\begin{aligned}
\mathbf{R}(\phi_i) = \begin{bmatrix}  \cos \phi_i & - \sin \phi_i \\ \sin \phi_i & \cos \phi_i \end{bmatrix}, \quad i = 1,\ldots, 4, 
\end{aligned}
\end{equation}
parameterizing a bistable planar kirigami design, just as in the main text.  These parameters are subject to equality and inequality constraints for a compatible design, which we now enumerate based on visual inspection of the figure.

The $\mathbf{s}_1, \ldots, \mathbf{v}_4$ vectors  must form closed loops when they are associated to the boundaries of a  panel or a slit. From Fig.\;\ref{fig:designSup}A and B, the eight total loop conditions associated to the four slits and four panels of the repeating cell are 
\begin{equation}
\begin{aligned}\label{eq:LoopRef}
 &\mathbf{s}_1 + \mathbf{s}_2 + \mathbf{s}_3 + \mathbf{s}_4 = \mathbf{0}, && \mathbf{t}_1 + \mathbf{t}_2 + \mathbf{t}_3 + \mathbf{t}_4 = \mathbf{0},  &&  \mathbf{u}_1 + \mathbf{u}_2 + \mathbf{u}_3 + \mathbf{u}_4 = \mathbf{0}, && \mathbf{v}_1 + \mathbf{v}_2 + \mathbf{v}_3 + \mathbf{v}_4 = \mathbf{0}, \\
&\mathbf{s}_1 - \mathbf{t}_1 - \mathbf{u}_1 + \mathbf{v}_1 = \mathbf{0},   && \mathbf{s}_2 -  \mathbf{t}_2 - \mathbf{u}_2 + \mathbf{v}_2 = \mathbf{0},  && \mathbf{s}_3 - \mathbf{t}_3 - \mathbf{u}_3 + \mathbf{v}_3 = \mathbf{0},   && \mathbf{s}_4 - \mathbf{t}_4 - \mathbf{u}_4 + \mathbf{v}_4 = \mathbf{0}.
\end{aligned}
\end{equation}
The second stable state is obtained by rotating the panels according to Fig.\;\ref{fig:designSup}C. These rotation transform the $\mathbf{s}_1, \ldots, \mathbf{v}_4$ vectors  to  $\mathbf{s}_i \mapsto \mathbf{R}(\phi_i) \mathbf{s}_i$, $\mathbf{t}_i \mapsto \mathbf{R}(\phi_i) \mathbf{t}_i$, $\mathbf{u}_i \mapsto \mathbf{R}(\phi_i) \mathbf{u}_i$, $\mathbf{v}_i  \mapsto \mathbf{R}(\phi_i) \mathbf{v}_i$, $i = 1,\ldots, 4$. These deformed vectors are also subject to  loop compatibility conditions. From Fig.\;\ref{fig:designSup}C and D, the  slit loop conditions for the second stable state are 
\begin{equation}
\begin{aligned}\label{eq:LoopDef}
&\mathbf{R}(\phi_1) \mathbf{s}_1 + \mathbf{R}(\phi_2)\mathbf{s}_2 + \mathbf{R}(\phi_3)\mathbf{s}_3 +  \mathbf{R}(\phi_4)\mathbf{s}_4 = \mathbf{0}, &&  \mathbf{R}(\phi_1) \mathbf{t}_1 +  \mathbf{R}(\phi_2)\mathbf{t}_2 +  \mathbf{R}(\phi_3)\mathbf{t}_3 +  \mathbf{R}(\phi_4) \mathbf{t}_4 = \mathbf{0},  \\
&\mathbf{R}(\phi_1)\mathbf{u}_1 + \mathbf{R}(\phi_2)\mathbf{u}_2 + \mathbf{R}(\phi_3)\mathbf{u}_3 + \mathbf{R}(\phi_4)\mathbf{u}_4 = \mathbf{0},  &&  \mathbf{R}(\phi_1) \mathbf{v}_1 + \mathbf{R}(\phi_2)\mathbf{v}_2 + \mathbf{R}(\phi_3)\mathbf{v}_3 + \mathbf{R}(\phi_4)\mathbf{v}_4 = \mathbf{0}.
\end{aligned}
\end{equation}
The  panel loop conditions for this state are trivially implied by their counterparts  in the second line of in Eq.\;[\ref{eq:LoopRef}] and are thus not a constraint.  Periodicity implies its own set of equality constraints on these parameters. We deduce from Fig.\;\ref{fig:designSup}A and B  that reference Bravais lattice vectors satisfy 
\begin{equation}
\begin{aligned}\label{eq:refLattice}
\boldsymbol{\ell}_1^R = \mathbf{u}_1 + \mathbf{u}_2 -\mathbf{v}_1 - \mathbf{v}_2 , \quad \boldsymbol{\ell}_2^R = \mathbf{u}_1 + \mathbf{u}_4 - \mathbf{s}_1 - \mathbf{s}_4.
\end{aligned}
\end{equation}
Similarly, from Fig.\;\ref{fig:designSup}C and D, we deduce that 
\begin{equation}
\begin{aligned}\label{eq:defLattice}
\boldsymbol{\ell}_1^D =\mathbf{R}(\phi_1) \mathbf{u}_1 + \mathbf{R}(\phi_2) \mathbf{u}_2 - \mathbf{R}(\phi_1) \mathbf{v}_1  - \mathbf{R}(\phi_2) \mathbf{v}_2 , \quad \boldsymbol{\ell}_2^D =\mathbf{R}(\phi_1) \mathbf{u}_1 + \mathbf{R}(\phi_4) \mathbf{u}_4  - \mathbf{R}(\phi_1) \mathbf{s}_1  - \mathbf{R}(\phi_4) \mathbf{s}_4.
\end{aligned}
\end{equation}
Eqs.\;[\ref{eq:LoopRef}-\ref{eq:defLattice}] complete the description of the equality constraints governing a compatible design for bistable kirigami. 

We now pursue a formulation of all inequality constraints associated to such designs. The basic point is that when four vectors close a loop, they need not always  form a convex quadrilateral. Whether they do or not depends on how the vectors are oriented in space relative to each other. Our designs require all the panels and slits to be convex quadrilaterals,  hence the need for additional constraints. 

The inequality constraints again follow  from visual inspection of Fig.\;\ref{fig:designSup} (by essentially taking a 2D version of the cross products associated with the vectors around slits and panels). The  convexity conditions for the four slits in the first stable state are 
\begin{equation}
\begin{aligned}\label{eq:slitRefConvexity}
&\mathbf{R}(\tfrac{\pi}{2}) \mathbf{s}_1 \cdot  \mathbf{s}_2 > 0,  && \mathbf{R}(\tfrac{\pi}{2}) \mathbf{s}_2 \cdot  \mathbf{s}_3 > 0,  && \mathbf{R}(\tfrac{\pi}{2}) \mathbf{s}_3 \cdot  \mathbf{s}_4 > 0,  && \mathbf{R}(\tfrac{\pi}{2}) \mathbf{s}_4 \cdot  \mathbf{s}_1 > 0,  \\
& \mathbf{t}_1 \cdot   \mathbf{R}(\tfrac{\pi}{2}) \mathbf{t}_2 > 0,  &&  \mathbf{t}_2 \cdot  \mathbf{R}(\tfrac{\pi}{2})\mathbf{t}_3 > 0,  && \mathbf{t}_3 \cdot \mathbf{R}(\tfrac{\pi}{2})  \mathbf{t}_4 > 0, &&  \mathbf{t}_4 \cdot  \mathbf{R}(\tfrac{\pi}{2})\mathbf{t}_1 > 0,  \\ 
& \mathbf{u}_1 \cdot   \mathbf{R}(\tfrac{\pi}{2}) \mathbf{u}_2 > 0, &&  \mathbf{u}_2 \cdot  \mathbf{R}(\tfrac{\pi}{2})\mathbf{u}_3 > 0,  && \mathbf{u}_3 \cdot \mathbf{R}(\tfrac{\pi}{2})  \mathbf{u}_4 > 0, &&  \mathbf{u}_4 \cdot  \mathbf{R}(\tfrac{\pi}{2})\mathbf{u}_1 > 0,  \\ 
&\mathbf{R}(\tfrac{\pi}{2}) \mathbf{v}_1 \cdot  \mathbf{v}_2 > 0,  && \mathbf{R}(\tfrac{\pi}{2}) \mathbf{v}_2 \cdot  \mathbf{v}_3 > 0,  && \mathbf{R}(\tfrac{\pi}{2}) \mathbf{v}_3 \cdot  {\mathbf{v}_4} > 0,  && \mathbf{R}(\tfrac{\pi}{2}) \mathbf{v}_4 \cdot  {\mathbf{v}_1} > 0.
\end{aligned}
\end{equation}
The  convexity conditions for the four panels in the first stable state are 
\begin{equation}
\begin{aligned}\label{eq:panelRefConvexity}
&\mathbf{R}(\tfrac{\pi}{2}) \mathbf{t}_1 \cdot  \mathbf{s}_1 > 0,  &&   \mathbf{R}(\tfrac{\pi}{2}) \mathbf{v}_1 \cdot  \mathbf{t}_1 > 0, &&  \mathbf{R}(\tfrac{\pi}{2}) \mathbf{u}_1 \cdot  \mathbf{v}_1 > 0, &&  \mathbf{R}(\tfrac{\pi}{2}) \mathbf{s}_1 \cdot  \mathbf{u}_1 > 0, \\
&\mathbf{R}(\tfrac{\pi}{2}) \mathbf{u}_2 \cdot  \mathbf{s}_2 > 0,  &&   \mathbf{R}(\tfrac{\pi}{2}) \mathbf{v}_2 \cdot  \mathbf{u}_2 > 0, &&  \mathbf{R}(\tfrac{\pi}{2}) \mathbf{t}_2 \cdot  \mathbf{v}_2 > 0, &&  \mathbf{R}(\tfrac{\pi}{2}) \mathbf{s}_2 \cdot  \mathbf{t}_2 > 0, \\
&\mathbf{R}(\tfrac{\pi}{2}) \mathbf{t}_3 \cdot  \mathbf{s}_3 > 0,  &&   \mathbf{R}(\tfrac{\pi}{2}) \mathbf{v}_3 \cdot  \mathbf{t}_3 > 0, &&  \mathbf{R}(\tfrac{\pi}{2}) \mathbf{u}_3 \cdot  \mathbf{v}_3 > 0, &&  \mathbf{R}(\tfrac{\pi}{2}) \mathbf{s}_3 \cdot  \mathbf{u}_3 > 0, \\
&\mathbf{R}(\tfrac{\pi}{2}) \mathbf{u}_4 \cdot  \mathbf{s}_4 > 0,  &&   \mathbf{R}(\tfrac{\pi}{2}) \mathbf{v}_4 \cdot  \mathbf{u}_4 > 0, &&  \mathbf{R}(\tfrac{\pi}{2}) \mathbf{t}_4 \cdot  \mathbf{v}_4 > 0, &&  \mathbf{R}(\tfrac{\pi}{2}) \mathbf{s}_4 \cdot  \mathbf{t}_4 > 0.
\end{aligned}
\end{equation}
 The last set of convexity conditions are for the four slits in the second stable state. These conditions are given by replacing the $\mathbf{s}_1,\ldots, \mathbf{v}_4$ vectors in Eq.\;[\ref{eq:slitRefConvexity}] with their deformed counterparts, i.e., 
 \begin{equation}
\begin{aligned}\label{eq:slitDefConvexity}
&\mathbf{R}(\tfrac{\pi}{2}) \mathbf{R}(\phi_1) \mathbf{s}_1 \cdot  \mathbf{R}(\phi_2) \mathbf{s}_2 > 0,  && \mathbf{R}(\tfrac{\pi}{2}) \mathbf{R}(\phi_2) \mathbf{s}_2 \cdot  \mathbf{R}(\phi_3) \mathbf{s}_3 > 0,  && \mathbf{R}(\tfrac{\pi}{2})  \mathbf{R}(\phi_3) \mathbf{s}_3 \cdot  \mathbf{R}(\phi_4)\mathbf{s}_4 > 0,  && \mathbf{R}(\tfrac{\pi}{2})\mathbf{R}(\phi_4) \mathbf{s}_4 \cdot \mathbf{R}(\phi_1) \mathbf{s}_1 > 0,  \\
&\mathbf{R}(\phi_1) \mathbf{t}_1 \cdot   \mathbf{R}(\tfrac{\pi}{2}) \mathbf{R}(\phi_2)\mathbf{t}_2 > 0,  && \mathbf{R}(\phi_2) \mathbf{t}_2 \cdot  \mathbf{R}(\tfrac{\pi}{2})\mathbf{R}(\phi_3)\mathbf{t}_3 > 0,  && \mathbf{R}(\phi_3)\mathbf{t}_3 \cdot \mathbf{R}(\tfrac{\pi}{2}) \mathbf{R}(\phi_4)  \mathbf{t}_4 > 0, && \mathbf{R}(\phi_4) \mathbf{t}_4 \cdot  \mathbf{R}(\tfrac{\pi}{2}) \mathbf{R}(\phi_1) \mathbf{t}_1 > 0,  \\ 
&\mathbf{R}(\phi_1) \mathbf{u}_1 \cdot   \mathbf{R}(\tfrac{\pi}{2})\mathbf{R}(\phi_2) \mathbf{u}_2 > 0, && \mathbf{R}(\phi_2)  \mathbf{u}_2 \cdot  \mathbf{R}(\tfrac{\pi}{2}) \mathbf{R}(\phi_3)\mathbf{u}_3 > 0,  && \mathbf{R}(\phi_3) \mathbf{u}_3 \cdot \mathbf{R}(\tfrac{\pi}{2}) \mathbf{R}(\phi_4)  \mathbf{u}_4 > 0, && \mathbf{R}(\phi_4)  \mathbf{u}_4 \cdot  \mathbf{R}(\tfrac{\pi}{2}) \mathbf{R}(\phi_1)\mathbf{u}_1 > 0,  \\ 
&\mathbf{R}(\tfrac{\pi}{2}) \mathbf{R}(\phi_1) \mathbf{v}_1 \cdot \mathbf{R}(\phi_2) \mathbf{v}_2 > 0,  && \mathbf{R}(\tfrac{\pi}{2})\mathbf{R}(\phi_2) \mathbf{v}_2 \cdot  \mathbf{R}(\phi_3)\mathbf{v}_3 > 0,  && \mathbf{R}(\tfrac{\pi}{2}) \mathbf{R}(\phi_3) \mathbf{v}_3 \cdot \mathbf{R}(\phi_4) {\mathbf{v}_4} > 0,  && \mathbf{R}(\tfrac{\pi}{2}) \mathbf{R}(\phi_4)\mathbf{v}_4 \cdot \mathbf{R}(\phi_1) {\mathbf{v}_1} > 0.
\end{aligned}
\end{equation}
Similar to the loop conditions, the convexity conditions of the deformed panels are implied by the reference versions in Eq.\;[\ref{eq:panelRefConvexity}]. While there are most certainly redundancies in the 48 inequalities in Eqs.\;[\ref{eq:slitRefConvexity}-\ref{eq:slitDefConvexity}], this issue is not so important as to be worth refining the equations.

 We now introduce a final set of  inequalities to address an important technical issue associated to the designs, namely that the second stable state should be distinct from the  first. This condition is achieved by the four inequality constraints 
\begin{equation}
\begin{aligned}\label{eq:distinctStable}
(\det [ \mathbf{R}(\phi_1) - \mathbf{R}(\phi_2)] )^2 > 0, \quad  ( \det [ \mathbf{R}(\phi_2) - \mathbf{R}(\phi_3)] )^2  > 0, \quad ( \det [ \mathbf{R}(\phi_3) - \mathbf{R}(\phi_4)] )^2  > 0,  \quad  ( \det [ \mathbf{R}(\phi_4) - \mathbf{R}(\phi_1)] )^2  > 0,
\end{aligned}
\end{equation}
imposing that rotations of neighboring panels in Fig.\;\ref{fig:designSup}C are distinct. At first glance, it might appear that we have assumed too much by these inequalities. However, if two neighboring panels rotate the same, then the corresponding sides of the central $(\mathbf{s}_1,\ldots, \mathbf{s}_4)$-slit are simply reoriented. Using the law of cosines, it follows that the other two sides of the convex slit must be reoriented in the exact same fashion since they cannot change their lengths. Thus, for a compatible design, all the panel rotations of the second state are the same  if any two adjacent panel rotations are the same. 

Eqs.\;[\ref{eq:slitRefConvexity}-\ref{eq:distinctStable}] enumerate all the inequality constraints associated to a compatible design. Overall, they form a list of 52 nonlinear inequalities of the form 
\begin{equation}
\begin{aligned}\label{eq:fIneq}
\mathbf{f}_{\text{ineq}}\big( \mathbf{s},  \mathbf{t} , \mathbf{u} ,  \mathbf{v}, \boldsymbol{\ell}, \boldsymbol{\phi} \big) > \mathbf{0},
\end{aligned}
\end{equation}
using the notation 
\begin{equation}
\begin{aligned}\label{eq:notation}
        \mathbf{s} = \begin{bmatrix}
            \mathbf{s}_1 \\ \mathbf{s}_2 \\ \mathbf{s}_3 \\ \mathbf{s}_4
        \end{bmatrix}, \quad \mathbf{t} = \begin{bmatrix}
            \mathbf{t}_1 \\ \mathbf{t}_2 \\ \mathbf{t}_3 \\ \mathbf{t}_4
        \end{bmatrix}, \quad \mathbf{u} = \begin{bmatrix}
            \mathbf{u}_1 \\ \mathbf{u}_2 \\ \mathbf{u}_3 \\ \mathbf{u}_4
        \end{bmatrix},  \quad \mathbf{v} = \begin{bmatrix}
            \mathbf{v}_1 \\ \mathbf{v}_2 \\ \mathbf{v}_3 \\ \mathbf{v}_4
        \end{bmatrix}, \quad \boldsymbol{\ell} = \begin{bmatrix}
            \boldsymbol{\ell}_1^R \\ \boldsymbol{\ell}_2^R \\ \boldsymbol{\ell}_1^D \\ \boldsymbol{\ell}_2^D
        \end{bmatrix}, \quad \boldsymbol{\phi} = (\phi_1,\phi_2, \phi_3, \phi_4)
    \end{aligned}
\end{equation}
for organizational purposes.  This list of inequalities,  combined with Eqs.\;[\ref{eq:LoopRef}-\ref{eq:defLattice}], furnish the necessary and sufficient conditions for a cell-based quad kirigami metamaterial with at least two stable states of specified lattice vectors $\boldsymbol{\ell}_1^R,\boldsymbol{\ell}_2^R, \boldsymbol{\ell}_1^D,\boldsymbol{\ell}_2^D$.

\subsection{Solving the equality constraints} We now derive the key design formula parameterizing the equality constraints in Eqs.\;[\ref{eq:LoopRef}-\ref{eq:defLattice}]. First eliminate $\mathbf{v}_1, \ldots, \mathbf{v}_4$ via the parameterizations  
\begin{equation}
\begin{aligned}\label{eq:sovleVi}
\mathbf{v}_1= - \mathbf{s}_1 +  \mathbf{t}_1 + \mathbf{u}_1, \quad \mathbf{v}_2= - \mathbf{s}_2 +  \mathbf{t}_2 + \mathbf{u}_2, \quad \mathbf{v}_3= - \mathbf{s}_3 +  \mathbf{t}_3 + \mathbf{u}_3, \quad \mathbf{v}_4= - \mathbf{s}_4 +  \mathbf{t}_4 + \mathbf{u}_4.
\end{aligned}
\end{equation}
Next observe that the conditions $\mathbf{v}_1 + \mathbf{v}_2 + \mathbf{v}_3 + \mathbf{v}_4 = \mathbf{0}$ and   $\mathbf{R}(\phi_1) \mathbf{v}_1 + \mathbf{R}(\phi_2)\mathbf{v}_2 + \mathbf{R}(\phi_3)\mathbf{v}_3 + \mathbf{R}(\phi_4)\mathbf{v}_4 = \mathbf{0} $ are implied by the remaining unsolved loop conditions
\begin{equation}
\begin{aligned}
&\sum_{i = 1,\ldots, 4} \mathbf{s}_i = \mathbf{0}, && \sum_{i = 1,\ldots, 4} \mathbf{t}_i = \mathbf{0},  &&  \sum_{i = 1,\ldots, 4} \mathbf{u}_i = \mathbf{0},  && \sum_{i = 1,\ldots, 4} \mathbf{R}(\phi_i) \mathbf{s}_i  = \mathbf{0} && \sum_{i = 1,\ldots, 4} \mathbf{R}(\phi_i) \mathbf{t}_i  = \mathbf{0}    && \sum_{i = 1,\ldots, 4} \mathbf{R}(\phi_i) \mathbf{u}_i  = \mathbf{0}. 
\end{aligned}
\end{equation}
The first three of these conditions are solved by 
\begin{equation}
\begin{aligned}\label{eq:solveSub4}
\mathbf{s}_4 = - \mathbf{s}_1 - \mathbf{s}_2 - \mathbf{s}_3, \quad \mathbf{t}_4 = - \mathbf{t}_1 - \mathbf{t}_2 - \mathbf{t}_3,  \quad \mathbf{u}_4 = - \mathbf{u}_1 - \mathbf{u}_2 - \mathbf{u}_3. 
\end{aligned}
\end{equation}
Substituting this parameterization into the final three loop conditions gives 
\begin{equation}
\begin{aligned}\label{eq:AfterManip1}
\boldsymbol{\Delta}_{14}  \mathbf{s}_1 + \boldsymbol{\Delta}_{24} \mathbf{s}_2 + \boldsymbol{\Delta}_{34} \mathbf{s}_3 = \mathbf{0}, \quad \boldsymbol{\Delta}_{14}  \mathbf{t}_1 + \boldsymbol{\Delta}_{24} \mathbf{t}_2 + \boldsymbol{\Delta}_{34} \mathbf{t}_3 = \mathbf{0}, \quad \boldsymbol{\Delta}_{14}  \mathbf{u}_1 + \boldsymbol{\Delta}_{24} \mathbf{u}_2 + \boldsymbol{\Delta}_{34} \mathbf{u}_3 = \mathbf{0}
\end{aligned}
\end{equation}
using the definitions $\boldsymbol{\Delta}_{ij} = \mathbf{R}(\phi_i) - \mathbf{R}(\phi_j)$.  Eqs.\;[\ref{eq:sovleVi}] and [\ref{eq:solveSub4}] turn the Bravais lattices conditions in Eqs.\;[\ref{eq:refLattice}] and [\ref{eq:defLattice}] into equations of the form 
\begin{equation}
\begin{aligned}\label{eq:AfterManip2}
&\boldsymbol{\ell}_1^R =  \mathbf{s}_1 + \mathbf{s}_2 - \mathbf{t}_1 - \mathbf{t}_2, && \boldsymbol{\ell}_2^R = \mathbf{s}_2 + \mathbf{s}_3 - \mathbf{u}_2 - \mathbf{u}_3 , \\
&\boldsymbol{\ell}_1^D =  \mathbf{R}(\phi_1) (\mathbf{s}_1 - \mathbf{t}_1)   + \mathbf{R}(\phi_2) ( \mathbf{s}_2 -  \mathbf{t}_2 ) , && \boldsymbol{\ell}_2^D =\boldsymbol{\Delta}_{14} \mathbf{u}_1 - \mathbf{R}(\phi_4) ( \mathbf{u}_2 + \mathbf{u}_3)  - \boldsymbol{\Delta}_{14}\mathbf{s}_1  + \mathbf{R}(\phi_4)( \mathbf{s}_2 + \mathbf{s}_3).
\end{aligned}
\end{equation}
Solving Eqs.\;[\ref{eq:AfterManip1}] and [\ref{eq:AfterManip2}] completes the description. 

Observe that there are 14 total constraints and 30 total DOFs in the 7 remaining 2D vector equations above. We solve these constraints by prescribing $\mathbf{t}_1, \mathbf{t}_2, \mathbf{u}_1, \mathbf{u}_2, \mathbf{s}_3,\mathbf{t}_3, \mathbf{u}_3$ and leaving $\mathbf{s}_1, \mathbf{s}_2, \boldsymbol{\ell}_1^R, \boldsymbol{\ell}_2^R, \boldsymbol{\ell}_1^D, \boldsymbol{\ell}_2^D, \phi_1, \ldots \phi_4$ as free DOFs. Note that  $\boldsymbol{\Delta}_{12}, \boldsymbol{\Delta}_{23}, \boldsymbol{\Delta}_{34}$, and $\boldsymbol{\Delta}_{41}$ are all invertible under the inequality constraints in Eq.\;[\ref{eq:distinctStable}] necessary for a compatible bistable design. Thus,  Eq.\;[\ref{eq:AfterManip1}] is solved by prescribing 
\begin{equation}
    \begin{aligned}\label{eq:AfterManip3}
        \mathbf{s}_3 = -\boldsymbol{\Delta}_{34}^{-1} \big(\boldsymbol{\Delta}_{14}  \mathbf{s}_1 + \boldsymbol{\Delta}_{24} \mathbf{s}_2 \big), \quad \mathbf{t}_3 = -\boldsymbol{\Delta}_{34}^{-1} \big(\boldsymbol{\Delta}_{14}  \mathbf{t}_1 + \boldsymbol{\Delta}_{24} \mathbf{t}_2 \big), \quad \mathbf{u}_3 = -\boldsymbol{\Delta}_{34}^{-1} \big(\boldsymbol{\Delta}_{14}  \mathbf{u}_1 + \boldsymbol{\Delta}_{24} \mathbf{u}_2 \big).
    \end{aligned}
\end{equation}
Direct substitution of Eq.\;[\ref{eq:AfterManip3}] into Eq.\;[\ref{eq:AfterManip2}] leads to the conditions  
\begin{equation}
\begin{aligned}\label{eq:AfterManip4}
\boldsymbol{\ell}_1^R &=  \mathbf{s}_1 - \mathbf{t}_1  + \mathbf{s}_2 -  \mathbf{t}_2,\\
\boldsymbol{\ell}_2^R &= \mathbf{s}_2  -\boldsymbol{\Delta}_{34}^{-1} \big(\boldsymbol{\Delta}_{14}  \mathbf{s}_1 + \boldsymbol{\Delta}_{24} \mathbf{s}_2 \big) - \mathbf{u}_2 +\boldsymbol{\Delta}_{34}^{-1} \big(\boldsymbol{\Delta}_{14}  \mathbf{u}_1 + \boldsymbol{\Delta}_{24} \mathbf{u}_2 \big) , \\
&=\boldsymbol{\Delta}_{34}^{-1}\boldsymbol{\Delta}_{14}(\mathbf{u}_1 -\mathbf{s}_1) + \boldsymbol{\Delta}_{34}^{-1}\boldsymbol{\Delta}_{23} (\mathbf{u}_2 - \mathbf{s}_2), \\
\boldsymbol{\ell}_1^D &=  \mathbf{R}(\phi_1) (\mathbf{s}_1 - \mathbf{t}_1)   + \mathbf{R}(\phi_2) ( \mathbf{s}_2 -  \mathbf{t}_2 ) \\
&= \mathbf{R}(\phi_1) \boldsymbol{\ell}_1^R - \boldsymbol{\Delta}_{12} ( \mathbf{s}_2 -  \mathbf{t}_2 ), \\
 \boldsymbol{\ell}_2^D &=\boldsymbol{\Delta}_{14} \mathbf{u}_1 - \mathbf{R}(\phi_4) ( \mathbf{u}_2  -\boldsymbol{\Delta}_{34}^{-1} \big(\boldsymbol{\Delta}_{14}  \mathbf{u}_1 + \boldsymbol{\Delta}_{24} \mathbf{u}_2 \big))  - \boldsymbol{\Delta}_{14}\mathbf{s}_1  + \mathbf{R}(\phi_4)( \mathbf{s}_2  -\boldsymbol{\Delta}_{34}^{-1} \big(\boldsymbol{\Delta}_{14}  \mathbf{s}_1 + \boldsymbol{\Delta}_{24} \mathbf{s}_2 \big)) \\
&= \mathbf{R}(\phi_3) \boldsymbol{\Delta}_{34}^{-1}\boldsymbol{\Delta}_{14}(\mathbf{u}_1 -\mathbf{s}_1) + \mathbf{R}(\phi_4)\boldsymbol{\Delta}_{34}^{-1}\boldsymbol{\Delta}_{23} (\mathbf{u}_2 - \mathbf{s}_2) \\
&= \mathbf{R}(\phi_3) \boldsymbol{\ell}_2^R - \boldsymbol{\Delta}_{23} (\mathbf{u}_2 - \mathbf{s}_2)
\end{aligned}
\end{equation}
after standard algebraic manipulations.  It follows that the last two equations in Eq.\;[\ref{eq:AfterManip4}] are solved by 
\begin{equation}
\begin{aligned}\label{eq:AfterManip5}
    \mathbf{t}_2 =\mathbf{s}_2 +  \boldsymbol{\Delta}_{12}^{-1} \boldsymbol{\ell}_1^D - \boldsymbol{\Delta}_{12}^{-1}\mathbf{R}(\phi_1) \boldsymbol{\ell}_1^R, \quad \mathbf{u}_2 = \mathbf{s}_2 - \boldsymbol{\Delta}_{23}^{-1} \boldsymbol{\ell}_2^D  + \boldsymbol{\Delta}_{23}^{-1} \mathbf{R}(\phi_3) \boldsymbol{\ell}_2^R.
\end{aligned}
\end{equation}
Plugging these formula  back into the first two equations in Eq.\;[\ref{eq:AfterManip4}] and rearranging terms gives
\begin{equation}
    \begin{aligned}\label{eq:AfterManip6}
        \mathbf{t}_1 &= \mathbf{s}_1 + \mathbf{s}_2  - \boldsymbol{\ell}_1^R -   \big(\mathbf{s}_2 +  \boldsymbol{\Delta}_{12}^{-1} \boldsymbol{\ell}_1^D - \boldsymbol{\Delta}_{12}^{-1}\mathbf{R}(\phi_1) \boldsymbol{\ell}_1^R\big) \\
        &=  \mathbf{s}_1  +    \boldsymbol{\Delta}_{12}^{-1}\mathbf{R}(\phi_2) \boldsymbol{\ell}_1^R - \boldsymbol{\Delta}_{12}^{-1} \boldsymbol{\ell}_1^D , \\
        \mathbf{u}_1 &=  \mathbf{s}_1 + \boldsymbol{\Delta}_{14}^{-1} \boldsymbol{\Delta}_{34} \boldsymbol{\ell}_{2}^R  +\boldsymbol{\Delta}_{14}^{-1} \boldsymbol{\Delta}_{23} (\mathbf{s}_2 - \big(\mathbf{s}_2 - \boldsymbol{\Delta}_{23}^{-1} \boldsymbol{\ell}_2^D  + \boldsymbol{\Delta}_{23}^{-1} \mathbf{R}(\phi_3) \boldsymbol{\ell}_2^R\big))   \\
        &= \mathbf{s}_1 - \boldsymbol{\Delta}_{14}^{-1}  \mathbf{R}(\phi_4) \boldsymbol{\ell}_{2}^R  +\boldsymbol{\Delta}_{14}^{-1}     \boldsymbol{\ell}_2^D . 
    \end{aligned}
\end{equation}
The parameterizations in Eqs.\;[\ref{eq:sovleVi}], [\ref{eq:solveSub4}], [\ref{eq:AfterManip3}], [\ref{eq:AfterManip5}] and [\ref{eq:AfterManip6}] can now all be written in terms of the desired free variables  $\mathbf{s}_1, \mathbf{s}_2, \boldsymbol{\ell}_1^R, \boldsymbol{\ell}_2^R, \boldsymbol{\ell}_1^D, \boldsymbol{\ell}_2^D, \phi_1, \ldots \phi_4$ through the system of equations
\begin{equation}
    \begin{aligned}\label{eq:theorem111}
        \begin{bmatrix}
        \mathbf{s}_{3} \\ \mathbf{s}_{4} \\ \mathbf{t}_{1} \\ \mathbf{t}_{2} \\
        \mathbf{t}_{3} \\ \mathbf{t}_{4} \\
        \mathbf{u}_{1} \\ \mathbf{u}_{2} \\
        \mathbf{u}_{3} \\ \mathbf{u}_{4} \\
        \mathbf{v}_{1} \\ \mathbf{v}_{2} \\
        \mathbf{v}_{3} \\ \mathbf{v}_{4} \\
    \end{bmatrix}
    =
    \begin{bmatrix}
        \mathbf{\Delta}_{34}^{-1} \mathbf{\Delta}_{41} & \mathbf{-I}-\mathbf{\Delta}_{34}^{-1} \mathbf{\Delta}_{23} & \mathbf{0} & \mathbf{0} & \mathbf{0} & \mathbf{0} \\
        \mathbf{-I-\Delta}_{34}^{-1} \mathbf{\Delta}_{41} & \mathbf{\Delta}_{34}^{-1} \mathbf{\Delta}_{23} & \mathbf{0} & \mathbf{0} & \mathbf{0} & \mathbf{0} \\
        \mathbf{I} & \mathbf{0} & \mathbf{\Delta}_{12}^{-1} \mathbf{R}(\phi_2) & \mathbf{0} & \mathbf{-\Delta}_{12}^{-1} & \mathbf{0} \\
        \mathbf{0} & \mathbf{I} & \mathbf{-\Delta}_{12}^{-1} \mathbf{R}(\phi_1) & \mathbf{0} & \mathbf{\Delta}_{12}^{-1} & \mathbf{0} \\
        \mathbf{\Delta}_{34}^{-1} \mathbf{\Delta}_{41} & \mathbf{-I-\Delta}_{34}^{-1} \mathbf{\Delta}_{23} & -\mathbf{\Delta}_{34}^{-1} \mathbf{R}(\phi_4) & \mathbf{0} & \mathbf{\Delta}_{34}^{-1} & \mathbf{0} \\
        \mathbf{-I-\Delta}_{34}^{-1} \mathbf{\Delta}_{41} & \mathbf{\Delta}_{34}^{-1} \mathbf{\Delta}_{23} & \mathbf{\Delta}_{34}^{-1} \mathbf{R}(\phi_3) & \mathbf{0} & \mathbf{-\Delta}_{34}^{-1} & \mathbf{0} \\
        \mathbf{I} & \mathbf{0} & \mathbf{0} & \mathbf{\Delta}_{41}^{-1} \mathbf{R}(\phi_4) & \mathbf{0} & \mathbf{-\Delta}_{41}^{-1} \\
        \mathbf{0} & \mathbf{I} & \mathbf{0} & \mathbf{\Delta}_{23}^{-1} \mathbf{R}(\phi_3) & \mathbf{0} & \mathbf{-\Delta}_{23}^{-1} \\
        \mathbf{\Delta}_{34}^{-1} \mathbf{\Delta}_{41} & \mathbf{-I-\Delta}_{34}^{-1} \mathbf{\Delta}_{23} & \mathbf{0} & \mathbf{-\Delta}_{23}^{-1} \mathbf{R}(\phi_2) & \mathbf{0} & \mathbf{\Delta}_{23}^{-1} \\
        \mathbf{-I-\Delta}_{34}^{-1} \mathbf{\Delta}_{41} & \mathbf{\Delta}_{34}^{-1} \mathbf{\Delta}_{23} & \mathbf{0} & \mathbf{-\Delta}_{41}^{-1} \mathbf{R}(\phi_1) & \mathbf{0} & \mathbf{\Delta}_{41}^{-1} \\
        \mathbf{I} & \mathbf{0} & \mathbf{\Delta}_{12}^{-1} \mathbf{R}(\phi_2) & \mathbf{\Delta}_{41}^{-1} \mathbf{R}(\phi_4) & \mathbf{-\Delta}_{12}^{-1} & \mathbf{-\Delta}_{41}^{-1} \\
        \mathbf{0} & \mathbf{I} & \mathbf{-\Delta}_{12}^{-1} \mathbf{R}(\phi_1) & \mathbf{\Delta}_{23} \mathbf{R}(\phi_3) & \mathbf{\Delta}_{12}^{-1} & -\mathbf{\Delta}_{23}^{-1} \\
        \mathbf{\Delta}_{34}^{-1} \mathbf{\Delta}_{41} & \mathbf{-I-\Delta}_{34}^{-1} \mathbf{\Delta}_{23} & \mathbf{-\Delta}_{34}^{-1} \mathbf{R}(\phi_4) & \mathbf{-\Delta}_{23}^{-1} \mathbf{R}(\phi_2) & \mathbf{\Delta}_{34}^{-1} & \mathbf{\Delta}_{23}^{-1} \\
        \mathbf{-I-\Delta}_{34}^{-1} \mathbf{\Delta}_{41} & \mathbf{\Delta}_{34}^{-1} \mathbf{\Delta}_{23} & \mathbf{\Delta}_{34}^{-1} \mathbf{R}(\phi_3) & \mathbf{-\Delta}_{41}^{-1} \mathbf{R}(\phi_1) & \mathbf{-\Delta}_{34}^{-1} & \mathbf{\Delta}_{41}^{-1}
    \end{bmatrix}
    \begin{bmatrix}
        \mathbf{s}_{1} \\ \mathbf{s}_{2} \\
        \boldsymbol\ell_{1}^{R} \\ \boldsymbol\ell_{2}^{R} \\
        \boldsymbol\ell_{1}^{D} \\ \boldsymbol\ell_{2}^{D} \\
    \end{bmatrix}.
    \end{aligned}
\end{equation}
Since each $\boldsymbol{\Delta}_{ij} = \mathbf{R}(\phi_i) - \mathbf{R}(\phi_j)$ depends only on the angles $\phi_i$, $\phi_j$, this equation is structurally of the form 
\begin{equation}
    \begin{aligned}\label{eq:designFormula}
        \begin{bmatrix} \mathbf{s}_3 \\ \mathbf{s}_4 \\ \mathbf{t} \\ \mathbf{u} \\ 
\mathbf{v} \end{bmatrix} = \mathbf{D}(\boldsymbol{\phi}) \begin{bmatrix} \mathbf{s}_1 \\ \mathbf{s}_2 \\ \boldsymbol{\ell} \end{bmatrix}, 
    \end{aligned}
\end{equation}
using the notation in Eq.\;[\ref{eq:notation}]. This result is the desired design formula, and is also reported in the main text.   

\subsection{Parameterization of the inequality constraints} The parameterization of all inequality constraints in  Eq.\;[\ref{eq:fIneq}] can now be simplified given the solution to the equality constraints as
\begin{equation}
    \begin{aligned}\label{eq:firstIneq1}
        \mathbf{g}^0_{\text{ineq}}(\mathbf{s}_1, \mathbf{s}_2, \boldsymbol{\ell}, \boldsymbol{\phi}) =  \big\{ \mathbf{f}_{\text{ineq}}(\mathbf{s}, \mathbf{t}, \mathbf{u}, \mathbf{v}, \boldsymbol{\ell}, \boldsymbol{\phi}) \;  \big| \; \mathbf{s}_3,\mathbf{s}_4,\mathbf{t}, \mathbf{u}, \mathbf{v} \text{ solve Eq.\;[\ref{eq:designFormula}]} \big\}  > \mathbf{0}.
    \end{aligned}
\end{equation}

We optimize kirigami designs subject to these inequality constraints.  In the optimization, we consider the Bravais lattice vectors as a set of specified designer inputs by treating $\boldsymbol{\ell} = \overline{\boldsymbol{\ell}}$ as a  fixed list. The version of the inequality constraints introduced in the main text is  then 
\begin{equation}
    \begin{aligned}\label{eq:gineqDef}
        \mathbf{g}_{\text{ineq}}(\mathbf{s}_1, \mathbf{s}_2, \boldsymbol{\phi})  = \mathbf{g}^0_{\text{ineq}}(\mathbf{s}_1, \mathbf{s}_2, \bar{\boldsymbol{\ell}}, \boldsymbol{\phi}) - \varepsilon \mathbf{1} \geq \mathbf{0}
    \end{aligned}
\end{equation}
where $\varepsilon \mathbf{1}$ is a 52 component array with every element taking a small value $0< \varepsilon \ll 1$. The purpose of this modification is twofold: 1) From a purely numerical perspective, we need to replace the strict "$>$" inequality in Eq.\;[\ref{eq:firstIneq1}] with one that allows the boundary case "$\geq$" to formulate a standard constrained optimization problem. 2) A positive $\varepsilon$ also prevents the slits from closing and the panels from converging to a line on optimization; it thus serves as a simple tuning parameter that can make optimized patterns more amenable to any limitations that might arise from an experimental fabrication process.

\section{Elastic energy and optimization}

\subsection{General setup} This paper tunes the designs of bistable kirigami patterns by optimizing objective functions of the form 
\begin{equation}\label{eq:generalSetup}
    \begin{aligned}
        \min \big\{ f_{\text{obj}}(\mathbf{s}_1, \mathbf{s}_2, \boldsymbol{\phi})  \; \big|  \; \mathbf{g}_{\text{ineq}}(\mathbf{s}_1, \mathbf{s}_2, \boldsymbol{\phi}) \geq \mathbf{0} \big\}.
    \end{aligned}
\end{equation}
Our primary interest is in objective functions that can assess and optimize features of the stored elastic energy of the kirigami. The challenge is that calculating an elastic energy based on  high fidelity modeling, like FEM or even bar-hinge based modeling, is not efficient and thus creates a huge bottleneck in the optimization process.  We instead develop an elastic model that can be implemented directly into Matlab and evaluated using its fast solvers. 

\begin{figure}[htb!]
\centering
\includegraphics[scale=1]{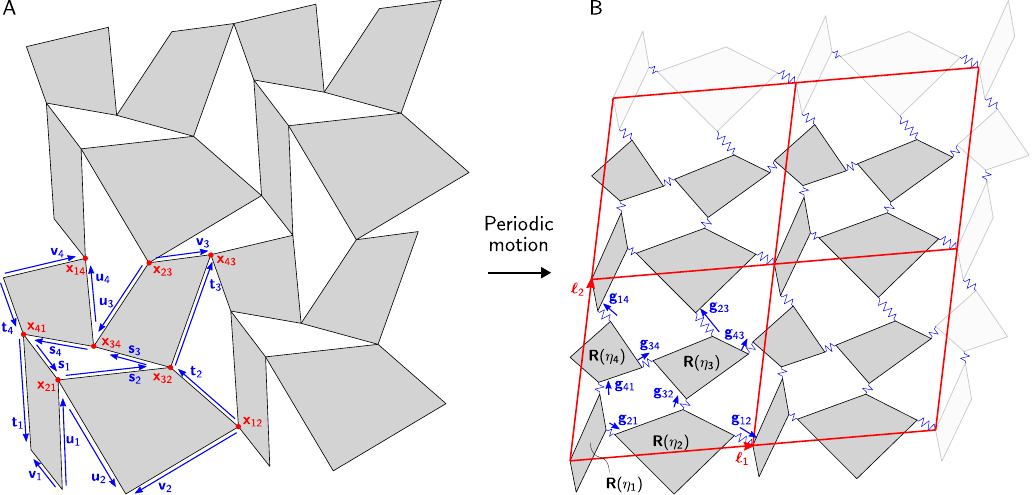}
\caption{Model for the stored elastic energy. Corner points are replaced by springs and the pattern is subject to a periodic motion composed of rigid rotations and translations of the panels.  (A) Labeling of points $\mathbf{x}_{ij}$ and sides $\mathbf{s}_1,\ldots, \mathbf{v}_4$ prior to deformation. (B) The  motion distorts each cell according to lattice vectors $\boldsymbol{\ell}_{1}$ and $\boldsymbol{\ell}_2$, which elongates the springs through eight repeated $\mathbf{g}_{ij}$ gap vectors.  }
\label{fig:elasticSup}
\end{figure}

Our approach is illustrated in Fig.\;\ref{fig:elasticSup}. We model the corner points of the kirigami pattern as linear springs of zero rest length and unit modulus and subject the pattern to the following simplified family of "elastic" distortions: First, we rigidly rotate and translate each panel in an individual unit cell through the 2D rotations $\mathbf{R}(\eta_i)$  and corresponding 2D translations $\mathbf{c}_i$, $i =1,\ldots,4$, as labeled. Then, we repeat these panel motions cell by cell along a set of Bravais lattice vectors $\boldsymbol{\ell}_1$ and $\boldsymbol{\ell}_2$ to achieve a bulk effective deformation of the pattern.  As a concrete illustration consistent with the notation in \ref{fig:elasticSup}A-B, the first panel in the cell has its corner points deformed as 
\begin{equation}
\begin{aligned}
    &\mathbf{x}_{21} \mapsto \mathbf{R}(\eta_1) \mathbf{x}_{21} + \mathbf{c}_1, && \mathbf{x}_{41} \mapsto \mathbf{R}(\eta_1) \mathbf{x}_{41} + \mathbf{c}_1, \\
    &\mathbf{x}_{41} + \mathbf{t}_1 \mapsto \mathbf{R}(\eta_1)( \mathbf{x}_{41} + \mathbf{t}_1) + \mathbf{c}_1, && \mathbf{x}_{21} - \mathbf{u}_1 \mapsto \mathbf{R}_1(\eta_1) ( \mathbf{x}_{21} - \mathbf{u}_1) + \mathbf{c}_1;
\end{aligned}
\end{equation}
the first panel in the adjacent cell to the right has its corner points deformed as 
\begin{equation}
\begin{aligned}
    &\mathbf{x}_{21} + \boldsymbol{\ell}_1^R \mapsto \mathbf{R}(\eta_1) \mathbf{x}_{21} + \mathbf{c}_1 + \boldsymbol{\ell}_1, && \mathbf{x}_{41}  + \boldsymbol{\ell}_1^R \mapsto \mathbf{R}(\eta_1) \mathbf{x}_{41} + \mathbf{c}_1 +\boldsymbol{\ell}_1, \\  &\mathbf{x}_{41} + \mathbf{t}_1 + \boldsymbol{\ell}_1^R \mapsto \mathbf{R}(\eta_1) (\mathbf{x}_{41} + \mathbf{t}_1) + \mathbf{c}_1 + \boldsymbol{\ell}_{1},
    &&  \mathbf{x}_{21} - \mathbf{u}_1  + \boldsymbol{\ell}_1^R \mapsto \mathbf{R}_1(\eta_1) ( \mathbf{x}_{21} - \mathbf{u}_1) + \mathbf{c}_1  + \boldsymbol{\ell}_1; 
\end{aligned}
\end{equation}
and so on.  Fig.\;\ref{fig:elasticSup}B sketches the output after all the panel are deformed by this procedure. Gaps typically form  between adjacent panels and are repeated periodically throughout the sample. They  are  quantified by eight gap vectors $\mathbf{g}_{ij}$, $ij \in \{12,23,34,14,21,32,43,41\}$, that satisfy 
\begin{equation}
    \begin{aligned}\label{eq:theGaps}
        \mathbf{g}_{ij} = \begin{cases}
        (\mathbf{R}(\eta_i) - \mathbf{R}(\eta_j)) \mathbf{x}_{ij}  + \mathbf{c}_{i} - \mathbf{c}_j  &   \text{ if } ij \in \{ 21,32,34,41\} ,\\ 
        (\mathbf{R}(\eta_i) - \mathbf{R}(\eta_j)) \mathbf{x}_{ij}  + \mathbf{c}_{i} - \mathbf{c}_j + \boldsymbol{\ell}_{1} - \mathbf{R}(\eta_i)\boldsymbol{\ell}_1^R & \text{ if } ij \in \{12,43\},  \\
        (\mathbf{R}(\eta_i) - \mathbf{R}(\eta_j)) \mathbf{x}_{ij}  + \mathbf{c}_{i} - \mathbf{c}_j + \boldsymbol{\ell}_{2} - \mathbf{R}(\eta_i)\boldsymbol{\ell}_2^R & \text{ if } ij \in \{23,14\}.
        \end{cases} 
    \end{aligned}
\end{equation}
The stored elastic energy per unit cell due to the assumed spring model is thus 
\begin{equation}
    \begin{aligned}\label{eq:sprEnergy}
       \text{spring energy per cell} &= \sum_{ij \in \{21,32,34,41\} } |\mathbf{g}_{ij}|^2 + |\mathbf{g}_{ji}|^2  \\
       &= \sum_{ij \in \{21,34\} } |(\mathbf{R}(\eta_i) - \mathbf{R}(\eta_j)) \mathbf{x}_{ij}  + \mathbf{c}_{i} - \mathbf{c}_j |^2 + |(\mathbf{R}(\eta_j) - \mathbf{R}(\eta_i)) \mathbf{x}_{ji}  + \mathbf{c}_{j} - \mathbf{c}_i + \boldsymbol{\ell}_{1} - \mathbf{R}(\eta_j)\boldsymbol{\ell}_1^R|^2 \\
        & \quad + \sum_{ij \in \{32,41\} } |(\mathbf{R}(\eta_i) - \mathbf{R}(\eta_j)) \mathbf{x}_{ij}  + \mathbf{c}_{i} - \mathbf{c}_j |^2 + |(\mathbf{R}(\eta_j) - \mathbf{R}(\eta_i)) \mathbf{x}_{ji}  + \mathbf{c}_{j} - \mathbf{c}_i + \boldsymbol{\ell}_{2} - \mathbf{R}(\eta_j)\boldsymbol{\ell}_2^R|^2 \\
        &= E^0_{\text{spr}}(\eta_1, \eta_2, \eta_3, \eta_4, \mathbf{c}_1,\mathbf{c}_2, \mathbf{c}_3, \mathbf{c}_4, \boldsymbol{\ell}_1, \boldsymbol{\ell}_2 ).
        \end{aligned}
\end{equation}
In the remainder of this section, we minimize the excess DOFs above, after accounting for invariances, to develop a stored elastic energy for the design that is a function of a single actuation angle.

\subsection{Minimizing out the translations and Bravais lattices} First we  develop an analytical expression for the solution to the minimization problem 
\begin{equation}
\begin{aligned}\label{eq:minTrans}
 E_{\text{spr}}(\eta_1,\eta_2, \eta_3, \eta_4) = \min_{\substack{\mathbf{c}_1,\ldots, \mathbf{c}_4 \in \mathbb{R}^2 \\  \boldsymbol{\ell}_1, \boldsymbol{\ell}_2 \in \mathbb{R}^2}} E^0_{\text{spr}}(\eta_1, \eta_2, \eta_3, \eta_4, \mathbf{c}_1,\mathbf{c}_2, \mathbf{c}_3, \mathbf{c}_4, \boldsymbol{\ell}_1, \boldsymbol{\ell}_2 ),
\end{aligned}
\end{equation}
leading to the result in Eq.\;[3] of the main text. 

The energy above is quadratic in each variable being minimized. As a result,  this minimization problem involves only the linear algebra of finding a critical point. Observe that 
\begin{equation}
\begin{aligned}\label{eq:derivativesTrans}
        \frac{\partial E_{\text{spr}}^0}{\partial \mathbf{c}_1} &= 2 \Big[ \frac{\partial \mathbf{g}_{12}}{\partial \mathbf{c}_1} \Big]^T \mathbf{g}_{12} + 2 \Big[ \frac{\partial \mathbf{g}_{21}}{\partial \mathbf{c}_1} \Big]^T \mathbf{g}_{21} + 2 \Big[ \frac{\partial \mathbf{g}_{14}}{\partial \mathbf{c}_1} \Big]^T \mathbf{g}_{14} + 2 \Big[ \frac{\partial \mathbf{g}_{41}}{\partial \mathbf{c}_1} \Big]^T  \mathbf{g}_{41} = 2\big( \mathbf{g}_{12} - \mathbf{g}_{21} + \mathbf{g}_{14} - \mathbf{g}_{41} \big), \\
        \frac{\partial E_{\text{spr}}^0}{\partial \mathbf{c}_2} &= 2 \Big[ \frac{\partial \mathbf{g}_{21}}{\partial \mathbf{c}_2} \Big]^T \mathbf{g}_{21} + 2 \Big[ \frac{\partial \mathbf{g}_{12}}{\partial \mathbf{c}_2} \Big]^T \mathbf{g}_{12} + 2 \Big[ \frac{\partial \mathbf{g}_{23}}{\partial \mathbf{c}_2} \Big]^T \mathbf{g}_{23} + 2 \Big[ \frac{\partial \mathbf{g}_{32}}{\partial \mathbf{c}_2} \Big]^T \mathbf{g}_{32} = 2 \big(\mathbf{g}_{21} - \mathbf{g}_{12} + \mathbf{g}_{23} - \mathbf{g}_{32} \big) ,\\
        \frac{\partial E_{\text{spr}}^0}{\partial \mathbf{c}_3} &= 2 \Big[ \frac{\partial \mathbf{g}_{32}}{\partial \mathbf{c}_3} \Big]^T \mathbf{g}_{32} + 2 \Big[ \frac{\partial \mathbf{g}_{23}}{\partial \mathbf{c}_3} \Big]^T \mathbf{g}_{23} + 2 \Big[ \frac{\partial \mathbf{g}_{34}}{\partial \mathbf{c}_3} \Big]^T \mathbf{g}_{34} + 2 \Big[ \frac{\partial \mathbf{g}_{43}}{\partial \mathbf{c}_3} \Big]^T \mathbf{g}_{43} = 2\big( \mathbf{g}_{32} - \mathbf{g}_{23} + \mathbf{g}_{34} - \mathbf{g}_{43} \big), \\
        \frac{\partial E_{\text{spr}}^0}{\partial \mathbf{c}_4} &= 2 \Big[ \frac{\partial \mathbf{g}_{43}}{\partial \mathbf{c}_4} \Big]^T \mathbf{g}_{43} + 2 \Big[ \frac{\partial \mathbf{g}_{34}}{\partial \mathbf{c}_4} \Big]^T \mathbf{g}_{34} + 2 \Big[ \frac{\partial \mathbf{g}_{41}}{\partial \mathbf{c}_4} \Big]^T \mathbf{g}_{41} + 2 \Big[ \frac{\partial \mathbf{g}_{14}}{\partial \mathbf{c}_4} \Big]^T \mathbf{g}_{14}  = 2 \big( \mathbf{g}_{43} - \mathbf{g}_{34} + \mathbf{g}_{41} - \mathbf{g}_{14} \big), \\
        \frac{\partial E_{\text{spr}}^0}{\partial \boldsymbol{\ell}_1} &= 2 \Big[ \frac{\partial \mathbf{g}_{12}}{\partial \boldsymbol{\ell}_1} \Big]^T \mathbf{g}_{12} + 2 \Big[ \frac{\partial \mathbf{g}_{43}}{\partial \boldsymbol{\ell}_1} \Big]^T \mathbf{g}_{43} = 2 \big( \mathbf{g}_{12} + \mathbf{g}_{43} \big),\\
        \frac{\partial E_{\text{spr}}^0}{\partial \boldsymbol{\ell}_2} &= 2 \Big[ \frac{\partial \mathbf{g}_{23}}{\partial \boldsymbol{\ell}_2} \Big]^T \mathbf{g}_{23} + 2 \Big[ \frac{\partial \mathbf{g}_{14}}{\partial \boldsymbol{\ell}_2} \Big]^T \mathbf{g}_{14}  = 2 \big( \mathbf{g}_{23}  + \mathbf{g}_{14} \big). 
    \end{aligned}
\end{equation}
using the linear dependence of the gap vectors in Eq.\;[\ref{eq:theGaps}] on the variables being differentiated.  The minimizers in Eq.\;[\ref{eq:minTrans}] are given by any collection of vectors   $\mathbf{c}_1,\ldots, \mathbf{c}_4, \boldsymbol{\ell}_1, \boldsymbol{\ell}_2$ that makes these derivatives vanish. Following a straightforward manipulation of the right side of Eq.\;[\ref{eq:derivativesTrans}], these six derivatives vanish if and only if the gap vectors satisfy the five vector equations
\begin{equation}
\begin{aligned}\label{eq:solveGaps}
    &\mathbf{g}_{43} = -\mathbf{g}_{12}  , \quad \mathbf{g}_{23} =  -\mathbf{g}_{14}, \quad \mathbf{g}_{32} = - \mathbf{g}_{41}, \quad  \mathbf{g}_{34} = - \mathbf{g}_{21} , \quad   \mathbf{g}_{12} - \mathbf{g}_{21} + \mathbf{g}_{14} - \mathbf{g}_{41} = \mathbf{0}.
\end{aligned}  
\end{equation}
That there are only five equations to solve for these six derivatives is fully expected, since the elastic energy is invariant under a translation of the pattern. In pursuit of a solution, observe that 
\begin{equation}
\begin{aligned}
&\mathbf{g}_{43} + \mathbf{g}_{12} = \boldsymbol{\Delta}_{43}^{\boldsymbol{\eta}} \mathbf{x}_{43} + \boldsymbol{\Delta}_{12}^{\boldsymbol{\eta}} \mathbf{x}_{12}  + \mathbf{c}_1 - \mathbf{c}_2 - \mathbf{c}_3  +\mathbf{c}_4  + 2 \boldsymbol{\ell}_1 - (\mathbf{R}(\eta_1) + \mathbf{R}(\eta_4)) \boldsymbol{\ell}_1^R, \\
&\mathbf{g}_{34} + \mathbf{g}_{21} =  \boldsymbol{\Delta}_{34}^{\boldsymbol{\eta}} \mathbf{x}_{34} + \boldsymbol{\Delta}_{21}^{\boldsymbol{\eta}} \mathbf{x}_{21} - \mathbf{c}_1 + \mathbf{c}_2 + \mathbf{c}_3 - \mathbf{c}_4,  \\ 
&\mathbf{g}_{23} + \mathbf{g}_{14} = \boldsymbol{\Delta}_{23}^{\boldsymbol{\eta}} \mathbf{x}_{23} + \boldsymbol{\Delta}_{14}^{\boldsymbol{\eta}} \mathbf{x}_{14} + \mathbf{c}_1 + \mathbf{c}_2 - \mathbf{c}_3 - \mathbf{c}_4 + 2 \boldsymbol{\ell}_2 - (\mathbf{R}(\eta_1) + \mathbf{R}(\eta_2)) \boldsymbol{\ell}_2^R,\\
&\mathbf{g}_{32} + \mathbf{g}_{41} =  \boldsymbol{\Delta}_{32}^{\boldsymbol{\eta}} \mathbf{x}_{32} + \boldsymbol{\Delta}_{41}^{\boldsymbol{\eta}} \mathbf{x}_{41}  - \mathbf{c}_1 - \mathbf{c}_2  + \mathbf{c}_3 + \mathbf{c}_4, \\
& \mathbf{g}_{12} - \mathbf{g}_{21} + \mathbf{g}_{14} - \mathbf{g}_{41} = \boldsymbol{\Delta}_{12}^{\boldsymbol{\eta}} (\mathbf{x}_{12} + \mathbf{x}_{21}) + 2(\mathbf{c}_1 - \mathbf{c}_2) + \boldsymbol{\ell}_{1} - \mathbf{R}(\eta_1)\boldsymbol{\ell}_1^R +  \boldsymbol{\Delta}_{14}^{\boldsymbol{\eta}} (\mathbf{x}_{14} + \mathbf{x}_{41}) + 2(\mathbf{c}_1 - \mathbf{c}_4) + \boldsymbol{\ell}_{2} - \mathbf{R}(\eta_1)\boldsymbol{\ell}_2^R
\end{aligned}
\end{equation}
using $\boldsymbol{\Delta}_{ij}^{\boldsymbol{\eta}} = \mathbf{R}(\eta_i) - \mathbf{R}(\eta_j)$ for short.   Setting these equations to zero and rearranging terms gives 
\begin{equation}
    \begin{aligned}\label{eq:getEverything}
    &\boldsymbol{\ell}_1 = \frac{1}{2} \Big( \mathbf{\Delta}_{12}^{\boldsymbol{\eta}}(\mathbf{x}_{21}-\mathbf{x}_{12})+\mathbf{\Delta}_{43}^{\boldsymbol{\eta}}(\mathbf{x}_{34} -\mathbf{x}_{43})   +  \big( \mathbf{R}(\eta_1) + \mathbf{R}(\eta_4) \big) \boldsymbol{\ell}_1^R \Big), \\
    &\boldsymbol{\ell}_2 = \frac{1}{2}\Big(  \boldsymbol{\Delta}_{14}^{\boldsymbol{\eta}}(\mathbf{x}_{41}-\mathbf{x}_{14}) + \boldsymbol{\Delta}_{23}^{\boldsymbol{\eta}} (\mathbf{x}_{32}-\mathbf{x}_{23}) +\big( \mathbf{R}(\eta_1) + \mathbf{R}(\eta_2) \big) \boldsymbol{\ell}_2^R\Big), \\
    &\mathbf{c}_{4}- \mathbf{c}_1 = \frac{1}{8}\Big( \boldsymbol{\Delta}_{14}^{\boldsymbol{\eta}} (5 \mathbf{x}_{41} + \mathbf{x}_{14} - \boldsymbol{\ell}_1^R)  + \boldsymbol{\Delta}_{12}^{\boldsymbol{\eta}} ( \mathbf{x}_{12} + \mathbf{x}_{21}  -\boldsymbol{\ell}_2^{R}) + \boldsymbol{\Delta}_{23}^{\boldsymbol{\eta}} (3 \mathbf{x}_{32} - \mathbf{x}_{23}) - \boldsymbol{\Delta}_{43}^{\boldsymbol{\eta}}(\mathbf{x}_{34} + \mathbf{x}_{43}) \Big),\\
    &\mathbf{c}_{3} - \mathbf{c}_4 = \frac{1}{8} \Big( \boldsymbol{\Delta}_{43}^{\boldsymbol{\eta}} ( 5\mathbf{x}_{34} + \mathbf{x}_{43}) + \boldsymbol{\Delta}_{23}^{\boldsymbol{\eta}} ( \mathbf{x}_{32} + \mathbf{x}_{23} ) + \boldsymbol{\Delta}_{12}^{\boldsymbol{\eta}} (3 \mathbf{x}_{21} - \mathbf{x}_{12} + \boldsymbol{\ell}_{2}^R) - \boldsymbol{\Delta}_{14}^{\boldsymbol{\eta}} (\mathbf{x}_{41} + \mathbf{x}_{14} - \boldsymbol{\ell}_{1}^R)\Big),  \\ 
    &\mathbf{c}_{2} - \mathbf{c}_1 = \frac{1}{8} \Big( \boldsymbol{\Delta}_{12}^{\boldsymbol{\eta}}( 5 \mathbf{x}_{21} + \mathbf{x}_{12} - \boldsymbol{\ell}_2^R) + \boldsymbol{\Delta}_{43}^{\boldsymbol{\eta}} (3 \mathbf{x}_{34} - \mathbf{x}_{43}) + \boldsymbol{\Delta}_{14}^{\boldsymbol{\eta}}( \mathbf{x}_{41} + \mathbf{x}_{14} - \boldsymbol{\ell}_1^R) - \boldsymbol{\Delta}_{23}^{\boldsymbol{\eta}} (\mathbf{x}_{32} + \mathbf{x}_{23})\Big).
    \end{aligned}
\end{equation}
Substituting this parameterization back into the gap vectors gives the minimizing vectors 
\begin{equation}
\begin{aligned}\label{eq:gapFormulas}
    \mathbf{g}_{12} = - \mathbf{g}_{43} &= \frac{1}{8}\Big( \boldsymbol{\Delta}_{12}^{\boldsymbol{\eta}}(3 \mathbf{x}_{12} - \mathbf{x}_{21} + \boldsymbol{\ell}_2^R) - \boldsymbol{\Delta}_{14}^{\boldsymbol{\eta}}(\mathbf{x}_{41} + \mathbf{x}_{14} +3 \boldsymbol{\ell}_1^R) + \boldsymbol{\Delta}_{43}^{\boldsymbol{\eta}} ( \mathbf{x}_{34} - 3 \mathbf{x}_{43})  + \boldsymbol{\Delta}_{23}^{\boldsymbol{\eta}} (\mathbf{x}_{32} + \mathbf{x}_{23}) \Big) \\
    &= \frac{1}{8}\Big(2\sum_{i=1,\ldots,4} \mathbf{R}(\eta_i)\mathbf{s}_i - \sum_{i=1,\ldots,4} \mathbf{R}(\eta_i)\mathbf{t}_i - 3\sum_{i=1,\ldots,4} \mathbf{R}(\eta_i)\mathbf{u}_i + 4 \sum_{i=1,\ldots,4} \mathbf{R}(\eta_i)\mathbf{v}_i \Big), \\ 
    \mathbf{g}_{21} = - \mathbf{g}_{34} &= \frac{1}{8} \Big(\boldsymbol{\Delta}_{12}^{\boldsymbol{\eta}}(-3 \mathbf{x}_{21} + \mathbf{x}_{12} - \boldsymbol{\ell}_{2}^R)  + \boldsymbol{\Delta}_{43}^{\boldsymbol{\eta}} (3 \mathbf{x}_{34} - \mathbf{x}_{43}) + \boldsymbol{\Delta}_{14}^{\boldsymbol{\eta}}( \mathbf{x}_{41} + \mathbf{x}_{14} - \boldsymbol{\ell}_1^R) - \boldsymbol{\Delta}_{23}^{\boldsymbol{\eta}} (\mathbf{x}_{32} + \mathbf{x}_{23})\Big) \\
    &= \frac{1}{8}\Big(-4\sum_{i=1,\ldots,4}  \mathbf{R}(\eta_i)\mathbf{s}_i +3 \sum_{i=1,\ldots,4} \mathbf{R}(\eta_i)\mathbf{t}_i + \sum_{i=1,\ldots,4} \mathbf{R}(\eta_i)\mathbf{u}_i - 2 \sum_{i=1,\ldots,4} \mathbf{R}(\eta_i)\mathbf{v}_i \Big), \\ 
    \mathbf{g}_{14} = - \mathbf{g}_{23} &= \frac{1}{8} \Big(\boldsymbol{\Delta}_{14}^{\boldsymbol{\eta}} (3 \mathbf{x}_{14} - \mathbf{x}_{41} + \boldsymbol{\ell}_1^R) - \boldsymbol{\Delta}_{12}^{\boldsymbol{\eta}} ( \mathbf{x}_{12} + \mathbf{x}_{21}  +3 \boldsymbol{\ell}_2^{R}) + \boldsymbol{\Delta}_{23}^{\boldsymbol{\eta}} ( \mathbf{x}_{32} - 3 \mathbf{x}_{23}) + \boldsymbol{\Delta}_{43}^{\boldsymbol{\eta}}(\mathbf{x}_{34} + \mathbf{x}_{43})\Big)\\
    &= \frac{1}{8}\Big(3\sum_{i=1,\ldots,4}  \mathbf{R}(\eta_i)\mathbf{s}_i -2 \sum_{i=1,\ldots,4} \mathbf{R}(\eta_i)\mathbf{t}_i -4 \sum_{i=1,\ldots,4} \mathbf{R}(\eta_i)\mathbf{u}_i + \sum_{i=1,\ldots,4} \mathbf{R}(\eta_i)\mathbf{v}_i \Big), \\ 
    \mathbf{g}_{41} = - \mathbf{g}_{32} &= \frac{1}{8} \Big( \boldsymbol{\Delta}_{14}^{\boldsymbol{\eta}} (-3 \mathbf{x}_{41} + \mathbf{x}_{14} - \boldsymbol{\ell}_1^R)  + \boldsymbol{\Delta}_{12}^{\boldsymbol{\eta}} ( \mathbf{x}_{12} + \mathbf{x}_{21}  -\boldsymbol{\ell}_2^{R}) + \boldsymbol{\Delta}_{23}^{\boldsymbol{\eta}} (3 \mathbf{x}_{32} - \mathbf{x}_{23}) - \boldsymbol{\Delta}_{43}^{\boldsymbol{\eta}}(\mathbf{x}_{34} + \mathbf{x}_{43}) \Big)  \\ 
    &= \frac{1}{8}\Big(-\sum_{i=1,\ldots,4}  \mathbf{R}(\eta_i)\mathbf{s}_i +4 \sum_{i=1,\ldots,4} \mathbf{R}(\eta_i)\mathbf{t}_i +2 \sum_{i=1,\ldots,4} \mathbf{R}(\eta_i)\mathbf{u}_i -3 \sum_{i=1,\ldots,4} \mathbf{R}(\eta_i)\mathbf{v}_i \Big), \\ 
\end{aligned}
\end{equation}
where the second equality in each formula is the result of a careful manipulation  using the correspondence between design vectors $\mathbf{s}_1,\ldots, \mathbf{v}_4$  and points $\mathbf{x}_{ij}$ in Fig.\;\ref{fig:designSup}A and Fig.\;\ref{fig:elasticSup}A, as well as the loop conditions in Eq.\;[\ref{eq:LoopRef}].

This derivation reveals an analytical expression for the minimization in Eq.\;[\ref{eq:minTrans}] of the form 
\begin{equation}
    \begin{aligned}\label{eq:springEnergyFinal}
        E_{\text{spr}}(\eta_1, \eta_2, \eta_3, \eta_4) = \begin{bmatrix}\sum_{i = 1,\ldots, 4} \mathbf{R}(\eta_i) \mathbf{s}_i \\  \sum_{i = 1,\ldots, 4} \mathbf{R}(\eta_i) \mathbf{t}_i \\  \sum_{i = 1,\ldots, 4} \mathbf{R}(\eta_i) \mathbf{u}_i \\ \sum_{i = 1,\ldots, 4} \mathbf{R}(\eta_i) \mathbf{v}_i \end{bmatrix} \cdot \underbrace{\begin{bmatrix} \frac{15}{16} \mathbf{I} & -\frac{3}{4} \mathbf{I} & -\frac{3}{4} \mathbf{I} & \frac{11}{16} \mathbf{I} \\ -\frac{3}{4} \mathbf{I} & \frac{15}{16} \mathbf{I} & \frac{11}{16} \mathbf{I} & - \frac{3}{4} \mathbf{I}  \\ -\frac{3}{4} \mathbf{I} & \frac{11}{16} \mathbf{I} & \frac{15}{16} \mathbf{I} & - \frac{3}{4} \mathbf{I} \\ \frac{11}{16} \mathbf{I} & -\frac{3}{4} \mathbf{I} & -\frac{3}{4} \mathbf{I} & \frac{15}{16} \mathbf{I} \end{bmatrix}}_{= \mathbf{G}}  \begin{bmatrix}\sum_{i = 1,\ldots, 4} \mathbf{R}(\eta_i) \mathbf{s}_i \\  \sum_{i = 1,\ldots, 4} \mathbf{R}(\eta_i) \mathbf{t}_i \\  \sum_{i = 1,\ldots, 4} \mathbf{R}(\eta_i) \mathbf{u}_i \\ \sum_{i = 1,\ldots, 4} \mathbf{R}(\eta_i) \mathbf{v}_i \end{bmatrix},
    \end{aligned}
\end{equation}
specifically, by combining  the first line of the definition of $E_{\text{spr}}^0$ in Eq.\;[\ref{eq:sprEnergy}] with the gaps formulas in Eq.\;[\ref{eq:gapFormulas}]. Eq.\;[\ref{eq:springEnergyFinal}] is Eq.\;[3] in the main text. The matrix $\mathbf{G} \in \mathbb{R}^{8 \times 8}$ defined above is symmetric and positive definite.

\subsection{Actuation energy and the energy barrier between stable states} The elastic energy in Eq.\;[\ref{eq:springEnergyFinal}] is formulated in terms of four angle variables, corresponding to the four panel rotations of each unit cell. We now formulate  an elastic energy for the pattern's motion between the two stable states through a combination of invariances and further energy minimization. This energy is written in terms of a single generalized displacement variable called the \textit{actuation angle}. 

Assume that Eq.\;[\ref{eq:theorem111}] holds for the design vectors $\mathbf{s}_1, \ldots, \mathbf{v}_4$ and angles $\phi_1,\ldots, \phi_4$. The energy in Eq.\;[\ref{eq:springEnergyFinal}] has the following properties 
\begin{equation}
\begin{aligned}
   (\text{invariance under rigid motion:})& \quad  E_{\text{spr}}(\eta_1 + \eta, \eta_2 + \eta, \eta_3 + \eta, \eta_4 + \eta) =  E_{\text{spr}}(\eta_1, \eta_2, \eta_3, \eta_4), \quad  \eta, \eta_1, \ldots, \eta_4 \in \mathbb{R}, \\
   (\text{first designed stable state:})& \quad E_{\text{spr}}(0,0,0,0) = 0,\\
   (\text{second designed stable state:})& \quad E_{\text{spr}}(\phi_1,\phi_2,\phi_3,\phi_4) = 0.
\end{aligned}
\end{equation}
Thus, one of the four angles is redundant on energy minimization since it describes an overall rotation of the pattern.
We therefore fix the first angle to set this overall rotation and  define the actuation energy  as 
\begin{equation}\label{eq:actuationEnergy}
    \begin{aligned}
        E_{\text{act}}(\xi) = \min_{\eta_3, \eta_4 \in \mathbb{R}} E_{\text{spr}}(\phi_1, \phi_1 + \xi, \eta_3, \eta_4).
    \end{aligned}
\end{equation}
The \textit{actuation angle} $\xi$ is the relative rotation angle between the first and second panel of a unit cell, after minimizing out all other DOFs under the assumed panel motions. It thus provides a quantitative measure for how the slits are opening/closing during the actuation from one stable state to another. Mathematically,  $E_{\text{act}}(\xi) =0$ when $\xi = 0$ and when $\xi = \phi_2 - \phi_1$, reflecting the two designed stable states.  The interval between these angles  furnishes the energy barrier to actuation via 
\begin{equation}
    \begin{aligned}
        E_b = \begin{cases}
            \underset{\xi \in (0, \phi_2- \phi_1)} {\max}{E_{\text{act}}(\xi)}
            \text{ if } \phi_2-\phi_1>0
            \\
            \underset{\xi \in (\phi_2- \phi_1, 0)} {\max}{E_{\text{act}}(\xi)}
            \text{ if } \phi_2-\phi_1<0.
        \end{cases}
    \end{aligned}
\end{equation}

 After imposing the bistability constraint in Eq.\;[\ref{eq:theorem111}] and prescribing a shape change through $\boldsymbol{\ell} = \overline{\boldsymbol{\ell}}$, the energies $E_{\text{spr}}^0$, $E_{\text{spr}}$, $E_{\text{act}}$ and $E_b$ depend on the design variables $\mathbf{s}_1,\mathbf{s}_2$ and $\boldsymbol{\phi}$, in addition to the kinematic variables noted in their derivation. We can tune these variables, and thereby the bistable kirigami design, by optimizing an objective function of the form 
\begin{equation}
    \begin{aligned}\label{eq:objEnergyBarrier}
        f_{\text{obj}}(\mathbf{s}_1, \mathbf{s}_2, \boldsymbol{\phi}) = |E_b(\mathbf{s}_1, \mathbf{s}_2, \boldsymbol{\phi}) - E_b^{\text{targ}}|^2
    \end{aligned}
\end{equation}
for a specified target energy barrier $E_b^{\text{targ}}$. The main text presents a detailed example of such tuning.

\subsection{Optimizing for stiffness} Another natural quantity to tune,  beyond the energy barrier between the two stable states, is the stiffness of these states. For simplicity, we  focus on the stiffness of the first stable state, as the other can be obtained by analogous reasoning. 

A typical stiffness measure is formulated with respect to some strain or stretch measure. In this exposition, we consider the stiffness with respect to a simple \textit{characteristic stretch} $\lambda(\xi)$, defined as follows. First note that a characteristic length of a kirigami cell in it first stable state is
\begin{equation}
    \begin{aligned}
        |\boldsymbol{\ell}_1^R| = |\mathbf{u}_1 -\mathbf{v}_1 + \mathbf{u}_2 - \mathbf{v}_2|
    \end{aligned}
\end{equation}
per Eq.\;[\ref{eq:refLattice}]. 
Its counterpart in the second stable state is  
\begin{equation}
    \begin{aligned}
        |\boldsymbol{\ell}_1^D| = |\mathbf{u}_1 -\mathbf{v}_1 + \mathbf{R}(\phi_2 - \phi_1) (\mathbf{u}_2 - \mathbf{v}_2)|
    \end{aligned}
\end{equation}
per Eq.\;[\ref{eq:defLattice}].  The ratio of these two quantities defines a stretch relating the stable states.  Moreover, the actuation angle $\xi$ is  zero in the first stable state and   $\phi_2 - \phi_1$ in the second.  Consequently, we take the characteristic stretch to be  
\begin{equation}
    \begin{aligned}
\lambda(\xi) = \frac{| \mathbf{u}_1 -\mathbf{v}_1 + \mathbf{R}(\xi) (\mathbf{u}_2 - \mathbf{v}_2)|}{|\mathbf{u}_1 -\mathbf{v}_1 + \mathbf{u}_2 - \mathbf{v}_2|}
    \end{aligned}
\end{equation}
during the actuation from $\xi = 0$ to $\xi = \phi_2-\phi_1$.

The stiffness with respect to this stretch is as follows. Set  $\hat{E}_{\text{act}}(\lambda(\xi)) = E_{\text{act}}(\xi)$ and  observe that 
\begin{equation}
\begin{aligned}
E'_{\text{act}}(\xi) =\hat{E}_{\text{act}}'(\lambda(\xi)) \lambda'(\xi), \quad 
 E''_{\text{act}}(\xi)  = \hat{E}_{\text{act}}''(\lambda(\xi)) (\lambda'(\xi))^2 + \hat{E}_{\text{act}}'(\lambda(\xi)) \lambda''(\xi).
\end{aligned}
\end{equation}
As the first stable state corresponds to  $(\xi, \lambda(\xi)) = (0,1)$ and is a minimizer of $E_{\text{act}}(\xi)$, we conclude that the desired stiffness is
\begin{equation}
    \begin{aligned}
        k_1 = \hat{E}_{\text{act}}''(1) = (\lambda'(0))^{-2} E_{\text{act}}''(0)
    \end{aligned}
\end{equation}
provided that $\lambda'(0)$ is nonzero. To derive an explicit formula for $k_1$, we first Taylor expand the spring energy in Eq.\;[\ref{eq:springEnergyFinal}] about the first stable state to obtain the leading order quadratic form associated to the spring energy in a neighborhood of this state. We then link these quadratic forms to the second derivatives of the actuation energy and supply a formula for $\lambda'(0)$ to complete the derivation.

The loop conditions in Eqs.\;[\ref{eq:LoopRef}] imply that, on Taylor expansion, 
the first  stable state satisfies 
\begin{equation}
    \begin{aligned}
        \begin{bmatrix}\sum_{i = 1,\ldots, 4} \mathbf{R}(\delta\eta_i) \mathbf{s}_i \\  \sum_{i = 1,\ldots, 4} \mathbf{R}(\delta \eta_i) \mathbf{t}_i \\  \sum_{i = 1,\ldots, 4} \mathbf{R}(\delta\eta_i) \mathbf{u}_i \\ \sum_{i = 1,\ldots, 4} \mathbf{R}(\delta\eta_i) \mathbf{v}_i \end{bmatrix} = \begin{bmatrix} \mathbf{R}(\tfrac{\pi}{2}) \mathbf{s}_1 & \mathbf{R}(\tfrac{\pi}{2})\mathbf{s}_2 & \mathbf{R}(\tfrac{\pi}{2})\mathbf{s}_3 & \mathbf{R}(\tfrac{\pi}{2})\mathbf{s}_4 \\ \mathbf{R}(\tfrac{\pi}{2}) \mathbf{t}_1 & \mathbf{R}(\tfrac{\pi}{2})\mathbf{t}_2 & \mathbf{R}(\tfrac{\pi}{2})\mathbf{t}_3 & \mathbf{R}(\tfrac{\pi}{2})\mathbf{t}_4  \\ \mathbf{R}(\tfrac{\pi}{2})\mathbf{u}_1 & \mathbf{R}(\tfrac{\pi}{2})\mathbf{u}_2 & \mathbf{R}(\tfrac{\pi}{2})\mathbf{u}_3 & \mathbf{R}(\tfrac{\pi}{2})\mathbf{u}_4 \\ 
        \mathbf{R}(\tfrac{\pi}{2})\mathbf{v}_1 & \mathbf{R}(\tfrac{\pi}{2})\mathbf{v}_2 & \mathbf{R}(\tfrac{\pi}{2})\mathbf{v}_3 & \mathbf{R}(\tfrac{\pi}{2})\mathbf{v}_4 \end{bmatrix} \underbrace{\begin{bmatrix} \delta \eta_1 \\ \delta \eta_2 \\ \delta \eta_3 \\ \delta \eta_4 \end{bmatrix}}_{=\boldsymbol{\delta}\boldsymbol{\eta}} + O( |\boldsymbol{\delta} \boldsymbol{\eta}|^2).
    \end{aligned}
\end{equation}
A Taylor expansion of the spring energy around this state gives that 
\begin{equation}
    \begin{aligned}
     &E_{\text{spr}}(\delta \eta_1, \ldots, \delta \eta_4 ) =  \begin{bmatrix} \delta \eta_1 \\ \delta \eta_2 \\ \delta \eta_3 \\ \delta \eta_4 \end{bmatrix} \cdot \underbrace{\begin{bmatrix}  \mathbf{s}_1 & \mathbf{s}_2 & \mathbf{s}_3 & 
 \mathbf{s}_4 \\ \mathbf{t}_1 & \mathbf{t}_2 & \mathbf{t}_3 & \mathbf{t}_4  \\ \mathbf{u}_1 & \mathbf{u}_2 & \mathbf{u}_3 & \mathbf{u}_4 \\ 
        \mathbf{v}_1 &\mathbf{v}_2 &\mathbf{v}_3 & \mathbf{v}_4 \end{bmatrix}^T \mathbf{G} \begin{bmatrix}  \mathbf{s}_1 & \mathbf{s}_2 & \mathbf{s}_3 & 
 \mathbf{s}_4 \\ \mathbf{t}_1 & \mathbf{t}_2 & \mathbf{t}_3 & \mathbf{t}_4  \\ \mathbf{u}_1 & \mathbf{u}_2 & \mathbf{u}_3 & \mathbf{u}_4 \\ 
        \mathbf{v}_1 &\mathbf{v}_2 &\mathbf{v}_3 & \mathbf{v}_4 \end{bmatrix}}_{=\frac{1}{2} \mathbf{K}_1}\begin{bmatrix} \delta \eta_1 \\ \delta \eta_2 \\ \delta \eta_3 \\ \delta \eta_4 \end{bmatrix}  + O(|\boldsymbol{\delta}\boldsymbol{\eta}|^3).
    \end{aligned}
\end{equation}
As a result, the actuation energy near the first stable state is 
\begin{equation}
    \begin{aligned}
        E_{\text{act}}(\delta \xi) = \min_{\delta \eta_3, \delta \eta_4 \in \mathbb{R}} E_{\text{spr}}(0,\delta \xi, \delta \eta_3, \delta \eta_4)  = \min_{\delta \eta_3, \delta \eta_4 \in \mathbb{R}} \left\{ \frac{1}{2}  \begin{bmatrix} 0 \\ \delta \xi \\ \delta \eta_3 \\ \delta \eta_4  \end{bmatrix}  \cdot \mathbf{K}_1 \begin{bmatrix} 0 \\ \delta \xi \\ \delta \eta_3 \\ \delta \eta_4  \end{bmatrix} \right\}  + O(\delta \xi^3).
    \end{aligned}
\end{equation}
To organize the calculation, we introduce the scalar, vector, and matrix defined as 
\begin{equation}
    \begin{aligned}
        k_1^0 = [\mathbf{K}_1]_{22}, \qquad \mathbf{k}_1^0 = \begin{bmatrix} [\mathbf{K}_1]_{23} \\ [\mathbf{K}_1]_{24} \end{bmatrix} = \begin{bmatrix} [\mathbf{K}_1]_{32} \\ [\mathbf{K}_1]_{43} \end{bmatrix} , \qquad \mathbf{K}_1^0  = \begin{bmatrix} [\mathbf{K}_1]_{33}  & [\mathbf{K}_1]_{34} \\ [\mathbf{K}_1]_{34} & [\mathbf{K}_1]_{44} \end{bmatrix},
    \end{aligned}
\end{equation}
where  $[\mathbf{K}_{1}]_{kl}$ indicates the $kl$ component of the $\mathbf{K}_1$ matrix. Minimizing out the last two variables in the quadratic form leads to 
\begin{equation}
    \begin{aligned}
        E_{\text{act}}(\delta \xi) = \min_{\delta \eta_3, \delta \eta_4 \in \mathbb{R}} \left\{ \frac{1}{2}  \begin{bmatrix}  \delta \xi \\ \delta \eta_3 \\ \delta \eta_4  \end{bmatrix}  \cdot \begin{bmatrix} k_1^0 & (\mathbf{k}_1^0)^T \\ \mathbf{k}_1^0 & \mathbf{K}_1^0  \end{bmatrix} \begin{bmatrix}  \delta \xi \\ \delta \eta_3 \\ \delta \eta_4  \end{bmatrix} \right\} + O(\delta \xi^3) =\frac{1}{2} \big(k_1^{0} - \mathbf{k}_1^0 \cdot \big(\mathbf{K}_1^0\big)^{-1} \mathbf{k}_1^0 \big) \delta \xi^2 + O(\delta \xi^3).
    \end{aligned}
\end{equation}
Thus $E''_{\text{act}}(0)  = k_1^{0} - \mathbf{k}_1^0 \cdot \big(\mathbf{K}_1^0\big)^{-1} \mathbf{k}_1^0$, which  in terms of the design vectors $\mathbf{s}_1, \ldots, \mathbf{v}_4$ is 
\begin{equation}
    \begin{aligned}\label{eq:k1Final}
        E''_{\text{act}}(0)  &= \underbrace{2\begin{bmatrix} \mathbf{s}_2 \\ \vdots  \\ \mathbf{v}_2 \end{bmatrix}\cdot  \mathbf{G}\begin{bmatrix} \mathbf{s}_2 \\ \vdots  \\ \mathbf{v}_2 \end{bmatrix}}_{= k_1^0}  - \underbrace{2 \begin{bmatrix} \mathbf{s}_3^T & \cdots & \mathbf{v}_3^T \\  \mathbf{s}_4^T & \cdots & \mathbf{v}_4^T\end{bmatrix} \mathbf{G}\begin{bmatrix} \mathbf{s}_2 \\ \vdots  \\ \mathbf{v}_2 \end{bmatrix}}_{=\mathbf{k}_1^0} \cdot \underbrace{\left(2 \begin{bmatrix} \mathbf{s}_3^T & \cdots & \mathbf{v}_3^T \\  \mathbf{s}_4^T & \cdots & \mathbf{v}_4^T\end{bmatrix} \mathbf{G} \begin{bmatrix}  \mathbf{s}_3 & \mathbf{s}_4 \\ \vdots & \vdots \\ \mathbf{v}_3 & \mathbf{v}_4 \end{bmatrix}  \right)^{-1}}_{= (\mathbf{K}_1^0)^{-1}}  \underbrace{2 \begin{bmatrix} \mathbf{s}_3^T & \cdots & \mathbf{v}_3^T \\  \mathbf{s}_4^T & \cdots & \mathbf{v}_4^T\end{bmatrix} \mathbf{G}\begin{bmatrix} \mathbf{s}_2 \\ \vdots  \\ \mathbf{v}_2 \end{bmatrix}}_{=\mathbf{k}_1^0}.
    \end{aligned}
\end{equation}

To complete the derivation, we provide a formula for $\lambda'(0)$ in terms of the design vectors. Observe that  
\begin{equation}
    \begin{aligned}
        \lambda(\delta \xi) =  \frac{|\mathbf{u}_1 - \mathbf{v}_1 + \mathbf{u}_2 - \mathbf{v}_2  + \delta \xi \mathbf{R}(\tfrac{\pi}{2})(\mathbf{u}_2 - \mathbf{v}_2) + O(\delta \xi^2)|}{|\mathbf{u}_1 - \mathbf{v}_1 + \mathbf{u}_2 - \mathbf{v}_2 |} = 1 + \delta \xi \frac{(\mathbf{u}_1 - \mathbf{v}_1) \cdot \mathbf{R}(\tfrac{\pi}{2}) (\mathbf{u}_2 - \mathbf{v}_2)}{|\mathbf{u}_1 - \mathbf{v}_1 + \mathbf{u}_2 - \mathbf{v}_2 |^2} + O(\delta \xi^2)
    \end{aligned}
\end{equation}  
  on  Taylor expansion,
since $|\mathbf{v} + \boldsymbol{\delta} \mathbf{v}| = |\mathbf{v}| +|\mathbf{v}|^{-1} \mathbf{v} \cdot \boldsymbol{\delta} \mathbf{v} + O(|\boldsymbol{\delta} \mathbf{v}|^2)$. So we conclude that 
\begin{equation}
    \begin{aligned}\label{eq:finalLambda}
\lambda'(0) = \frac{(\mathbf{u}_1 - \mathbf{v}_1) \cdot \mathbf{R}(\tfrac{\pi}{2}) (\mathbf{u}_2 - \mathbf{v}_2)}{|\mathbf{u}_1 - \mathbf{v}_1 + \mathbf{u}_2 - \mathbf{v}_2 |^2}.
    \end{aligned}
\end{equation}
The formulas in Eq.\;[\ref{eq:k1Final}] and Eq.\;[\ref{eq:finalLambda}] furnish an explicit description of the stiffness $k_1 = (\lambda'(0))^{-2} E_{\text{act}}''(0)$.

 Since this stiffness is subject to the bistability constraint in Eq.\;[\ref{eq:theorem111}], it depends only on the tuning variables $\mathbf{s}_1, \mathbf{s}_2$ and $\boldsymbol{\phi}$ after applying this constraint to the above formulas for a fixed shape change $\boldsymbol{\ell} = \overline{\boldsymbol{\ell}}$, i.e., 
\begin{equation}
\begin{aligned}
    k_1 \equiv k_1(\mathbf{s}_1, \mathbf{s}_2, \boldsymbol{\phi}).
\end{aligned}
\end{equation}
In the main text, we present examples where we optimzie both for stiffness and a target energy barrier  by considering objective functions of the form  
\begin{equation}
    \begin{aligned}\label{eq:fogjExample}
        f_{\text{obj}}(\mathbf{s}_1, \mathbf{s}_2, \boldsymbol{\phi}) = c_b | E_b(\mathbf{s}_1, \mathbf{s}_2, \boldsymbol{\phi}) - E_b^{\text{targ}}|^2 + c_1 | k_1(\mathbf{s}_1, \mathbf{s}_2, \boldsymbol{\phi}) - k_{1}^{\text{targ}}| ^2 ,
    \end{aligned}
\end{equation}
for a specified target stiffness $k_1^{\text{targ}} \geq 0$ in the first  stable state and target energy barrier $E_b^{\text{targ}} \geq0$. The weights of the parameters $c_b \geq 0 $ and $c_1 \geq 0$ are  carefully assigned to best achieve the multiple objectives.

\begin{figure}[htb!]
\centering
\includegraphics[scale=0.7]{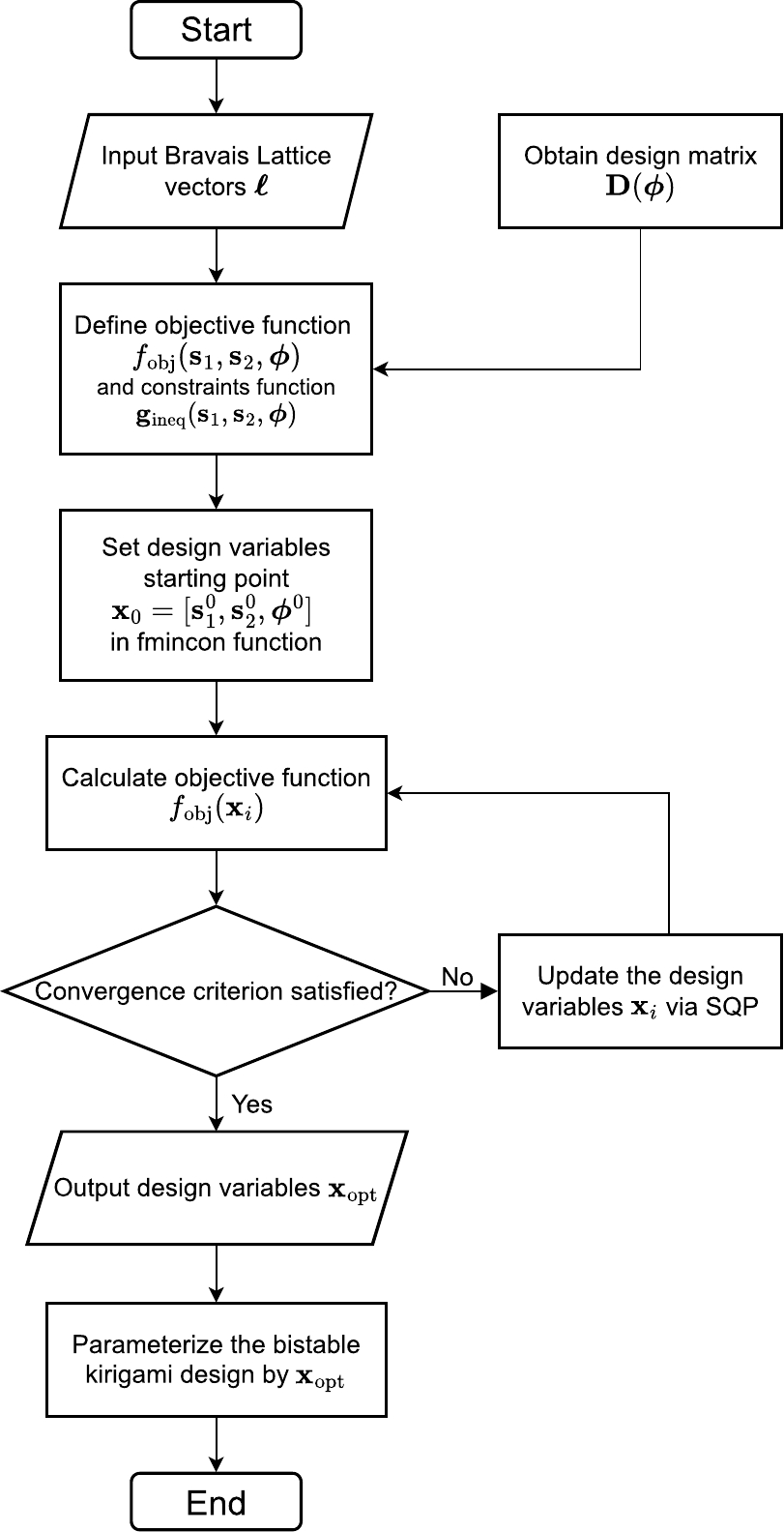}
\caption{ {\color{black}Flowchart of   numerical aspects to the optimization framework.}}
\label{fig:SM-optimization flowchart}
\end{figure}

\subsection{Numerical implementation}\label{ssec:NumImplement}

The numerical framework for the constrained optimization problem  employs the \verb+fmincon+ toolbox to perform a sequential quadratic programming (SQP) algorithm in Matlab (R2023a). {\color{black}A flowchart of the procedure is illustrated in Fig\;\ref{fig:SM-optimization flowchart}. To enable the optimization, we prescribe the Bravais lattice vectors $\boldsymbol{\ell}$ for two stable states and constrain the design vectors $\mathbf{s}_3, \ldots, \mathbf{v}$ according to Eq.\;(\ref{eq:designFormula}). The eight remaining variables --- the design vectors  $\mathbf{s}_1$ and $\mathbf{s}_2$ and four rotation angles in the array $\boldsymbol{\phi}$ --- are then optimized based on an assigned objective function  $f_{\text{obj}}(\mathbf{s}_1, \mathbf{s}_2, \boldsymbol{\phi})$ (for instance Eq.\;(\ref{eq:fogjExample})) and the inequality constraints $\mathbf{g}_{\text{ineq}}(\mathbf{s}_1, \mathbf{s}_2, \boldsymbol{\phi})$ in Eq.\;(\ref{eq:gineqDef}). Beginning with an initial guess $\mathbf{x}_0 = (\mathbf{s}_1^0, \mathbf{s}_2^0, \boldsymbol{\phi}^0)$, these  variables undergo  iterative update $\mathbf{x}_i = (\mathbf{s}_1^i, \mathbf{s}_2^i, \boldsymbol{\phi}^i)$ using the SQP algorithm until a local minimum of the energy is achieved. This procedure transforms the kirigami geometry from its initial guess $\mathbf{x}_0$ to an optimal design $\mathbf{x}_{\text{opt}}= (\mathbf{s}_1^{\text{opt}}, \mathbf{s}_2^{\text{opt}}, \boldsymbol{\phi}^{\text{opt}})$ that accomplishes the target properties and meets the nonlinear constraints.}

Given the non-convex nature of the objective functions and constrained sets, the optimized geometry obtained by each simulation is dependent upon the choice of the starting point $\mathbf{x}_0$.  As illustrated by way of an example in Fig.\;4A (main text), different initial guesses for the design can yield different optimized designs that achieve the specified overall deformation and target energy barrier. Also, in some of the cases of extreme shape change in Fig.\;4A-B (main text), we are not able to achieve the target energy barrier, so we test several initial guesses to ensure that the deviation from the target is robust. We have also explored other objective functions consistent with minimizing the same objective, e.g., by replacing $|\cdot|^2$ terms with $|\cdot|$ terms, and found that such modifications can sometimes lead to improvements in the algorithms ability to find designs that get closer to the overall objective. To avoid \emph{ad hoc} procedures, we only present results  for cases where the objective function and its fitting parameters are consistent across all optimized designs being compared.

\subsection{Supporting information on the design exploration}

%We will provide further information to support the richness of our design framework and more discussion about the parallelogram slits/symmetry. (invite curiosity?)

%The objective function in our cases is defined as the squared difference between the target energy barrier and the actual value. Defining the objective function differently will yield distinct optimization results.

%The Bravais lattice vectors in two stable states and the target physical properties are the user-defined parameters to facilitate the creation of bistable kirigami designs exhibiting preferred deformations.

%By employing the proposed framework, it is feasible to efficiently generate the optimal configuration featuring specific stable states. The geometrical results of a single unit cell at both stable states and the corresponding energy plots are presented. Successful optimizations usually converge to minima that meet the given constraints within one hundred iterations, typically taking only several minutes.

The main text explores the design space of bistable kirigami  by featuring a wide range patterns that exhibit axial and shearing shape changes. One of the more intriguing findings is that extreme bistable designs seem to limit to mechanism-based designs with parallelogram slits, even though we optimize for a finite energy barrier. Here we lend further evidence to this finding. 

\begin{figure}[htb!]
\centering
\includegraphics[scale=1.0]{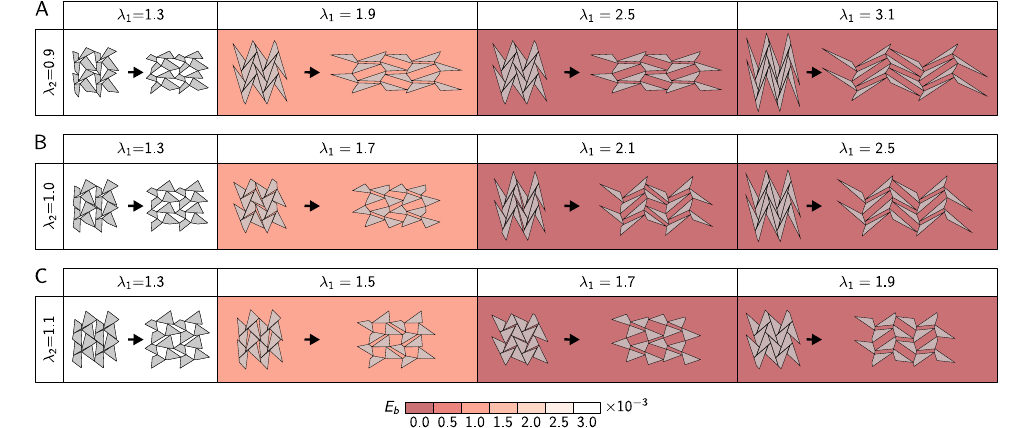}
\caption{Pushing our design exploration to the limit. Designs obtained by increasing $\lambda_1$, keeping $\lambda_2$ constant and equal to:  (A) $\lambda_2=0.9$, (B) $\lambda_2=1.0$,  (C) $\lambda_2=1.1$.}
\label{fig:SM-extreme exploration}
\end{figure}

Focusing on the axial setting (parameterized by the axial stretches $\lambda_1$ and $\lambda_2$ from the main text),  we fix $\lambda_2$ and progressively increase the value of $\lambda_1$ to generate optimized patterns until the algorithm is unable to find a design that meets the given constraints. Fig\;\ref{fig:SM-extreme exploration}A-C presents designs obtained in this setting by optimizing the energy barrier using $c_b = 1, c_1 = 0$ and $E_b^{\text{targ}} = 0.003$. The coloring scheme reflects the energy barrier of the design achieved on optimization. We see that,  on increasing $\lambda_1$ from  its largest value in main text $(\lambda_1 = 1.3)$  to its  approximate maximum at a fixed $\lambda_2$, a mechanism-based design  with parallelogram slits emerges. This observation adds further credence to the suggestion that there is perhaps a non-trivial universal relationship between bistable designs and mechanism-based ones. Informally, our basic conjecture is that the boundary of the set of all bistable planar kirigami designs is a set of mechanism-based designs.

%Designs obtained by optimizing the energy barrier, using$c_b = 1, c_1 = 0$ and $E_b^{\text{targ}} = 0.003$ for different values of $\lambda_{1}$
%, are presented in Fig\;\ref{fig:SM-extreme exploration}A-C. We see that,  in increasing $\lambda_1$ from $1.3$ -- its largest value in main text --  to its  approximate maximum at a fixed $\lambda_2$, a mechanism-based design  with parallelogram slits emerges. This observation is consistent with assertions made in the main text. 

%When the designs are subjected to extreme axial cases, the energy barriers associated with them are significantly reduced. It is observed that the optimized designs with extreme axial deformation exhibit a low energy barrier close to zero and possess parallelogram slits. These patterns demonstrate a single degree of freedom (DOF) mechanism-based motion between the two states. In our simulations, we consistently observe that the optimized patterns with extreme deformation converge to designs featuring parallelogram slits. 

%Aside from the slits, it is shown that the unit cell of the extreme designs is comprised of one single unique panel. By rotating or mirroring one of the four panels within each unit cell, the entire pattern can be obtained by symmetry. 

%It is predictable that gradually pushing the lattice vectors to their limits makes the algorithm's design space for locating local minima more and more restricted. Additionally, it is plausible that the designs highlighted in dark red in the figures represent the boundaries of the design space. 

\section{Heterogeneous shape change}

\subsection{The design procedure} 
\begin{figure}[!htb]
\centering
\includegraphics[scale=1]{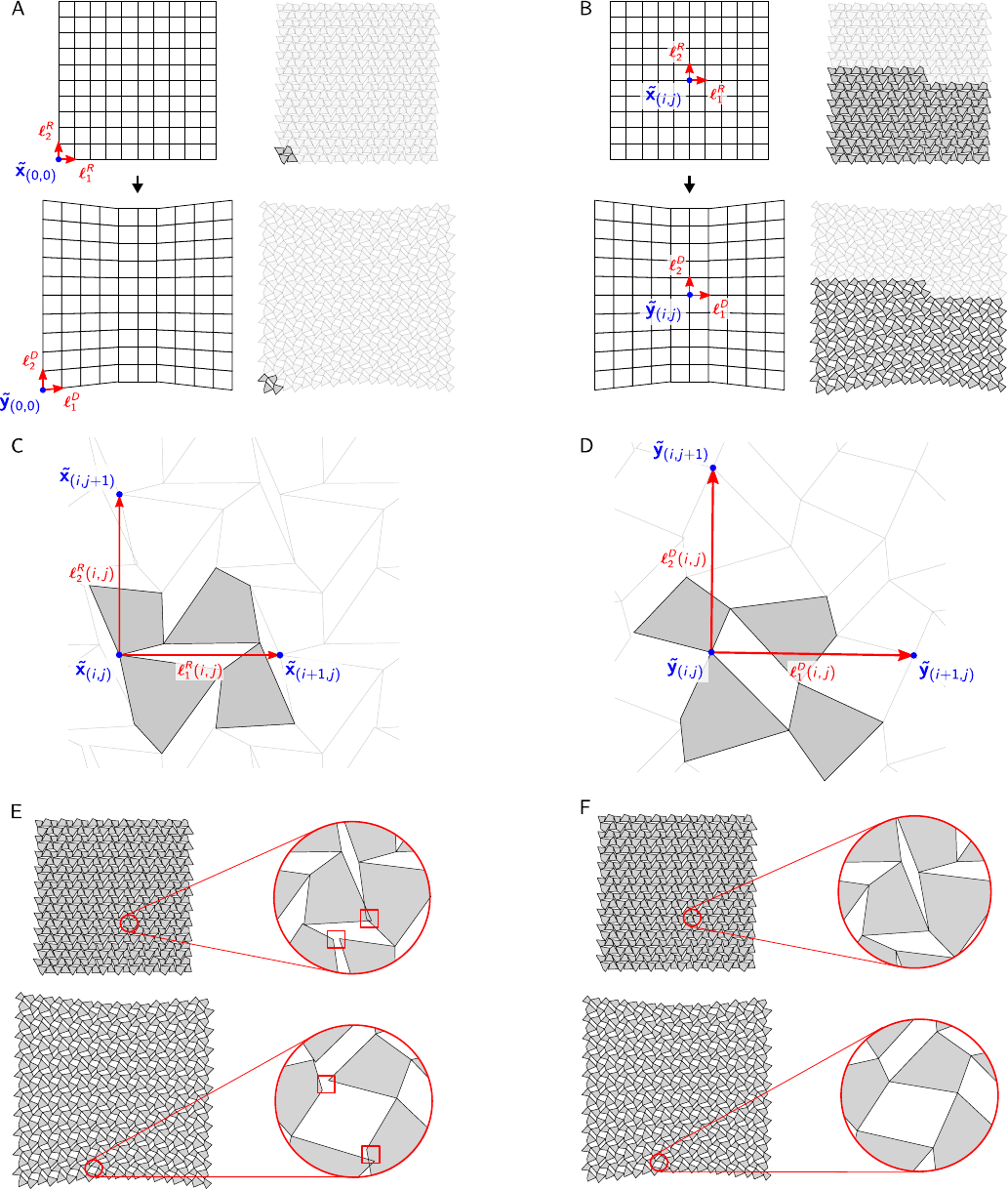}
\caption{Marching algorithm for bistable kirigami with target shape change. (A-B) Starting with bijective pair of  quad meshes, kirigami cells are filled in iteratively using the cell level  points and lattice vectors in  (C-D) and design variables from the previously computed cell. (E) Intercell incompatibilties during the marching process lead to gaps or overlaps between cells. (F) The final design is obtained by averaging out the incompatibilites.}
\label{fig:SM-Bowtie-ave}
\end{figure}

This section develops a numerical method to obtain bistable planar kirigami designs with specified heterogeneous shape change.  We use the example of a "square-to-bowtie" transformation in Fig.\;\ref{fig:SM-Bowtie-ave}  to illustrate the method. We also present the details in four distinct steps.  

\begin{itemize}
    \item \textbf{Step 1: Creating a quad mesh for the two stable states.} The initial step entails creating a one-to-one, regular quad mesh mapping from the reference state to the deformed state in 2D. The left part of Fig.\;\ref{fig:SM-Bowtie-ave}A shows an example of this type of  meshing for the square-to-bowtie transformation.  Such meshes can be achieved for a wide variety of target shapes. For a suitable meshing of the two states, we label the mesh points in the first state as $\Tilde{\mathbf{x}}(i,j)$ such that neighboring $(i,j)$ pairs correspond to neighboring $\Tilde{\mathbf{x}}(i,j)$ points. We also assume that the  mesh points of the second state, labeled $\Tilde{\mathbf{y}}(i,j)$, can be described by a sufficiently smooth mapping of the $\Tilde{\mathbf{x}}(i,j)$ points, i.e., that $\Tilde{\mathbf{y}}(i,j) = \boldsymbol{\varphi}(\Tilde{\mathbf{x}}(i,j))$ for all $(i,j)$ for some sufficiently smooth  2D vector field $\boldsymbol{\varphi}(\mathbf{x})$. These assumptions simply formalize what is more-or-less the intuitive way to relate the two  meshes bijectively.  We then define the lattice vectors as 
    \begin{equation}
    \begin{aligned}
    &\boldsymbol{\ell}_1^R(i,j)=\Tilde{\mathbf{x}}(i+1,j) - \Tilde{\mathbf{x}}(i,j), &&
        \boldsymbol{\ell}_2^R(i,j)=\Tilde{\mathbf{x}}(i,j+1) - \Tilde{\mathbf{x}}(i,j), \\
        &\boldsymbol{\ell}_1^D(i,j)=\Tilde{\mathbf{y}}(i+1,j) - \Tilde{\mathbf{x}}(i,j), && 
        \boldsymbol{\ell}_2^D(i,j)=\Tilde{\mathbf{y}}(i,j+1) - \Tilde{\mathbf{x}}(i,j).
    \end{aligned}
    \end{equation}
Fig\;\ref{fig:SM-Bowtie-ave}A-D highlight various aspects of the labeling of these meshes.

    \item \textbf{Step 2: Initializing the design.}  The second step involves seeding the kirigami design starting from an initial cell. In the example in Fig.\;\ref{fig:SM-Bowtie-ave}A, we seed the first cell from   the lower left corner quad  with the  Bravais lattice vectors $\overline{\boldsymbol{\ell}_i^R} = \boldsymbol{\ell}_{i}^R(0,0)$ and $\overline{\boldsymbol{\ell}_i^D} = \boldsymbol{\ell}_{i}^D(0,0)$, $i =1,2$. To do so, we  solve the minimization problem in Eq.\;[\ref{eq:generalSetup}], wherein we typically use the objective function  $f_{\text{obj}}(\mathbf{s}_1, \mathbf{s}_2, \boldsymbol{\phi}) =-|E_b(\mathbf{s}_1, \mathbf{s}_2, \boldsymbol{\phi})|^2$ with the goal of producing an optimized cell design with a high energy barrier between the two states. (The objective function  can be defined differently if needed, to help facilitate the design of a particular set of target shapes.) Once the optimization is solved, we use the optimized variables $\mathbf{s}_1(0,0)$, $\mathbf{s}_2(0,0)$, $\boldsymbol{\phi}(0,0)$  to construct the two stable states of the cell as follows: Let  $\mathbf{s}_3(0,0), \ldots, \mathbf{v}_4(0,0)$ denote the design vectors obtained from the  compatibility conditions in Eq.\;[\ref{eq:theorem111}] for the now given $\mathbf{s}_1(0,0)$, $\mathbf{s}_2(0,0)$, $\boldsymbol{\phi}(0,0)$, $\boldsymbol{\ell}_{1,2}^R(0,0)$ and $\boldsymbol{\ell}_{1,2}^D(0,0)$.  A direct  comparison of  the notation of a single cell in Fig.\;\ref{fig:designSup} and that of Fig.\;\ref{fig:SM-Bowtie-ave} furnishes a parameterization of the four corner points of the lower left panel in the  cell as  
    \begin{equation}
        \begin{aligned}
            &\underline{\text{in the first state:}} \\
            & \big\{ \tilde{\mathbf{x}}(0,0), \tilde{\mathbf{x}}(0,0) + \mathbf{v}_1(0,0), \tilde{\mathbf{x}}(0,0) + \mathbf{u}_1(0,0) , \tilde{\mathbf{x}}(0,0) + \mathbf{u}_1(0,0) - \mathbf{s}_1(0,0) \big\},  \\
            &\underline{\text{in the second state:}} \\ 
            &\big\{ \tilde{\mathbf{y}}(0,0), \tilde{\mathbf{y}}(0,0) + \mathbf{R}(\phi_1(0,0)) \mathbf{v}_1(0,0), \tilde{\mathbf{y}}(0,0)+ \mathbf{R}(\phi_1(0,0)) \mathbf{u}_1(0,0) , \tilde{\mathbf{y}}(0,0)+ \mathbf{R}(\phi_1(0,0))[ \mathbf{u}_1(0,0) - \mathbf{s}_1(0,0)]\big\}.
        \end{aligned}
    \end{equation}
    The other three panels of the cell are prescribed in an analogous fashion. The two stable states of the kirigami cell are emphasized in Fig.\;\ref{fig:SM-Bowtie-ave}A.

    \item \textbf{Step 3: Marching to obtain the overall design.}
    The next step involves  generating a complete preliminary pattern of the two target shapes by marching from cell-to-cell and solving a local optimization problem for each unit cell. In this procedure, we choose to minimize the difference in design variables between neighboring cells by introducing an objective function $f_{\text{obj}}(\mathbf{s}_1, \mathbf{s}_2, \boldsymbol{\phi}) = \big|(\mathbf{s}_1, \mathbf{s}_2, \boldsymbol{\phi}) - (\mathbf{s}^{\text{prev}}_1, \mathbf{s}^{\text{prev}}_2, \boldsymbol{\phi}^{\text{prev}})\big|^2$,  where $(\mathbf{s}^{\text{prev}}_1, \mathbf{s}^{\text{prev}}_2, \boldsymbol{\phi}^{\text{prev}})$ is the optimized result computed from a neighboring cell in the previous iteration. The goal is to produce cell designs whose parameters vary slowly from cell to cell.  In the square-to-bowtie example in Fig.\;\ref{fig:SM-Bowtie-ave}B, we march along the rows as indicated. At a given $(i,j)$ cell in the interior of the domain, we input the lattice vectors as $\overline{\boldsymbol{\ell}_k^R} = \boldsymbol{\ell}_{k}^R(i,j)$ and $\overline{\boldsymbol{\ell}_k^D} = \boldsymbol{\ell}_{k}^D(i,j)$, $k =1,2$, and set $(\mathbf{s}^{\text{prev}}_1, \mathbf{s}^{\text{prev}}_2, \boldsymbol{\phi}^{\text{prev}}) = (\mathbf{s}_1(i-1,j), \mathbf{s}_2(i-1,j), \boldsymbol{\phi}(i-1,j))$. (For a cell in the left most column, the input for $(\mathbf{s}^{\text{prev}}_1, \mathbf{s}^{\text{prev}}_2, \boldsymbol{\phi}^{\text{prev}})$ is the $(i,j-1)$ cell parameters.) The optimization then furnishes $\mathbf{s}_1(i,j), \mathbf{s}_2(i,j), \boldsymbol{\phi}(i,j)$, while Eq.\;[\ref{eq:generalSetup}] furnishes the remaining design variables $\mathbf{s}_3(i,j), \ldots, \mathbf{v}_4(i,j)$.  We construct the two stable states of the cell exactly as we did for the intialized one. Specifically,  we take the four corner points of the lower left panel in this cell as 
    \begin{equation}
        \begin{aligned}
            &\underline{\text{in the first state:}} \\
            & \big\{ \tilde{\mathbf{x}}(i,j), \tilde{\mathbf{x}}(i,j) + \mathbf{v}_1(i,j), \tilde{\mathbf{x}}(i,j) + \mathbf{u}_1(i,j) , \tilde{\mathbf{x}}(i,j) + \mathbf{u}_1(i,j) - \mathbf{s}_1(i,j) \big\},  \\
            &\underline{\text{in the stable state:}} \\ 
            &\big\{ \tilde{\mathbf{y}}(i,j), \tilde{\mathbf{y}}(i,j) + \mathbf{R}(\phi_1(i,j)) \mathbf{v}_1(i,j), \tilde{\mathbf{y}}(i,j)+ \mathbf{R}(\phi_1(i,j)) \mathbf{u}_1(i,j) , \tilde{\mathbf{y}}(i,j)+ \mathbf{R}(\phi_1(i,j))[ \mathbf{u}_1(i,j) - \mathbf{s}_1(i,j)]\big\},
        \end{aligned}
    \end{equation}
    and so-on for the other three panels. Iterating over the entirety of the quad meshes produces the two preliminary patterns that achieve the target shapes.

    \item \textbf{Step 4: Averaging gaps by gluing together.}
    Although we have identified  overall patterns with the desired shapes, the procedure is completely local and cell-based. As a result,  the panels within each cell are perfectly compatible, but there are  intercell incompatibilities in the form of gaps or overlaps between neighboring unit cells. Fig.\;\ref{fig:SM-Bowtie-ave}E shows examples of these incompatibilites for the square-to-bowtie transformation. The final step of the procedure is to average the intercell incompatibilities to produce the two planar kirigami states that approximate the target shape with a fully connected set of panels and slits. Fig.\;\ref{fig:SM-Bowtie-ave}F shows  the two state obtained after averaging for the square-to-bowtie transformation.

\end{itemize}

  Let's assume that the first state is the manufactured one and thus stress free. 
    Averaging the intercell incompatibilities in the last step leads to panels in the second state that are not necessarily rigid deformations of their counterparts in the first, and thus incur panel strain not present in the periodic setting of bistable designs. Even so,   certain heuristics suggest that these patterns should still be bistable in the typical case. Specifically,  our meshing and optimization is designed to produce variables $\boldsymbol{\ell}_{1,2}^R(i,j),\boldsymbol{\ell}_{1,2}^D(i,j),\mathbf{s}_1(i,j), \mathbf{s}_2(i,j),$ and $\boldsymbol{\phi}(i,j)$ that vary slowly from cell to cell. Our previous work homogenizing mechanism-based planar kirigami \cite{zheng2022continuum} shows that slow variations very similar to the type in this procedure lead quantitatively to average panel strains that scale as $O(l/L)$, where $L$ is the characteristic length of the total domain and $l$ is that of a single unit cell. Thus, for fine meshes with many slowly varying quad cells, we expect the strain in the second designed state to be very small (albeit not zero). This suggests that the second state, or something very nearby to it, should be stable.

It is also possible to numerically investigate whether or not the patterns obtained by this procedure are geometrically bistable through bar-hinge modeling. By way of illustration, we have performed such an investigation for the square-to-bowtie transformation in Fig.\;\ref{fig:SM-Bowtie-ave} and verified its bistabiltiy.  As anticipated, the  second stable state takes a bowtie shape and is  slightly internally stressed (see the main text for a more detailed discussion).   The bar-hinge model used for the verification is an "in-house" Matlab code formulated in  Section \ref{sec:NumSec}.

\begin{figure}[htb!]
\centering
\includegraphics[scale=0.7]{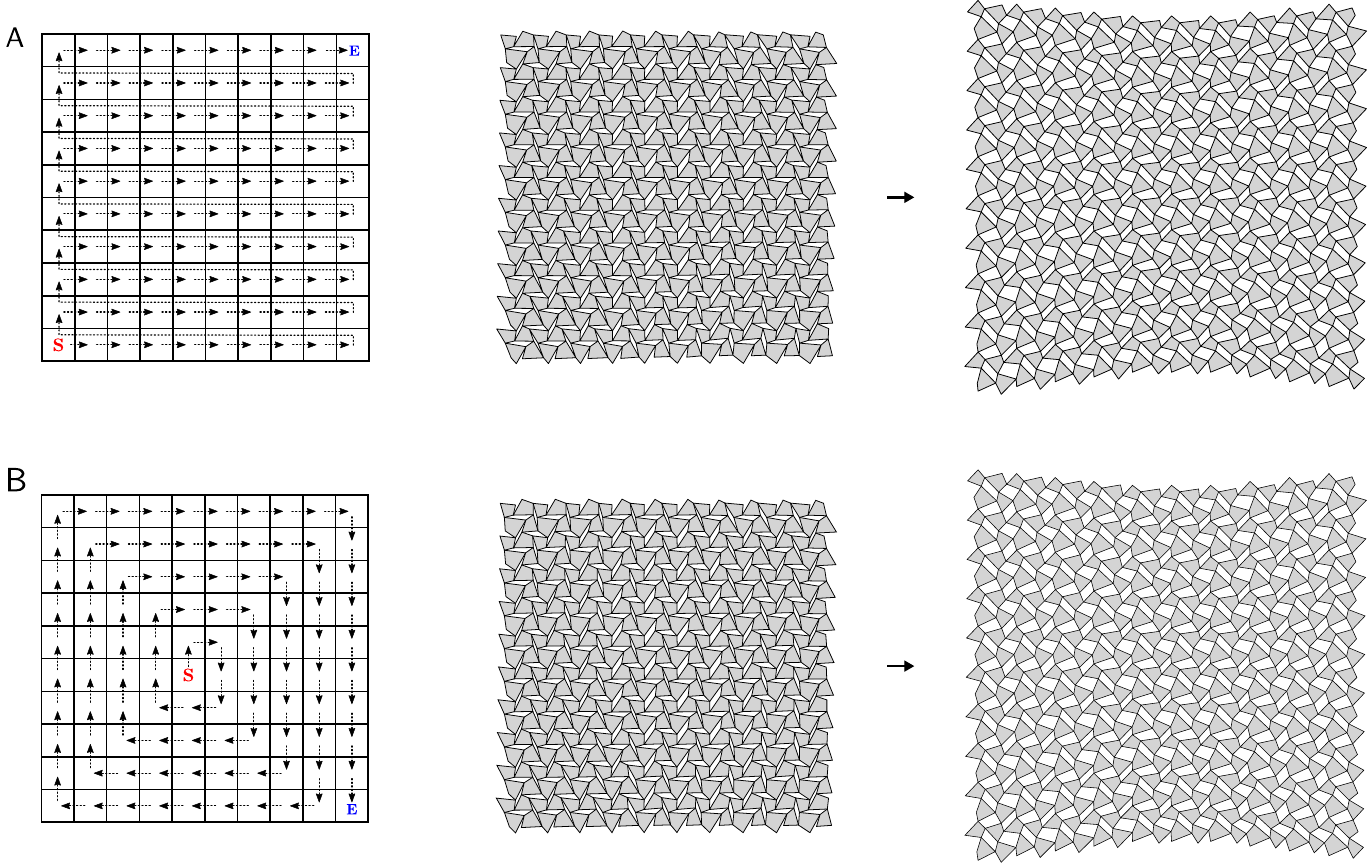}
\caption{Two marching algorithm procedures and the corresponding heterogeneous configuration results (Letters 'S' in red and 'E' in blue denote the starting and ending cell). (A) From the lower-left corner to the upper-right (algorithm shown in the previous section). (B) Spiral marching from the center towards the right-bottom corner.}
\label{fig:SM-heterogeneous marching algorithms}
\end{figure}

\subsection{Robustness to marching strategy}

{\color{black} This section discusses the robustness of examples of heterogeneous shape change to the choice of  marching  strategy. One strategy entails starting from lower-left unit cell and marching in a zig-zag fashion as  depicted in Fig.\;\ref{fig:SM-heterogeneous marching algorithms}-A. Another  involves initializing the algorithm from the central cell  and marching by spiraling out as shown in Fig.\;\ref{fig:SM-heterogeneous marching algorithms}-B. In the case of the square-to-bowtie transformation, we have investigated both ways of marching, keeping all other details of the procedure fixed. The right part of Fig.\;\ref{fig:SM-heterogeneous marching algorithms} shows the examples that emerge  from these two ways marching. The design variables do change slightly from cell-to-cell, as indicated by the table below. However, the overall designs are qualitatively quite similar to each other. This result suggests that the marching algorithm is robust to small changes in the way the algorithm is initialized. In other words, the design space for heterogeneous shape change seems to be highly constrained rather than highly degenerate. 

\begin{center}
\begin{tabular}{|c | c|} 
 \hline
 Design variables & $\underset{\{i,j\}} {\max}\frac{|\mathbf{x}^{A}(i,j)-\mathbf{x}^{B}(i,j)|}{|\mathbf{x}^{A}(i,j)|}$ \\ [1ex] 
 \hline
$\mathbf{x}=[\mathbf{s}_1,\mathbf{s}_2,\boldsymbol{\phi}]$ &   $1.057\%$\\
 \hline
\end{tabular}
\end{center}

}

% To be specific, we seed the first cell from the middle quad indicated by the red letter 'S' with the  Bravais lattice vectors $\overline{\boldsymbol{\ell}_i^R} = \boldsymbol{\ell}_{i}^R(4,4)$ and $\overline{\boldsymbol{\ell}_i^D} = \boldsymbol{\ell}_{i}^D(4,4)$, $i =1,2$ in the second marching algorithm. We extract the solution to the design variables for the central cell obtained by the first marching algorithm and assign it as the initial point of the second algorithm.  Then, we solve the minimization problem in Eq.\;[\ref{eq:generalSetup}] to maximize the energy barrier between two states, which is the same as the original marching algorithm. Subsequently, we generate the whole design by marching from cell-to-cell and solving a local optimization problem for each unit cell. By introducing the same objective function $f_{\text{obj}}(\mathbf{s}_1, \mathbf{s}_2, \boldsymbol{\phi}) = \big|(\mathbf{s}_1, \mathbf{s}_2, \boldsymbol{\phi}) - (\mathbf{s}^{\text{prev}}_1, \mathbf{s}^{\text{prev}}_2, \boldsymbol{\phi}^{\text{prev}})\big|^2$,  where $(\mathbf{s}^{\text{prev}}_1, \mathbf{s}^{\text{prev}}_2, \boldsymbol{\phi}^{\text{prev}})$ is the optimized result from the previous cell, we obtain the structures with the different marching approach. The input for $(\mathbf{s}^{\text{prev}}_1, \mathbf{s}^{\text{prev}}_2, \boldsymbol{\phi}^{\text{prev}})$ is the parameters from the previous cell in the spiral sequential order.

\subsection{Supporting information on examples}

This section provides additional details about two complex heterogeneous examples of shape change, namely, the
”beating heart” pattern and the square-to-disc pattern from the main text.

\begin{itemize}

\begin{figure}[htb!]
\centering
\includegraphics[scale=1.0]{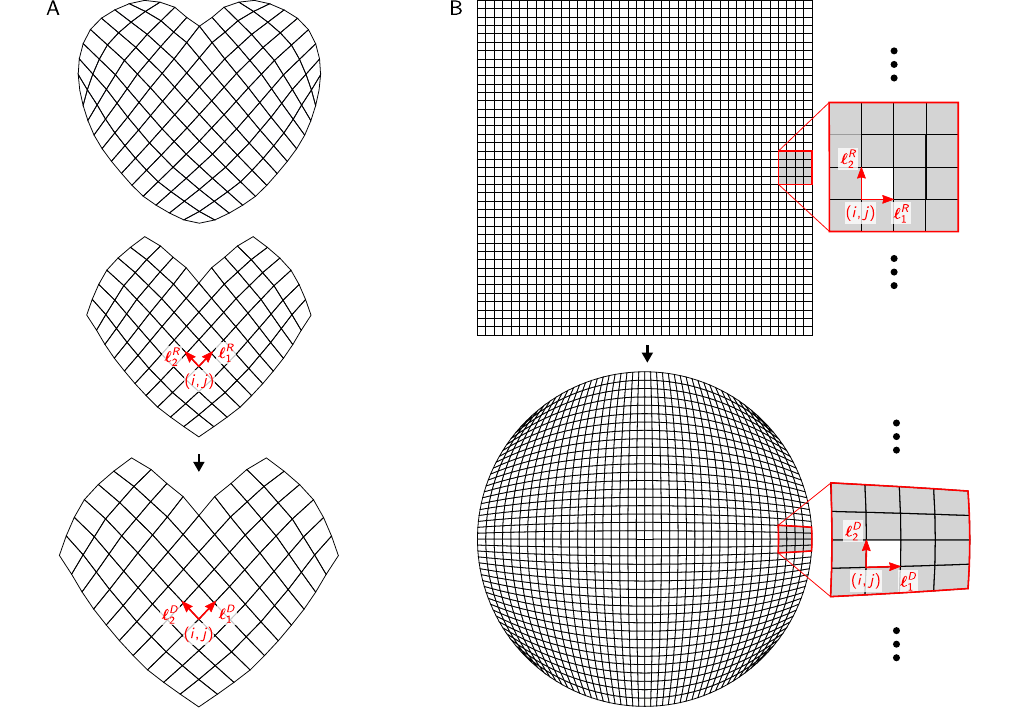}
\caption{Heterogeneous quad mapping for two examples presented in the main text. (A) From top to bottom, original mesh grid, tailored mesh grid of the first stable state and of the second stable state of the "beating heart" pattern. (B) Mesh grids of the two stable states of the "square to disc" pattern.}
\label{fig:SM-heterogeneous mapping}
\end{figure}

    \item  \textbf{Beating heart pattern.} 
    To simulate the expansion of the heart, we first draw out the boundary of the heart design  via  the 2D implicit function $(x^2 + y^2 - 1)^3 - x^2y^3 = 0$. We then mesh this domain via the `Spline' option  in the meshing software  Gmsh (version 4.11.1), selecting  in  the `Tools' option the `Quasi-Structured Quad (experiment)' 2D algorithm since it exclusively meshes the shape using quadrilateral elements. By tuning the `Element size factor' and `Subdivision algorithm' in the options, we can generate meshes with different levels of refinement and number of elements. In some cases, the meshes obtained may not be suitable for the marching algorithm to create an optimal design because they are not well-arranged for further iterative optimization. Hence, we generate a feasible mesh by utilizing the `Refine by splitting' method in Gmsh. Applying this approach ensures that the quadrilateral meshes are well-organized in rows and columns, thus providing each unit cell with specific neighboring cells that facilitate the implementation of our marching algorithm. Fig.\;\ref{fig:SM-heterogeneous mapping}(top) displays a feasible mesh generated in this fashion, while Fig.\;\ref{fig:SM-heterogeneous mapping}(middle) illustrates the mesh used for the first stable state of the heart in the main text. The latter  mesh  is modified from the auto-generated former one  to  exclude the outer layer, as   this layer has  significant variations in the neighboring lattice vectors and proved difficult for our marching algorithm to handle. The second stable state is obtained by dilating the modified mesh by a factor of $1.25$, as shown in Fig.\;\ref{fig:SM-heterogeneous mapping}(bottom).  Having generated the quad mesh for the two stable states, we  proceed with the same procedure as the square-to-bowtie transformation, thereby obtaining the whole design for the beating heart presented in the main text.
    
    \item  \textbf{Square-to-disc pattern.} The Elliptical grid mapping $(x,y) \mapsto (x\sqrt{1 - y^2/2},y\sqrt{1-x^2/2})$ is one type of transformation that can square the disc. We use this mapping to generate a one-to-one quad mesh between the two states and  the corresponding lattice vectors necessary for the marching algorithm. Specifically, we  create the desired mesh for the square reference using a $40 \times 40$ set of square unit cells uniformly distributed on $(-1,1)^2$. We then apply the elliptical grid mapping to this mesh to obtain the second stable state of quad cells. Fig.\;\ref{fig:SM-heterogeneous mapping}-B presents the one-to-one quad mesh mapping between the first and second stable states. The elements in the center experience the least distortion, whereas the quads near the four corners of the square collapse the most, resulting in the formation of four circular arcs in the deformed plane. The mapping smoothly distributes the mesh distortions from the interior to the exterior layers. The lattice vectors are represented by the adjacent sides in each quadrilateral. The full kirigami design is generated from these lattice vectors using the marching algorithm in approximately 50 seconds on a standard laptop.

\end{itemize}

\section{Fabrication}

In this section, we provide additional details on the design and fabrication of all types of specimens used in this work, from the monolithic and bi-material ones, designed for in-plane morphing, to the out-of-plane morphing mono-material ones.

\subsection{Monolithic kirigami}

Planar kirigami found in the literature are typically monolithic, and feature panels connected via elastic hinges. A common way to fabricate such metamaterials is by laser cutting rubber. While this strategy works well for metamaterials featuring low energy modes of deformation, it does not work well for bistable metamaterials; in fact, there is a limit to how thick of a rubber sheet one can laser cut, and thin sheets can lead to low out-of-plane stiffness and, potentially, to loss of bistability. Thus, we adopt a strategy previously employed by Wu and Pasini~\cite{wu2023situ}, where thick monolithic metamaterials are 3D printed out of thermoplastic polyurethane (TPU). In particular, we use a TPU-95A filament on an Ultimaker 2+ printer, \textcolor{black}{since this is the softest material of the Ultimaker material palette}. The steps we follow to design and fabricate these specimens are reported below. We let $\ell_1^R = |\boldsymbol{\ell}_1^R|$ and $\ell_2^R= |\boldsymbol{\ell}_2^R|$ characterize the lengths of the physical specimens.
\begin{figure}[!htb]
\centering
\includegraphics[scale=1.0]{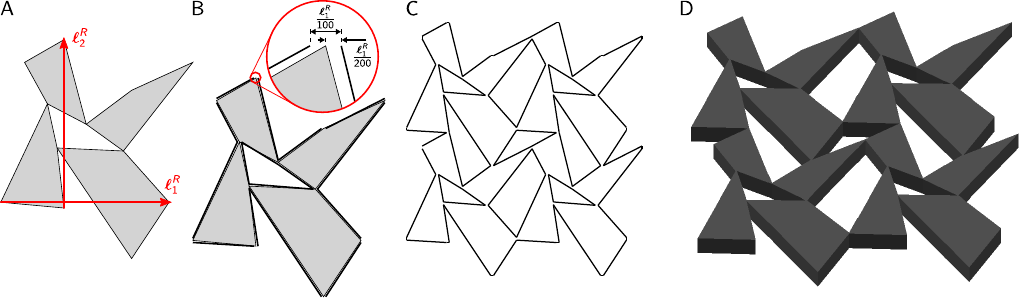}
\caption{Design workflow for monolithic specimens. (A) Unit cell from the theoretical design process, with its lattice vectors. (B) Superposition of the theoretical unit cell with a modified unit cell, where we introduce a gap at each corner, as shown in the detail. (C) Assembly of four cells as to create the outline of one of our specimens. (D) CAD drawing of the specimen to be printed.}
\label{fig:fabrication_3}
\end{figure}
\begin{itemize}
    \item \textbf{Step 1}: We import the coordinates of the vertices of each panel of a unit cell (Fig.~\ref{fig:fabrication_3}A) in Matlab, and scale it to the desired dimensions ($\ell_1^R=\ell_2^R=7$\,cm for all specimens). In order to introduce elastic connectors between panels, we duplicate each corner node of each panel, and set each couple of nodes apart by a desired distance (Fig.~\ref{fig:fabrication_3}B). In all our samples, we set this distance to $\ell_1^R/100$. We then export the geometry outline of four unit cells (Fig.~\ref{fig:fabrication_3}C).
    \item \textbf{Step 2}: We import the geometry outline in AutoCAD and extrude the pattern, setting an out-of-plane thickness of 1~cm (Fig.~\ref{fig:fabrication_3}D). The whole structure is exported as a STL file.
    \item \textbf{Step 3}: We import the file into UltiMaker Cura. We print using a 0.4\,mm nozzle and set up the following printing parameters: i) a layer height of 0.1 mm; ii) a wall thickness of 0.8 mm; iii) a top and bottom thickness of 1.2 mm; iv) 100\% infill density; v) a ``lines'' infill pattern; and vi) Ultimaker-preset values for printing temperature, build plate temperature, and print speed. Once printed, the specimens are ready to be tested and do not require additional post-processing.
\end{itemize}

\subsection{Bi-material kirigami} \label{s:bimatfab}

To avoid dealing with hinge energy, which can prevent patterns from exhibiting bistability, we resort to a design for kirigami where panels are connected via actual pin joints. To try and create specimens that deform in plane, we \textcolor{black}{consider relatively-thick specimens where each panel is skeletal in nature}, i.e., we replace each panel by an assembly of eight bars and five hinges. To make sure that deformations are not localized at the hinges \textcolor{black}{and to avoid excessive dissipation caused by pins rotating against soft materials}, we fabricate hinge regions out of stiff Nylon and bars out of softer TPU-95A. Each panel is fabricated in a single shot, leveraging the dual-extrusion nature of our Ultimaker 3 printer. The steps we follow to design and fabricate these bi-material specimens are reported below.
\begin{figure}[!htb]
\centering
\includegraphics[scale=1.0]{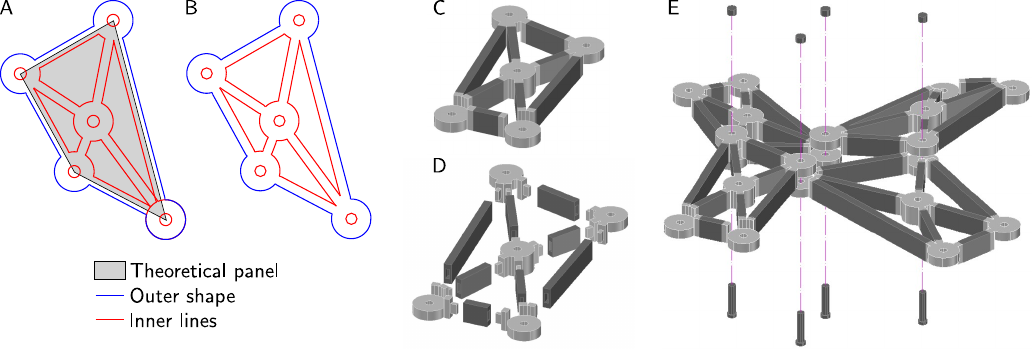}
\caption{Design workflow for bi-material specimens. (A) Outline of a theoretically-designed panel (gray quadrilateral) and Matlab-generated ``rough'' outline of the skeletal version of the same panel (blue and red lines); note the presence of unwanted red lines at the bottom-right hinge. (B) Actual panel outline after clean up in Inkscape. (C) Bi-material panel featuring nylon hinges and TPU bars, as created in AutoCAD. Note that the height of the hinges is half of the bars' height, so that the structure has a constant thickness when assembled. (D) Exploded view of the panel, showing how the hinges protrude into the bars to improve adhesion. (E) Assembly of a unit cell from four panels with pins and caps.}
\label{fig:fabrication_1}
\end{figure}
\begin{itemize}
    \item \textbf{Step 1}:  We import the coordinates of the vertices of each panel of a unit cell (Fig.~\ref{fig:fabrication_3}A) in Matlab, and draw the outline of a skeletal version of each panel, which features hinge regions (outer radius = 5 mm, inner radius = 1.5 mm) at each vertex location and at the center of the panel, and bars (width = 2.4 mm) connecting these hinges (Fig.~\ref{fig:fabrication_1}A). At this stage, we do not yet worry about the fact that our algorithms do not produce a continuous outline of each panel. 
    \item \textbf{Step 2}: We manipulate these files in Inkscape to remove some overlapping paths and create an actual panel outline (Fig.~\ref{fig:fabrication_1}B), and scale the geometry to the desired experimental size ($\ell_1^R=\ell_2^R=10$\,cm). 
    \item \textbf{Step 3:} In AutoCAD, we use the \verb+presspull+ command, starting from the panel outlines, to extrude the bars and hinges to a height of 6.35 mm while maintaining height clearance at the hinges where each panel connects with another panel (Fig.~\ref{fig:fabrication_1}C). We also extrude rectangular sections on the sides of the hinges and create subsequent cutouts at the ends of the bars to accommodate for the protrusion, as illustrated in the exploded view of Fig.~\ref{fig:fabrication_1}D; this is done to improve bonding between the two material phases. Once the 3D model is ready, we export hinges and bars as separate STL files.
    \item \textbf{Step 4}: We import these files into UltiMaker Cura, where the hinges and bars are automatically matched up, and set the following printing parameters: i) a layer height of 0.15 mm, the smallest recommended value for TPU; ii) a wall thickness of 0.4 mm; iii) zero thickness for top and bottom layers, since delamination between these top layers and the infill layers occurs when these layers have a non-zero value; iv) 100\% infill density; v) a ``zigzag'' infill pattern, for its ability to fill the most in between the outline of the walls of the bars; and vi) Ultimaker-preset values for printing temperature, build plate temperature, as well as print speed.
    \item \textbf{Step 5}: Once all the panels are 3D printed, we clear them of any filament residue and assemble four of them into one unit cell using metallic captive pins (McMaster-Carr part number: 95648A350) and caps 3D printed out of PLA (Fig.~\ref{fig:fabrication_1}E). When all four unit cells are assembled, we connect them together to form a 2x2 structure.
\end{itemize}

\subsection{Mono-material kirigami}
Our final physical incarnation of the theoretically-designed bistable patterns features thin mono-material panels laser cut out of PETG and assembled with push-in rivets. \textcolor{black}{We choose PETG since it provides the right amount of compliance with respect to more classical laser cuttable polymers like acrylic; the thickness is chosen to be small to avoid high stresses that could lead to breaking of the panels near the relatively-weak hinge regions.} These specimens are designed to bend out of plane as they transition between stable states, and the steps we follow to design and fabricate them are reported below.
\begin{figure}[!htb]
\centering
\includegraphics[scale=1.0]{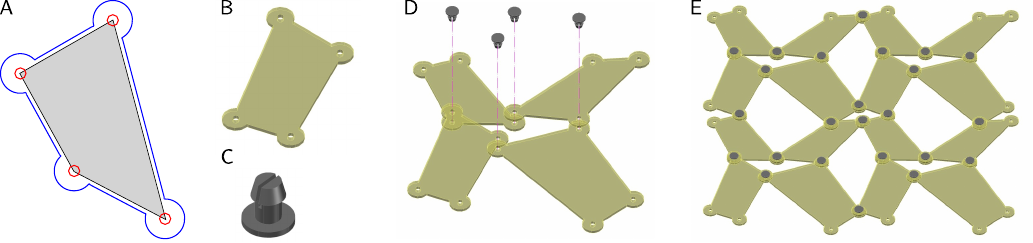}
\caption{Design workflow for mono-material specimens. (A) Outline of a theoretically-designed panel and Matlab-generated outline of the pin-jointed version of the same panel. Illustration of (B) one of the panels and (C) one of the push-in rivets. (D) How four panels are assembled in a unit cell via rivets; note that panels are arranged on two layers only. (E) Assembled 2x2 structure.}
\label{fig:fabrication_2}
\end{figure}
\begin{itemize}
    \item \textbf{Step 1}: Same as Step 1 for the bi-material specimens but, in this case, the Matlab script generates ready-to-cut panel outlines (Fig.~\ref{fig:fabrication_2}A). 
    \item \textbf{Step 2}: In Inkscape, we replace the inner circles of all hinges -- a step that is necessary since the circles created in Matlab are not actual circular paths but assemblies of short straight lines.
    \item \textbf{Step 3}: We use these drawings to laser cut the individual panels (Fig.\;\ref{fig:fabrication_2}B) out of a 1.3\,mm-thick PETG sheet (McMaster-Carr part number: 9513K92). We set our 80W laser cutter (Epilog FusionPro 30) to 30\% power, 10\% speed and 100\% frequency.
    \item \textbf{Step 4}: When all the panels of a unit cell are laser cut and ready, we assemble them using push-in rivets (Fig.\;\ref{fig:fabrication_2}C, McMaster-Carr part number: 90136A630) as illustrated in Fig.\;\ref{fig:fabrication_2}D. Key to this assembly process is making sure that cells are assembled on two-layers only. Once all four unit cells are assembled, we connect them together using the push-in rivets to form a 2x2 structure as shown in Fig.\;\ref{fig:fabrication_2}E.
\end{itemize}

\section{Experimental setup and procedures}
All of our quantitative experiments (used to obtain force-displacement curves and demonstrate bistability) are carried out using an Instron 68FM-100 Universal Testing System (UTS) equipped with a 10\,kN load cell and by actuating the specimens in tension, via custom fixtures, at a rate of 1\,mm/s. In this section, we provide details on the custom fixtures, which have distinct features depending on the physical incarnation of kirigami structures they are used for. In all cases, we choose to apply boundary conditions and loading at hinge locations; specifically, we always anchor the specimen at the leftmost hinge (connecting the top-left and bottom-left unit cells), and always pull on the specimen from the rightmost hinge (connecting the top-right and bottom-right unit cells).

\subsection{Monolithic kirigami}
\begin{figure}[!htb]
\centering
\includegraphics[scale=1.0]{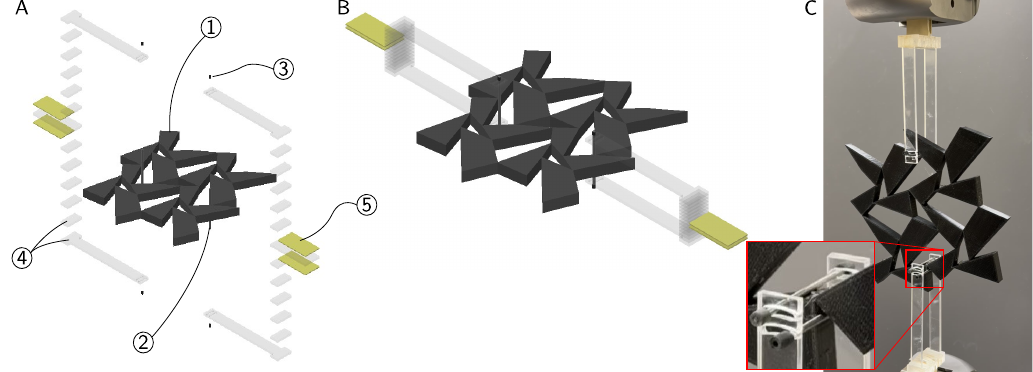}
\caption{Experimental setup for monolithic kirigami. (A) Exploded view of the specimen and of the parts that make up the fixtures. Specific parts include: (1) specimen; (2) metal dowel bars; (3) rubber stoppers to keep the bars and specimen in place; (4) laser-cut acrylic parts that make up the fixtures to hold the dowel bars and connect them to the UTS; and (5) natural rubber pads. (B) Experimental setup after the fixtures are assembled. (C) Photo of the specimen and fixtures, mounted on the UTS, with detail of how the hinge is loosely gripped by two free-to-translate dowel bars.}
\label{fig:ExpSetup_3}
\end{figure}

The experimental setup for the monolithic structure is constructed by assembling the components shown in Fig.\;\ref{fig:ExpSetup_3}A,B. (1): 3D printed 2x2 structure; (2): 50 mm long metallic dowel bar (McMaster-Carr part number: 91595A353); (3): small pieces of rubber tube (McMaster-Carr part number: 9776T26) (4): anchoring and actuation fixtures built from 3 mm thick acrylic laser cut pieces; (5): 1.5 mm thick rubber pads.
Laser cut pieces are used to construct C-shaped fixtures, which are connected to the specimen via dowel bars. Rubber stoppers are used to keep the dowel bars in place while allowing them to rotate freely. The fixtures are connected to the UTS grippers via rubber pads, to improve contact and avoid slippage.

Since our specimens feature elastic hinges, fixtures are designed to loosely grip these hinges, while trying not to obstruct the natural deformation of each pattern. For this reason, we design a gripper such that the hinge is sandwiched between two dowel bars. The metallic bars' motion is guided by slots located on the T-shaped laser cut pieces, as shown in the detail of Fig.\;\ref{fig:ExpSetup_3}C, which determine their minimum and maximum relative distance; this detail is necessary since the shape of the hinge region, and the fact that this shape changes drastically during deformation, makes it challenging to design fixed grippers that don't obstruct the natural deformation of the patters. As an unwanted consequence, our grippers provide complex boundary conditions that are difficult to simulate, as discussed Section~\ref{sec:NumSec}. Additionally, since these rods will move differently for all specimens due to large shape differences near the hinge regions, they will effectively provide different boundary conditions for each specimen. Thus, the comparison between energies computed from experiments on different samples should be taken with a grain of salt.

\subsection{Bi-material kirigami}
\begin{figure}[!htb]
\centering
\includegraphics[scale=1.0]{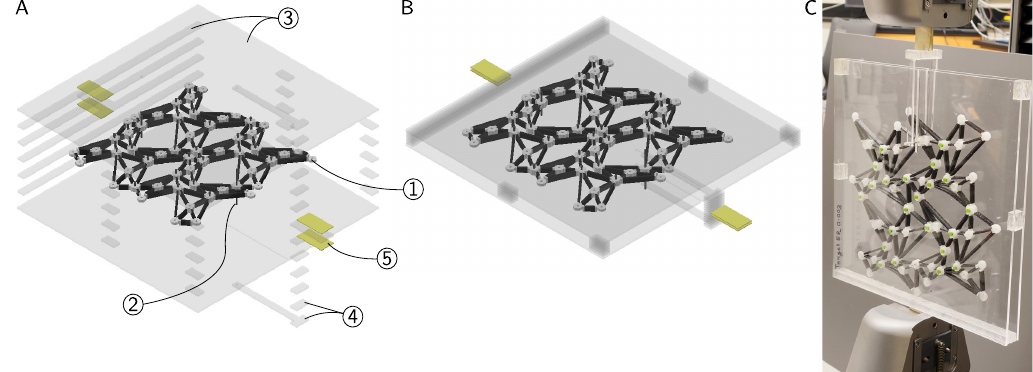}
\caption{Experimental setup for bi-material kirigami. (A) Exploded view of the specimen and of the parts that make up the fixtures. Specific parts include: (1) specimen; (2) metal dowel bars; (3) laser-cut acrylic parts and confining plates to anchor the specimen to the bottom gripper of the UTS, and to prevent out-of-plane deformation; (4) laser-cut acrylic parts that make up the actuation fixture, then connected to the top gripper of the UTS; and (5) natural rubber pads. (B) Experimental setup after the fixtures are assembled. (C) Photo of the specimen and fixtures, mounted on the UTS.}
\label{fig:ExpSetup_1}
\end{figure}

The experimental setup for the bi-material specimens is simpler than the previous one. In fact, specimens with pin-jointed panels can be actuated by inserting actuating rods in place of selected pins. This specific setup is built by assembling the components shown in Fig.~\ref{fig:ExpSetup_1}A,B. (1): assembled and pinned 2x2 structure; (2): 30\,mm- and 50\,mm-long, 2\,mm-diameter metallic dowel bars (McMaster-Carr part number: 91595A043 and 91595A353); (3): acrylic confining plates (292.5 mm x 292.5 mm) and other acrylic parts for anchoring; (4): actuation fixture built from various 3 mm thick acrylic laser cut pieces; (5): 1.5 mm thick rubber pads.
Laser cut pieces are used to construct a C-shaped actuation fixture, connected to the specimen via a dowel bar and secured by rubber stoppers, and an anchoring fixture which also prevents out-of-plane displacements. The anchoring fixture features two confining acrylic plates; these plates have 2.5\,mm-diameter holes to keep the anchoring dowel pin in place, and 2.2\,mm-wide slots to allow the actuating fixture to move along a straight line. These fixtures, connected to the UTS grippers via rubber pads, subject the specimen to a pin boundary condition at one end and to a roller boundary condition at the other end.

\subsection{Mono-material kirigami}
\begin{figure}[!htb]
\centering
\includegraphics[scale=1.0]{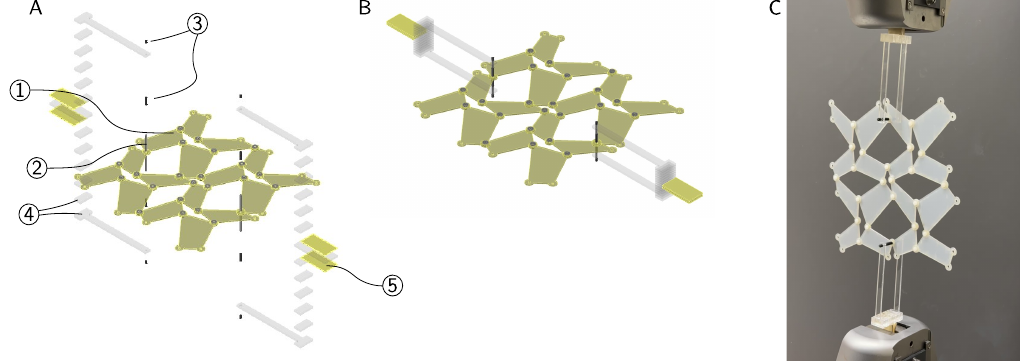}
\caption{Experimental setup for mono-material kirigami. (A) Exploded view of the specimen and of the parts that make up the fixtures. Specific parts include: (1) specimen; (2) metal dowel bars; (3) rubber stoppers to keep the bars and specimen in place; (4) laser-cut acrylic parts that make up the fixtures to hold the bars and connect to the UTS; and (5) natural rubber pads. (B) Experimental setup after the fixtures are assembled. (C) Photo of the specimen and fixtures, mounted on the UTS.}
\label{fig:ExpSetup_2}
\end{figure}

The experimental setup for the mono-material structures is very similar to the one for the bi-material ones, and is constructed by assembling the components shown in Fig.\;\ref{fig:ExpSetup_2}A,B: (1): assembled and riveted 2x2 structure; (2): 50 mm long metallic dowel bars; (3): small pieces of rubber tube (McMaster-Carr part number: 9776T26); (4): anchoring and actuation fixtures built from 3 mm thick acrylic laser cut pieces; (5): 1.5 mm thick rubber pads.
Here, the only difference with the bi-material case is that we use laser cut pieces to construct identical actuating and anchoring fixtures, and that we do not use constraining plates but rather embrace out-of-plane deformations. The rubber stoppers allow some out-of-plane rotation of the panels at the anchoring points, as discussed more in detail in Section~\ref{sec:NumSec}.

\begin{figure}[!htb]
\centering
\includegraphics[height=0.91\textheight]{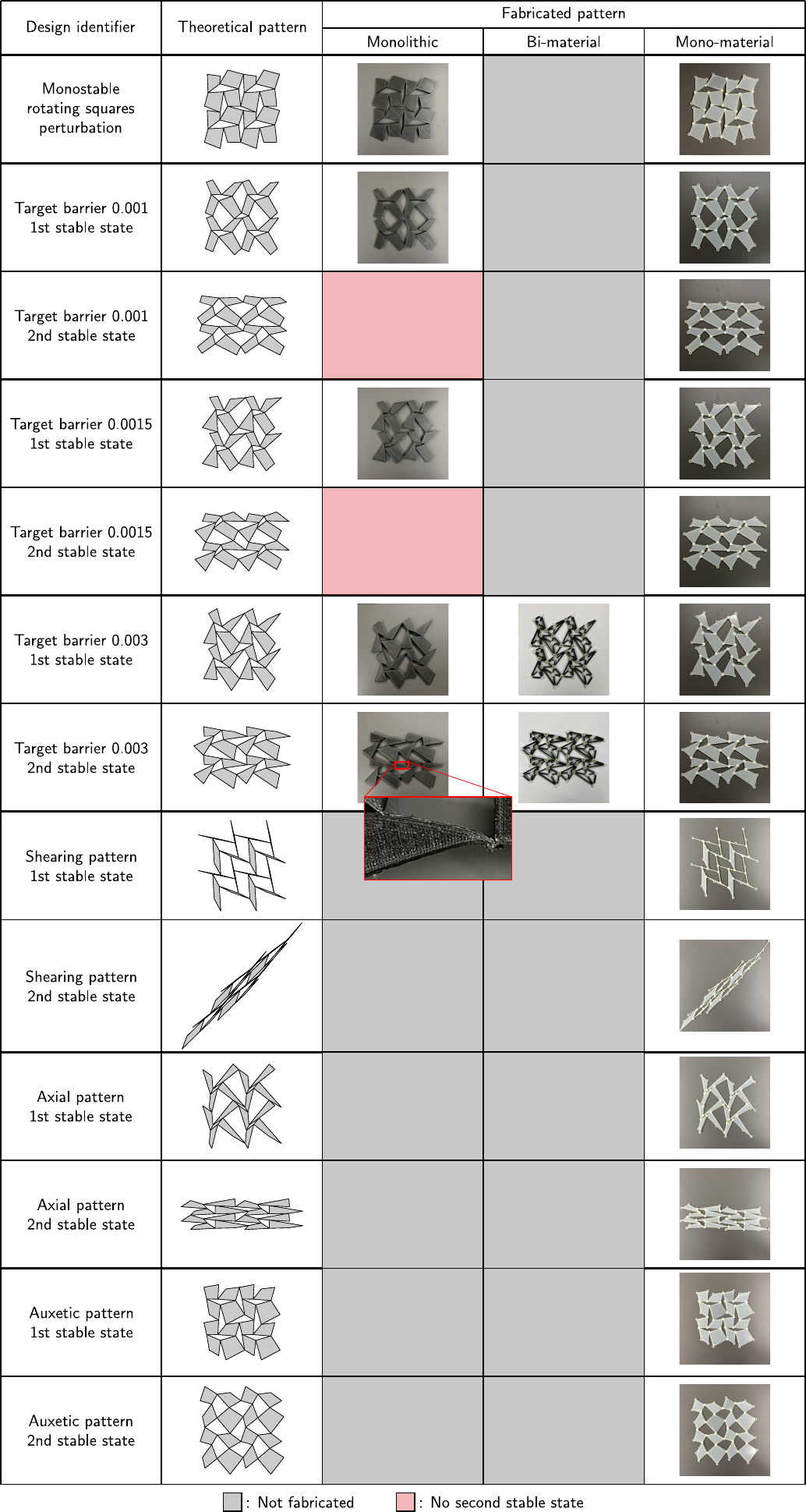}
\caption{Summary of the first and second stable states (if present) of all specimens fabricated in this study. Some of these images are not shown in the main text.}
\label{fig:ExpAdd}
\end{figure}

\section{Additional images and videos of experimental results}
Here, we show additional images of our specimens in their first and second (if present) stable states. We feel that this is needed to emphasize how, in those cases where our specimens do display a second stable state, the morphology of this second stable state closely resembles or is even identical the one predicted by theory. These images are displayed in Fig.~\ref{fig:ExpAdd}.

Monostable specimens do not display a second stable state regardless of their physical realization. As discussed in the main text, monolithic specimens are not always bistable. In fact, only the one with target energy barrier 0.003 displays clear bistability; in the other cases we fabricated (target energies 0.001 and 0.0015), the hinge energies are too high and prevent the specimens from displaying a second stable state. In the monolithic case that displays a second stable state, we can see that the morphology of the second stable state is very close to the theoretical one but not identical. In particular, one can notice that the most slender panels display some bending (as shown in the inset), thus indicating that this state is not stress-free. 

To provide further information on the behavior of all specimens, we report videos of their actuation. A brief description of each video is given in the following:
\begin{itemize}
    \item \verb+ActuationVideo_Monolithic_0d001.MOV+ : Video of the monolithic, 0.001 barrier specimen. The specimen is clearly not bistable.
    \item \verb+ActuationVideo_Monolithic_0d0015.MOV+ : Video of the monolithic, 0.0015 barrier specimen. The specimen seems to display some bistability initially but then returns to its original state.
    \item \verb+ActuationVideo_Monolithic_0d003.MOV+ : Video of the monolithic, 0.003 barrier specimen. This specimen shows clear and "long term" bistability.
    \item \verb+ActuationVideo_Bimaterial_0d003.MOV+ : Video of the bi-material, 0.003 barrier specimen. Note how the beams buckle in plane as the specimen is actuated, and how the specimen is clearly stable in its second state.
    \item \verb+ActuationVideo_Monomaterial_Monostable.MOV+ : Video of the mono-material, monostable specimen. When actuated, the specimen deforms out of plane and even buckles, but does not display a stress free second stable state.
    \item \verb+ActuationVideo_Monomaterial_0d001.MOV+ : Video of the mono-material, 0.001 barrier specimen. The specimen is clearly bistable.
    \item \verb+ActuationVideo_Monomaterial_0d0015.MOV+ : Video of the mono-material, 0.0015 barrier specimen. The specimen is clearly bistable.
    \item \verb+ActuationVideo_Monomaterial_0d003.MOV+ : Video of the mono-material, 0.003 barrier specimen. The specimen is clearly bistable.
    \item \verb+ActuationVideo_Monomaterial_ExtremeShear.MOV+ : Video of a mono-material specimen designed to undergo an extreme shear-like shape transformation. While the specimen is clearly bistable, since the panels bend between stress-free stable states, the energy barrier is not very significant, as is the case for most extreme designs.
    \item \verb+ActuationVideo_Monomaterial_ExtremeRectangle.MOV+ : Video of a mono-material specimen designed to undergo an extreme square-to-rectangle shape transformation. The specimen is clearly bistable.
    \item \verb+ActuationVideo_Monomaterial_ExtremeAuxetic.MOV+ : Video of a mono-material specimen designed to undergo an extreme auxetic shape transformation. This specimen is bistable but challenging to actuate, and requires adjusting the position of one of the corner panels to avoid self-intersection.
\end{itemize}

\section{Numerical simulations: additional results and details}\label{sec:NumSec}
In this section, we provide numerical insight into some of our experimental results via high-fidelity finite element simulations (using the software Abaqus). By showcasing the challenges encountered in modeling the nuanced behavior of our fabricated structures with high fidelity, we believe that we strengthen the case for the use of simplified energy metrics at the early stages of design. In this section, we also provide details on the reduced-order truss models used to validate the bistability of the heterogeneous patterns.

\subsection{Numerical model for monolithic specimens}
\begin{figure}[!htb]
\centering
\includegraphics[scale=1.0]{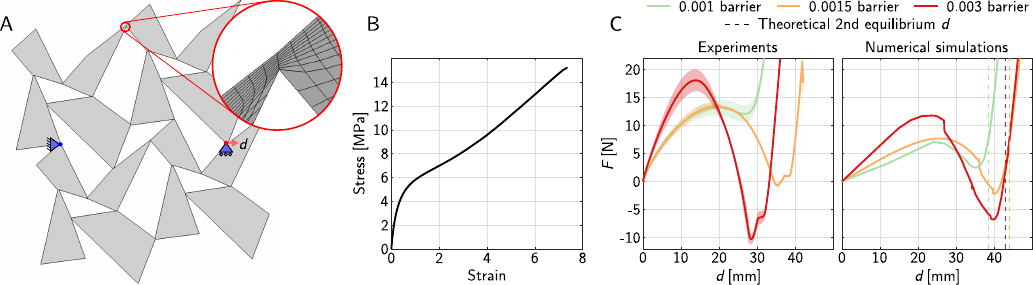}
\caption{Numerical modeling of monolithic specimens. (A) One of the modeled structures, with details on the applied boundary conditions and with a zoom-in of the mesh at one of the elastic hinges. (B) Measured stress-strain curve of a dogbone specimen 3D printed out of TPU-95, which we use to calibrate the nonlinear material model in Abaqus. (C) Comparison between experimental and numerical results in terms of force-displacement curves, for specimens with three target energy barriers.}
\label{fig:Num_3}
\end{figure}

Our monolithic specimens are simulated in Abaqus using two-dimensional plane-strain quadrilateral elements (S8R), as to create an ensemble of continuously connected panels. The geometry of our model is shown in Fig.\;\ref{fig:Num_3}A, where we also show a detail of the structured mesh at one of the elastic hinges.
The mesh is generated by adding lines separating the panels at each hinge, and by ``seeding'' those lines with a desired number of elements (here chosen to be 10 after a mesh convergence study). Boundary conditions (a pin at the left and a horizontal roller at the right, plus a concentrated displacement $d$ at the right roller) are then applied to the central nodes of the two hinges marked in Fig.\;\ref{fig:Num_3}A. The analysis is carried out in displacement control mode, allowing for geometric nonlinearity. The displacement-control mode is needed to capture the negative stiffness parts of the force-displacement curve. Due to the high strain expected in the elastic hinge regions, we also consider a nonlinear material model and fit an Ogden potential of order 3 to the experimental stress-strain curve shown in Fig.~\ref{fig:Num_3}B, obtained by testing a 3D printed dogbone specimen made of TPU-95A. We then run a displacement-control, static, geometrically-nonlinear analysis using a 0.0002 stabilization magnitude and an adaptive damping ratio of 0.05.

A comparison between numerical and experimental results for the theoretically-bistable patterns with target energy barriers 0.001, 0.0015 and 0.003 is shown in Fig.~\ref{fig:Num_3}C. We immediately notice that the numerical curves present lower forces with respect to the experimental ones. We ascribe this discrepancy to the material properties we used in our numerical model. In fact, the slope of the curve in the linear regime of Fig.~\ref{fig:Num_3}B amounts to $\approx 25$\,MPa, which is much lower than the $67$\,MPa given in Ultimaker material data sheets, possibly due to filament degradation throughout our experimental campaign. Despite this discrepancy, we notice that the numerical model captures the experimental trends quite well: the specimen with 0.003 barrier is the only clearly-bistable one, while the 0.0015 is borderline bistable and the 0.001 one is not bistable at all. Another sign of the accuracy of our simulations is the fact that the displacement corresponding to the second stable state, for the specimens that display bistability, matches the theoretical value, highlighted by vertical, color-coded dashed lines in Fig.~\ref{fig:Num_3}C. Please note that a one-to-one comparison between the experimental curves to the numerical ones is not appropriate, due to the complex boundary conditions imposed by our fixtures and reflected in our experimental results. Finally, we also observe that the numerical curves display sudden vertical jumps in the low stiffness region; this is most likely due to localized buckling mechanics that get smoothened out in our experiments due to viscoelasticity.

\subsection{Numerical model for bi-material specimens}
\begin{figure}[!htb]
\centering
\includegraphics[scale=1.0]{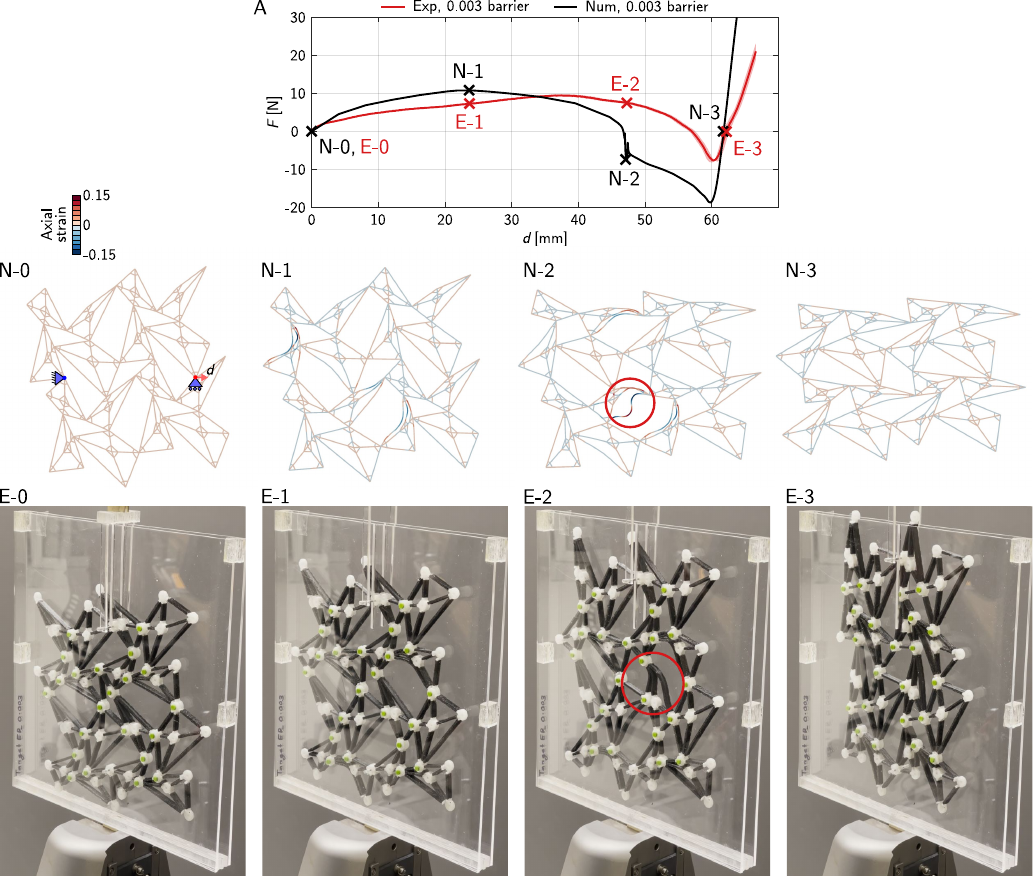}
\caption{Numerical modeling of bi-material specimens. (A) Comparison between the experimental and numerical results in terms of force-displacement curves, for the sample with a 0.003 energy barrier. (N-0)-(N-3) Numerical snapshots of the deformation at a few key points marked in (A). (E-0)-(E-3) Experimental snapshots of the deformation at a few key points marked in (A). The circled regions of the snapshots are intended to highlight discrepancies between numerical and experimental results.}
\label{fig:Num_1}
\end{figure}

The bi-material structures are simulated in Abaqus using beam elements, to replicate the skeletal structure of the panels shown in Fig.~\ref{fig:fabrication_1} at a reduced computational cost. The geometry of the model for the 0.003 barrier pattern is shown in Fig.\;\ref{fig:Num_1}N-0, together with the boundary conditions and an arrow indicating the direction of the applied displacement $d$. 
To capture the fact that hinge regions are stiff, we place cross bars at each hinge location. We define linear elastic materials TPU-95 (Young's Modulus = $67$ MPa, Poisson's Ratio = 0.37) and Nylon (Young's Modulus = $2336$ MPa, Poisson's Ratio = 0.4) and assign appropriate sections to the bars and hinges, respectively. Note that the material properties are taken from Ultimaker data sheets. We assign connector sections with properties join and rotation at each hinge location to simulate the in-plane rotation of the structure; to bias the structure towards in-plane rotation, we assign large values to all other rotational and translational stiffnesses at the joints. In particular, the connector rotational stiffnesses are set to D11 = 1, D22 = 1, and D33 =  $1\times10^{-6}$, where D$ii$ is to be interpreted as the rotational stiffness about $\mathbf{e}_i$. We mesh the model using B31H elements and mesh size $5\times10^{-4}$. We then run a displacement-control, static, geometrically-nonlinear analysis using a 0.0005 stabilization magnitude and an adaptive damping ratio of 0.05. 

The simulation results for the structure with 0.003 target energy barrier are presented in Fig.\;\ref{fig:Num_1} and compared with the experimental ones. From the force-displacement curves in Fig.\;\ref{fig:Num_1}A, we can see that results agree well in terms of critical force and in terms of the displacement corresponding to the second stable equilibrium point ($\approx60$\,mm). However, other features of the curves are less similar. To clarify these discrepancies, we compare numerical (Fig.\;\ref{fig:Num_1}N-1-N-3) and experimental deformed shapes (Fig.\;\ref{fig:Num_1}E-1-E-3), at various displacement values labeled in Fig.\;\ref{fig:Num_1}A. In Fig.\;\ref{fig:Num_1}N-1 and E-1, we can see that the beams that buckle are consistent between experiments and numerics, as indicated by the arrows. As we reach a displacement near 47\,mm, we can see that the numerical curve experiences a vertical jump, while the experimental one does not. By looking at Fig.\;\ref{fig:Num_1}E-2 and comparing the deflected shape to the one in N-2, we can see that this jump is associated with the appearance of a higher order buckling mode in the numerics that is not seen in the experiments (see the shape of the beams in the highlighted regions). Beyond this displacement, experiments and numerics follow two different paths but meet again at the second, stress-free equilibrium point (illustrated in Fig.\;\ref{fig:Num_1}N-3 and E-3). These results highlight that the challenge in modeling kirigami structures made of such skeletal panels is that these panels have a rich array of buckling modes, and that understanding which buckling mode to expect is far from simple.

\subsection{Numerical model for mono-material specimens}

\begin{figure}[!htb]
\centering
\includegraphics[scale=1.0]{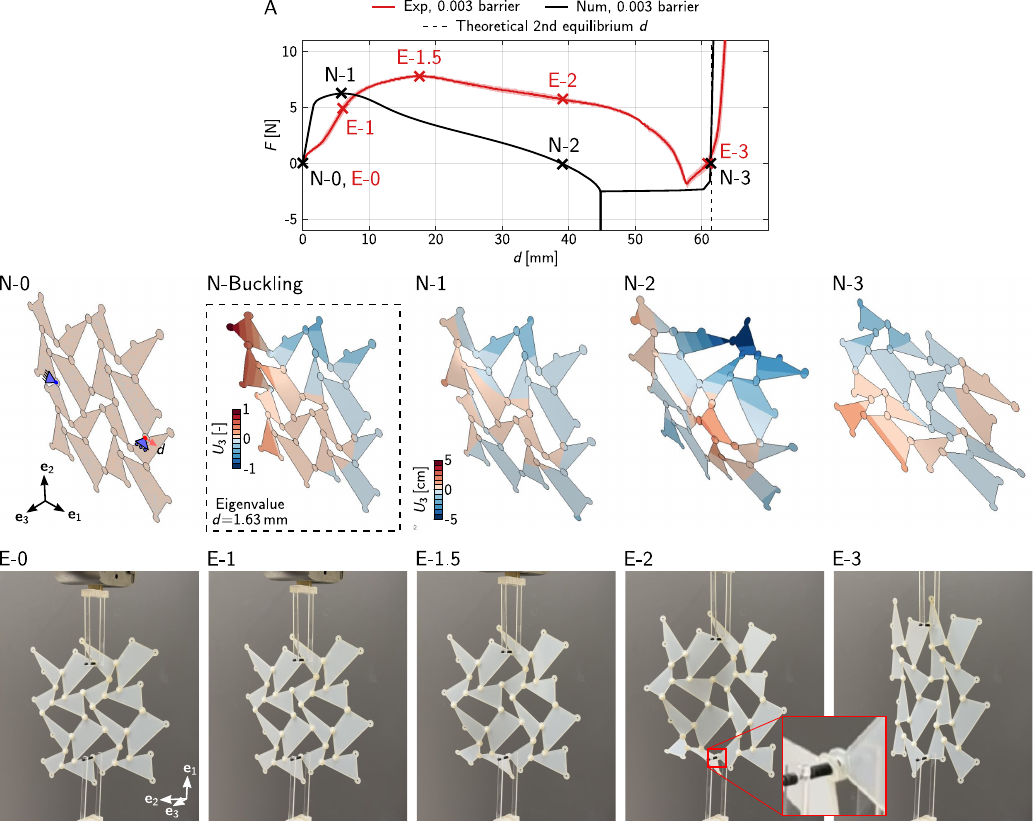}
\caption{Numerical modeling of mono-material specimens. (A) Comparison between the experimental and numerical results in terms of force-displacement curves, for the sample with 0.003 energy barrier. (N-0)-(N-3) Numerical snapshots of the deformation at a few key points marked in (A). (N-Buckling) indicates the first tensile buckling mode, whose eigenvector is used as seed for our post-buckling analysis. (E-0)-(E-3) Experimental snapshots of the deformation at a few key points marked in (A).}
\label{fig:Num_2}
\end{figure}

Finally, the mono-material specimens are also simulated in Abaqus, this time by means of S4R shell elements. The geometry of the model for the 0.003 barrier structure is shown in Fig.\;\ref{fig:Num_2}N-0, together with the boundary conditions and an arrow indicating the direction of the applied displacement $d$. 
Panels are created in Abaqus without the holes for the rivets, while we preserve the circular regions at the panels' corners. Overlapping circles from neighboring panels are connected via connector elements at their central nodes. In particular, we set rotational stiffnesses D11 = 1, D22 = 1 and D33 = $1\times10^{-6}$ for the joints. The material is considered to be linear, with a Young's modulus of $\approx$1 GPa, extracted from our own experiments on PETG dogbone samples. Again, the geometrically nonlinear analysis is performed in displacement-control mode with 0.0002 stabilization magnitude and damping factor stabilization method.

From the experiments, it is clear that mono-material structures buckle out of plane during the transition between stable states, but we do not know \emph{a priori} which shape they are supposed to assume. Thus, we perform an initial linear buckling analysis step and extract the first tensile buckling mode for the specimen, illustrated in Fig.\;\ref{fig:Num_2}N-Buckling. The eigenvector is then provided as seed for the geometrically nonlinear analysis. We choose to only use a single buckling mode since choosing multiple tensile buckling modes did not cause our results to vary significantly, and since we were unsure on the weight to be assigned to each mode.

A comparison between experimental and numerical results for the 0.003 specimen is shown in Fig.\;\ref{fig:Num_2}; in particular, force-displacement curves are shown in Fig.\;\ref{fig:Num_2}A. Numerical and experimental curves show a similar trend, including a similar change in slope in the negative stiffness region. Moreover, numerical and experimental critical forces are similar in magnitude, and the displacements corresponding to the second equilibrium coincide with the theoretical prediction. However, the curves also show some discrepancies, discussed in the following. The initial stiffness of the sample is much larger in the numerical simulation than the experiments. We ascribe this fact to some unavoidable play at the rivets in the experimental sample. The fact that the experimental curve shows higher forces throughout the 10 to 60\,mm region can be ascribed to the presence of significant friction between riveted panels, which is unavoidable as the panels undergo large out-of-plane bending and push on the head and tail of the rivet. 

To provide additional insight, we compare numerical (Fig.\;\ref{fig:Num_2}N-1-N-3) and experimental deformed shapes (Fig.\;\ref{fig:Num_2}E-1-E-3), at various displacement values labeled in Fig.\;\ref{fig:Num_2}A. We can see that these shapes are similar, especially at the large deformation level corresponding to snapshots N-2 and E-2. From the zoom-in in E-2 we can also see that our fixtures allow some relative rotations between panels about the $\mathbf{e}_1$ axis, something that our model does not capture and that could represent an additional source of discrepancies. 

\subsection{Bar-hinge model for heterogeneous kirigami}
\begin{figure}[!htb]
\centering
\includegraphics[scale=1.0]{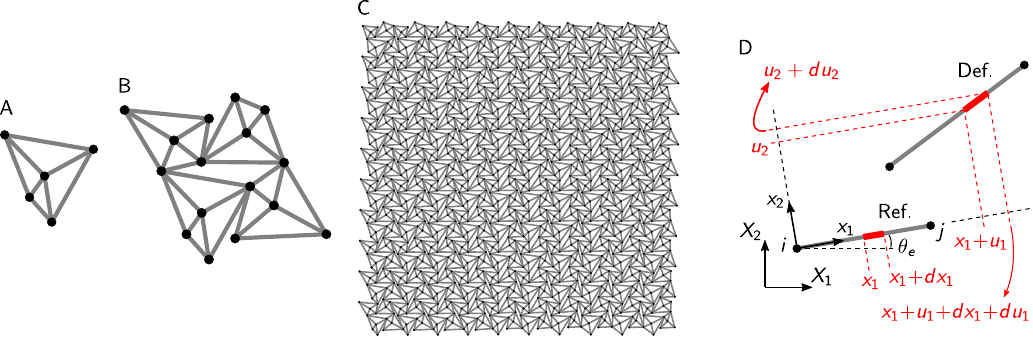}
\caption{Bar-hinge modeling details. (A) A single panel of our kirigami structures, made of 5 hinges and 8 bars. (B) Unit cell and (C) entire heterogeneous ``bowtie'' structure. (D) A single bar in its reference and deformed configurations, with all the geometrical quantities of interest.}
\label{fig:Bar&hinge}
\end{figure}

To verify the bistability of the heterogeneous kirigami designs, we perform  2D simulations of a simplified truss model, also called a "bar-hinge" model,  based on the origami-focused work of Liu and Paulino~\cite{Liu2017}. We choose this reduced-order model since heterogeneous designs comprise a large number of unit cells  for which   Abaqus simulations  would be highly computationally intensive and impractical. Our truss model for kirigami treats each panel as an assembly of pin-jointed bars that can only undergo tension, compression and finite rotations and translations. One such panel, featuring 5 nodes and 8 bars, is illustrated in Fig.\;\ref{fig:Bar&hinge}A; an assembly of panels that form a  unit cell is shown in Fig.\;\ref{fig:Bar&hinge}B, while an entire structure is shown in Fig.\;\ref{fig:Bar&hinge}C. It is important to note that, unlike classical bar-hinge models, ours does not feature any rotational stiffness at the hinges and is only appropriate for structures that do not feature pure mechanism modes of deformation. 

Our model is based on bar elements whose matrices are derived from a Green strain-based formulation. In the following, we show how to obtain formulas for the tangent stiffness matrix and internal force vectors used in our codes, drawing inspiration from the book of Crisfield [MA Crisfield, Non-linear finite element analysis of solids and structures, volume 1: Essentials (1991)] and the work of Liu and Paulino~\cite{Liu2017}. Note that the notation used in this section is independent from the rest of the article.

We consider a single bar element of initial length $L_e$ and cross sectional area $A_e$ in its reference and deformed configurations, as illustrated in Fig.~\ref{fig:Bar&hinge}D. Its nodes are labeled $i$ and $j$. In the figure, we define both a global ($X_i$ with $i=1,2$) and a local coordinate system ($x_i$ with $i=1,2$), with the latter being aligned with the element itself. The virtual internal work for the element can be written as
\begin{equation}
\delta W_{\text{int}\,e}=\int_{V_e}S_e\delta E_e\,dV,
\label{eq:bh_en1}
\end{equation}
where $E_e$ is the Green-Lagrange strain for the bar element, and $S_e$ is its work conjugate, the second Piola-Kirchhoff stress; both quantities are scalars in these one-dimensional elements. The $\delta$ symbol indicates virtual quantities. $V_e$ is the bar volume in its reference configuration. Leveraging the one-dimensional nature of the element, we can rewrite the energy as
\begin{equation}
\delta W_{\text{int}\,e}=\int_{0}^{L_e} A_e S_e\delta E_e\,dx_1.
\label{eq:bh_en2}
\end{equation}

Differential expressions for $E_e$ and $\delta E_e$ can be found by considering an infinitesimal slice of bar and by tracking its elongation in the local coordinate system. In particular, calling $\ell_0$ the initial length of the slice and $\ell$ its length in the deformed configuration, and by replacing these lengths with the coordinates indicated in Fig.~\ref{fig:Bar&hinge}D, it follows that
\begin{equation}
E_e=\frac{\ell^2-\ell_0^2}{2\ell_0^2}=\frac{du_1}{dx_1}+\frac{1}{2}\left(\frac{du_1}{dx_1}\right)^2+\frac{1}{2}\left(\frac{du_2}{dx_1}\right)^2.
\label{eq:bh_gl1} 
\end{equation}
In turn, the virtual work can be computed by taking the first variation of this quantity. Recalling that $E_e=E_e(u_1,u_2)$, this first variation can be computed by replacing $u_1$ and $u_2$ with $u_1+\alpha \delta u_1$ and $u_2+\alpha \delta u_2$ in Eq.~\ref{eq:bh_gl1}, where $\alpha$ is a constant, by taking a derivative with respect to said $\alpha$ and by finally setting $\alpha=0$:
\begin{equation}
\delta E_e=\left[\frac{dE_e\left( u_1+\alpha \delta u_1,u_2+\alpha \delta u_2 \right)}{d \alpha}\right]_{\alpha=0}.
\label{eq:bh_glvar1} 
\end{equation}
This operation yields
\begin{equation}
\delta E_e=\frac{d\delta u_1}{dx_1}+\frac{du_1}{dx_1}\frac{d\delta u_1}{dx_1}+\frac{du_2}{dx_1}\frac{d\delta u_2}{dx_1}.
\label{eq:bh_glvar} 
\end{equation}

Prior to substitution in Eq.~\ref{eq:bh_en2}, we introduce a finite element discretization with linear shape functions, such that displacement variables are written as
\begin{equation}
u_1=\left[ \begin{matrix} 1-\frac{x_1}{L_e} & \frac{x_1}{L_e} \end{matrix} \right]\left[ \begin{matrix} u_{1i} \\ u_{1j} \end{matrix} \right]=\mathbf{N}\,\mathbf{u}_{1e}, \quad
u_2=\left[ \begin{matrix} 1-\frac{x_1}{L_e} & \frac{x_1}{L_e} \end{matrix} \right]\left[ \begin{matrix} u_{2i} \\ u_{2j} \end{matrix} \right]=\mathbf{N}\,\mathbf{u}_{2e},
\end{equation}
where $\mathbf{N}$ is a matrix of shape functions, and their derivatives are written as 
\begin{equation}
\frac{du_1}{dx_1}=\left[ \begin{matrix} -\frac{1}{L_e} & \frac{1}{L_e} \end{matrix} \right]\left[ \begin{matrix} u_{1i} \\ u_{1j} \end{matrix} \right]=\mathbf{B}\,\mathbf{u}_{1e}, \quad \frac{du_2}{dx_1}=\left[ \begin{matrix} -\frac{1}{L_e} & \frac{1}{L_e} \end{matrix} \right]\left[ \begin{matrix} u_{2i} \\ u_{2j} \end{matrix} \right]=\mathbf{B}\,\mathbf{u}_{2e},
\end{equation}
where $\mathbf{B}$ contains derivatives fo the shape functions. We can then combine these expressions into compact ones capturing all displacements and their derivatives:
\begin{equation}
\mathbf{u}=\left[ \begin{matrix} u_1 \\ u_2 \end{matrix}\right]= \left[ \begin{matrix} 1-\frac{x_1}{L_e} & 0 & \frac{x_1}{L_e} & 0\\0 & 1-\frac{x_1}{L_e} & 0 & \frac{x_1}{L_e} \end{matrix} \right] \left[ \begin{matrix} u_{1i} \\ u_{2i} \\ u_{1j} \\ u_{2j} \end{matrix} \right] =\mathbf{N}_e \mathbf{u}_e,
\end{equation}
\begin{equation}
\frac{d\mathbf{u}}{dx_1}=\left[ \begin{matrix} \frac{du_1}{dx_1} \\ \frac{du_2}{dx_1} \end{matrix}\right]= \frac{1}{L_e}\left[ \begin{matrix} -1 & 0 & 1 & 0\\0 & -1 & 0 & 1 \end{matrix} \right] \left[ \begin{matrix} u_{1i} \\ u_{2i} \\ u_{1j} \\ u_{2j} \end{matrix} \right] =\mathbf{B}_e \mathbf{u}_e,
\end{equation}
where $\mathbf{N}_e$ and $\mathbf{B}_e$ capture shape functions and their derivatives for all discrete degrees of freedom in a bar. We also apply the same discretization to virtual quantities. For convenience, we also define the derivative of $u_1$ as a function of the vector of all nodal displacements:
\begin{equation}
\frac{du_1}{dx_1}=\frac{1}{L_e}\left[ \begin{matrix} -1 & 0 & 1 & 0 \end{matrix} \right] \left[ \begin{matrix} u_{1i} \\ u_{2i} \\ u_{1j} \\ u_{2j} \end{matrix} \right] = \mathbf{B}_1 \mathbf{u}_e.
\end{equation}

We can now substitute these expressions into the definition of strain, such that
\begin{equation}
E_e=\frac{du_1}{dx_1}+\frac{1}{2}\left(\frac{du_1}{dx_1}\right)^2+\frac{1}{2}\left(\frac{du_2}{dx_1}\right)^2=\frac{du_1}{dx_1}+\frac{1}{2}\left( \frac{d\mathbf{u}}{dx_1} \right)^{T}\frac{d\mathbf{u}}{dx_1}=\mathbf{B}_1 \mathbf{u}_e+\frac{1}{2} \mathbf{u}_e^T \mathbf{B}_e^T \mathbf{B}_e \mathbf{u}_e.
\end{equation}
In Liu and Paulino, the matrix in the second term is compactly defined as $\mathbf{B}_2=\mathbf{B}_e^T \mathbf{B}_e$, such that 
\begin{equation}
{E}_e = \mathbf{B}_1\mathbf{u}_e + \frac{1}{2}\mathbf{u}_e^T\mathbf{B}_2\mathbf{u}_e.
\label{eq:bh_gld}
\end{equation}
Similarly, the virtual strain can be discretized as follows:
\begin{equation}
\delta E_e=\frac{d\delta u_1}{dx_1}+\frac{du_1}{dx_1}\frac{d\delta u_1}{dx_1}+\frac{du_2}{dx_1}\frac{d\delta u_2}{dx_1}= \delta\mathbf{u}_e^T\mathbf{B}_1^T+ \delta\mathbf{u}_e^T \mathbf{B}_2 \mathbf{u}_e,
\label{eq:bh_gldvar}
\end{equation}
where the first term of the last expression has been transposed for convenience.

Plugging Eq.~\ref{eq:bh_gldvar} into Eq.~\ref{eq:bh_en2}, and by noticing that discrete variables don't depend on the axial coordinate, we obtain 
\begin{equation}
\delta W_{\text{int}\,e}=\int_{0}^{L_e} A_e S_e \delta\mathbf{u}_e^T \left(\mathbf{B}_1^T + \mathbf{B}_2 \mathbf{u}_e \right) \,dx_1 = \delta\mathbf{u}_e^T \, A_e S_e L_e \left(\mathbf{B}_1^T + \mathbf{B}_2 \mathbf{u}_e \right),
\label{eq:bh_en3}
\end{equation}
where we can identify the internal force vector for the element in the local coordinate system:
\begin{equation}
\mathbf{f}_{\text{int}\,e} = A_e L_e S_e \left(\mathbf{B}_1^T + \mathbf{B}_2 \mathbf{u}_e \right).
\label{eq:bh_fint}
\end{equation}

The definition of tangent stiffness matrix comes from the linearization of the nonlinear equilibrium equation with Newton-Raphson or any other root finding algorithm, and can be found by taking derivatives of the internal force vector. For a single element, in the local coordinate system, we can write
\begin{equation}
\mathbf{k}_{te}=\frac{d\mathbf{f}_{\text{int}\,e}}{d\mathbf{u}_e}=A_e L_e \left( \frac{d S_e}{d\mathbf{u}_e} \left(\mathbf{B}_1^T + \mathbf{B}_2 \mathbf{u}_e \right)^T +S_e \frac{d \left(\mathbf{B}_1^T + \mathbf{B}_2 \mathbf{u}_e \right)}{d\mathbf{u}_e}  \right).
\label{eq:bh_kt1}
\end{equation}
Completing this differentiation requires the definition of a material model. In our case, we use a St.\ Venant-Kirchhoff material (a linear material model in a nonlinear mechanics framework), such that $S_e=C E_e$ and $\frac{d S_e}{d\mathbf{u}_e}=C\frac{d E_e}{d\mathbf{u}_e}$. Differentiating Eq.~\ref{eq:bh_gld}, substituting it in the stress expression and computing the derivatives in Eq.~\ref{eq:bh_kt1}, yields 
\begin{equation}
\mathbf{k}_{te}=CA_eL_e\mathbf{B}_1^T\mathbf{B}_1+CA_eL_e\left( \mathbf{B}_1^T\left( \mathbf{B}_2 \mathbf{u}_e \right)^T+ \left( \mathbf{B}_2 \mathbf{u}_e \right)\mathbf{B}_1 \right) + {C}{A}_e{L}_e(\mathbf{B}_2\mathbf{u}_e)(\mathbf{B}_2\mathbf{u}_e)^T + {S}_e{A}_e{L}_e\mathbf{B}_2.
\label{eq:bh_kt}
\end{equation}

The definitions of internal force vector and tangent stiffness matrix in Eq.~\ref{eq:bh_fint} and \ref{eq:bh_kt} still refer to the local coordinate system. In order to appropriately assemble vectors and matrices for the whole structure, we first need to express such quantities in a global coordinate system. To perform this conversion, we rotate the local displacement vector $\mathbf{u}_e$ for the whole element as to obtain its counterpart in the global coordinate system, $\mathbf{U}_e$. Calling $\theta_e$ the angle between the local and global coordinate systems, we can write
\begin{equation}
\mathbf{u}_e= \left[ \begin{matrix} u_{1i} \\ u_{2i} \\ u_{1j} \\ u_{2j} \end{matrix} \right] = \left[ \begin{matrix} \cos{\theta_e} & \sin{\theta_e} & 0 & 0 \\ -\sin{\theta_e} & \cos{\theta_e} & 0 & 0 \\ 0 & 0 & \cos{\theta_e} & \sin{\theta_e} \\ 0 & 0 & -\sin{\theta_e} & \cos{\theta_e} \end{matrix} \right] \left[ \begin{matrix} U_{1i} \\ U_{2i} \\ U_{1j} \\ U_{2j} \end{matrix} \right] =\mathbf{T}_e\mathbf{U}_e.
\end{equation}
By making this substitution in Eq.~\ref{eq:bh_en3} and by carrying out the same computations shown above, we derive
\begin{equation}
\mathbf{F}_{\text{int}\,e} = A_e L_e S_e \left(\left(\mathbf{B}_1 \mathbf{T}_e \right)^T + \mathbf{T}_e^T \mathbf{B}_2 \mathbf{T}_e \mathbf{u}_e \right) = A_e L_e S_e \left(\left(\mathbf{B}_1 \mathbf{T}_e \right)^T + \mathbf{B}_2 \mathbf{u}_e \right),
\label{eq:bh_fintg}
\end{equation}
and
\begin{equation}
\mathbf{K}_{te}=CA_eL_e\left(\mathbf{B}_1 \mathbf{T}_e \right)^T\left(\mathbf{B}_1 \mathbf{T}_e \right)+CA_eL_e\left( \left(\mathbf{B}_1 \mathbf{T}_e \right)^T\left( \mathbf{B}_2 \mathbf{u}_e \right)^T+ \left( \mathbf{B}_2 \mathbf{u}_e \right)\left(\mathbf{B}_1 \mathbf{T}_e \right) \right) + {C}{A}_e{L}_e(\mathbf{B}_2\mathbf{u}_e)(\mathbf{B}_2\mathbf{u}_e)^T + {S}_e{A}_e{L}_e\mathbf{B}_2.
\label{eq:bh_ktg}
\end{equation}
In Liu and Paulino, $\mathbf{B}_1 \mathbf{T}_e$ is defined as $\tilde{\mathbf{B}}_1$ and is written as an explicit function of the position vectors of the nodes of the element:
\begin{equation}
\tilde{\mathbf{B}}_1=\frac{1}{L_e^2}\left[ \begin{matrix} -\left( X_{1j}-X_{1i} \right) & -\left( X_{2j}-X_{2i} \right) & X_{1j}-X_{1i} & X_{2j}-X_{2i} \end{matrix} \right].
\end{equation}

By assembling the internal force vectors $\mathbf{F}_{\text{int}\,e}$ and tangent stiffness matrices $\mathbf{K}_{te}$ for each element, we form the total internal force vector $\mathbf{F}_{\text{int}}$ and tangent stiffness matrix $\mathbf{K}_{t}$ for the whole structure. After imposing the required boundary conditions, we use a full Newton-Raphson solution technique to obtain the force-displacement curve of the structure. This force-controlled procedure, chosen over arc length procedures to avoid the convergence issues we expect in such complex-shaped bistable structures, does not allow to trace negative-sloped portions of the force-displacement curve. Thus, to identify the presence of a second stable equilibrium configuration, we load the specimen beyond the critical load and, when a desired force value is reached, we apply an incremental negative force to simulate unloading and verify whether the unloading curve crosses the zero-force axis at a point different from the origin -- which we interpret as a sign of bistability.

\end{document}